\documentclass[reprint,onecolumn,notitlepage,showpacs,preprintnumbers,nofootinbib,amsmath,amssymb,aps,prd,floatfix]{revtex4-1}
\pdfoutput=1
\usepackage{graphicx}
\usepackage{bm}
\usepackage{subfigure}
\usepackage{url}
\usepackage[hyperindex]{hyperref}
\usepackage{color}
\usepackage[ddmmyy,24hr]{datetime}
\usepackage{bigdelim}
\usepackage{booktabs}
\usepackage{dcolumn}
\usepackage{multirow}
\usepackage{subfigure}
\usepackage{cancel}
\usepackage{stackrel}
\usepackage{paralist}
\usepackage{xspace}
\usepackage{slashed}
\newcommand{\nua}[1]{\ensuremath{\rlap{\kern-2.5pt\ensuremath{\overset{\scriptscriptstyle(-)}{\phantom{\nu}}}}{\ensuremath{{\nu}_{#1}}}}}
\newcommand{\vet}[1]{\ensuremath{\hskip-1pt\vec{\hskip1pt#1}}}
\usepackage{enumitem}
\newcommand{\bluechange}[1]{{#1}}%{{\color{blue} #1}}
\definecolor{darkgreen}{rgb}{0,0.5,0}
\newcommand{\greenchange}[1]{{#1}}%{{\color{darkgreen} #1}}
\definecolor{cyan}{rgb}{0,0.5,0.5}
\newcommand{\cyanchange}[1]{{#1}}%{{\color{cyan} #1}}

\begin{document}

\title{General COHERENT Constraints on Neutrino Non-Standard Interactions}

\author{C. Giunti}
\email{carlo.giunti@to.infn.it}
\affiliation{Istituto Nazionale di Fisica Nucleare (INFN), Sezione di Torino, Via P. Giuria 1, I--10125 Torino, Italy}

%\date{\dayofweekname{\day}{\month}{\year} \ddmmyydate\today, \currenttime}
\date{5 February 2020}

\begin{abstract}
We present the results of a systematic study of the
constraints on neutrino neutral-current non-standard interactions (NSI)
that can be obtained from the
analysis of the COHERENT spectral and temporal data.
First, we consider the general case in which all the ten relevant
neutral-current NSI parameters are considered as free.
We show that they are very weakly constrained by the COHERENT data
because of possible cancellations between the up and down quark contributions.
However,
the up-down average parameters are relatively well constrained
and the strongest constraints are obtained for an
appropriate linear combination of up and down NSI parameters.
We also consider the case in which there are only NSI with either up or down quarks,
and we show that the LMA-Dark fit of solar neutrino data is excluded at
$5.6\sigma$
and
$7.2\sigma$
respectively,
for NSI with up and down quark.
We finally present the tight constraints that can be obtained on
each NSI parameter if it is the dominant one,
assuming that the effects of the others are negligible.
\end{abstract}

%\pacs{}

\maketitle

\section{Introduction}
\label{sec:introduction}

Coherent elastic neutrino-nucleus scattering
(CE$\nu$NS)
has been observed recently for the first time in the COHERENT experiment~\cite{Akimov:2017ade},
many years after its prediction~\cite{Freedman:1973yd,Freedman:1977xn,Drukier:1983gj}.
Several analyses of the COHERENT data provided interesting information on
nuclear physics~\cite{Cadeddu:2017etk,Papoulias:2019lfi,Huang:2019ene,Papoulias:2019txv,Khan:2019mju,Cadeddu:2019eta},
neutrino properties and interactions~\cite{Coloma:2017ncl,Liao:2017uzy,Kosmas:2017tsq,Denton:2018xmq,AristizabalSierra:2018eqm,Cadeddu:2018dux,Esteban:2018ppq,Papoulias:2019txv,Khan:2019mju,Cadeddu:2019eta},
weak interactions~\cite{Cadeddu:2018izq,Huang:2019ene,Papoulias:2019txv,Khan:2019mju,Cadeddu:2019eta}, and
physics beyond the Standard Model~\cite{Dutta:2019eml,Dutta:2019nbn}.
In particular, several authors constrained the parameters of
neutral-current neutrino non-standard interactions
(NSI)~\cite{Coloma:2017ncl,Liao:2017uzy,Kosmas:2017tsq,Denton:2018xmq,Esteban:2018ppq,Papoulias:2019txv,Khan:2019mju}.
However,
in those studies the constraints have been derived by considering only one or two of the
NSI parameters as non-vanishing~\cite{Liao:2017uzy,Kosmas:2017tsq,Denton:2018xmq,Papoulias:2019txv,Khan:2019mju},
or by considering only NSI interactions with either up or down quarks~\cite{Coloma:2017ncl},
\bluechange{%
%\marginnote{\bf\color{red}$\leftarrow$B1}%
or assuming that the ratio of NSI interactions with up and down quarks
is independent of the neutrino flavor~\cite{Esteban:2018ppq}.%
}
In this paper we present the general COHERENT constraints
on the relevant NSI parameters
obtained with a fit of the COHERENT data
in which all the NSI parameters are considered as free.

Our calculations implement the improved quenching factor in Ref.~\cite{Collar:2019ihs}\footnote{%
\bluechange{%
%\marginnote{\bf\color{red}$\leftarrow$A1}%
The quenching factor in Ref.~\cite{Collar:2019ihs} is, however,
not supported by the COHERENT collaboration
[private communication received after the completion of this work].%
}
}.
and use both the spectral and temporal information
given in the COHERENT data release~\cite{Akimov:2018vzs}.
In particular, as already shown in Refs.~\cite{Cadeddu:2018dux,Cadeddu:2019eta},
the combined spectral and temporal information of the COHERENT data
allows us a better determination of the different interactions
of $\nu_{e}$ and $\nu_{\mu}$
than the spectral data alone, which are used in some analyses,
or the total number of event data,
which is used in the simplest analyses.
This is due to the fact that
in the Oak Ridge Spallation Neutron Source
muon neutrinos are produced
from $\pi^+$ decays at rest
($\pi^+\to \mu^+ + \nu_\mu$)
and arrive at the COHERENT detector as a prompt monochromatic signal
with energy
$ ( m_{\pi}^2 - m_{\mu}^2 ) / 2 m_{\pi} \simeq 29.8 \, \text{MeV} $,
within about
$1.5 \, \mu\text{s}$ after protons-on-target.
On the other hand,
muon antineutrinos and electron neutrinos
are produced by $\mu^{+}$ decays at rest
($\mu^{+} \to e^{+} + \nu_{e} + \bar\nu_{\mu}$)
and arrive at the detector with continuous spectra extending up to
$( m_{\mu} - m_{e} ) / 2 \simeq 52.8 \, \text{MeV}$
in a longer time interval of about
$10 \, \mu\text{s}$ after protons-on-target.

Since previous studies that obtained the constraints on NSI
assuming interactions with either up or down quarks
did not consider the complete spectral and temporal information of the COHERENT data
and used the old quenching factor in Ref.~\cite{Akimov:2017ade},
we present also the updated values of these constraints.
This is particularly interesting for testing the
LMA-Dark~\cite{Miranda:2004nb} fit of solar neutrino data
under the assumption of NSI interactions with either up or down quarks~\cite{Coloma:2017ncl}.
\bluechange{%
%\marginnote{\bf\color{red}$\leftarrow$B1}%
The study in Ref.~\cite{Esteban:2018ppq} disfavored the LMA-Dark fit of solar neutrino data
for a wide range of linear combinations of NSI interactions with up or down quarks
assuming that the ratio is independent of the neutrino flavor
and combining the COHERENT fit with constraints from a global analysis of neutrino oscillation data.
However,
since there is no information in Ref.~\cite{Esteban:2018ppq} on the fit of the COHERENT data alone,
we cannot compare the results.%
}

Finally, we also present the constraints on each
individual NSI parameter considered as the only non-vanishing one.
This is the simplest approach for the analysis of the data and has been adopted
by some authors.
Although it is a very special case,
it is physically possible if for some reason
one of the NSI parameters is much larger than the other ones,
whose effects are negligible in the analysis of the COHERENT data.

The plan of the paper is as follows.
In Section~\ref{sec:nsi}
we review the contribution of NSI to
coherent neutrino-nucleus elastic scattering,
setting our conventions and notation.
\bluechange{%
In Section~\ref{sec:analysis}
we describe our method of analysis of the COHERENT data.%
}
In Section~\ref{sec:general}
we present the general constraints on NSI from the COHERENT data.
In Section~\ref{sec:ud}
we discuss the constraints on NSI from the COHERENT data
assuming only interactions with either up or down quarks.
In Section~\ref{sec:dominant}
we present the constraints on each
individual NSI parameter considered as the only effectively non-vanishing one.
At the end,
in Section~\ref{sec:conclusions}
we summarize our results.

\section{NSI in CE$\nu$NS}
\label{sec:nsi}

We consider vector neutral-current neutrino non-standard interactions
generated by a heavy mediator and described by the effective four-fermion
interaction Lagrangian
(see the reviews in Refs.~\cite{Ohlsson:2012kf,Miranda:2015dra,Farzan:2017xzy,Dev:2019anc})\footnote{%
\bluechange{%
In general,
neutral-current neutrino non-standard interactions can have also
axial components,
but their effect is negligible in coherent elastic scattering of neutrinos with heavy
nuclei~\cite{Barranco:2005yy}.
}
}
\begin{equation}
\mathcal{L}_{\text{NSI}}^{\text{NC}}
=
- 2 \sqrt{2} G_{\text{F}}
\sum_{\alpha,\beta=e,\mu,\tau}
\left( \overline{\nu_{\alpha L}} \gamma^{\rho} \nu_{\beta L} \right)
\sum_{f=u,d}
\varepsilon_{\alpha\beta}^{fV}
\left( \overline{f} \gamma_{\rho} f \right)
,
\label{lagrangian}
\end{equation}
where $G_{\text{F}}$ is the Fermi constant.
The parameters
$\varepsilon_{\alpha\beta}^{fV}$
describe the size of non-standard interactions relative to standard neutral-current weak interactions.
From the hermiticity of the Lagrangian,
we have
$\varepsilon_{\alpha\beta}^{fV} = \varepsilon_{\beta\alpha}^{fV*}$.
\bluechange{%
%\marginnote{\bf\color{red}$\leftarrow$B9}%
As explained in Ref.~\cite{Esteban:2018ppq},
the mediator of the NSI must be heavier than about 10 MeV.%
}

\bluechange{%
%\marginnote{\bf\color{red}$\leftarrow$A3}%
The strongest constraints on
vector neutral-current neutrino non-standard interactions
have been obtained from the analysis of the data
of solar and atmospheric neutrino oscillation experiments
through their contribution to the matter effect
(see the reviews in Refs.~\cite{Ohlsson:2012kf,Miranda:2015dra,Farzan:2017xzy,Dev:2019anc}
and the recent results in
Refs.~\cite{Coloma:2017ncl,Esteban:2018ppq}).
Here we discuss only the constraints on
vector neutral-current NSI that can be obtained from the
analysis of the COHERENT CE$\nu$NS data.
}

The differential cross section
for coherent elastic scattering of a $\nu_{\alpha}$ with energy $E$
and a nucleus $\mathcal{N}$
with $Z$ protons, $N$ neutrons, and mass $M$, is given by
(see Ref.~\cite{Barranco:2005yy})
\begin{equation}
\dfrac{d\sigma_{\nu_{\alpha}\text{-}\mathcal{N}}}{d T}
(E,T)
=
\dfrac{G_{\text{F}}^2 M}{\pi}
\left(
1 - \dfrac{M T}{2 E^2}
\right)
Q_{\alpha}^2
,
\label{cs}
\end{equation}
where $T$ is the nuclear recoil kinetic energy and
\begin{align}
Q_{\alpha}^2
=
\null & \null
\left[
\left( g_{V}^{p} + 2 \varepsilon_{\alpha\alpha}^{uV} + \varepsilon_{\alpha\alpha}^{dV} \right)
Z
F_{Z}(|\vet{q}|^2)
+
\left( g_{V}^{n} + \varepsilon_{\alpha\alpha}^{uV} + 2 \varepsilon_{\alpha\alpha}^{dV} \right)
N
F_{N}(|\vet{q}|^2)
\right]^2
\nonumber
\\
\null & \null
+
\sum_{\beta\neq\alpha}
\left|
\left( 2 \varepsilon_{\alpha\beta}^{uV} + \varepsilon_{\alpha\beta}^{dV} \right)
Z
F_{Z}(|\vet{q}|^2)
+
\left( \varepsilon_{\alpha\beta}^{uV} + 2 \varepsilon_{\alpha\beta}^{dV} \right)
N
F_{N}(|\vet{q}|^2)
\right|^2
,
\label{Qalpha2}
\end{align}
with
\begin{equation}
g_{V}^{p}
=
\dfrac{1}{2} - 2 \sin^2\!\vartheta_{W}
,
\qquad
g_{V}^{n}
=
- \dfrac{1}{2}
.
\label{gV}
\end{equation}
Here
$\vartheta_{W}$ is the weak mixing angle,
given by
$\sin^2\!\vartheta_{W} = 0.23857 \pm 0.00005$
at low energies~\cite{Tanabashi:2018oca}.
\greenchange{%
Since for antineutrinos both the weak and the NSI couplings change sign,
antineutrinos have the same cross section as neutrinos.
}

In Eq.~(\ref{Qalpha2}),
$F_{Z}(|\vet{q}|^2)$
and
$F_{N}(|\vet{q}|^2)$
are, respectively, the form factors of the proton and neutron distributions in the nucleus,
that depend on the three-momentum transfer $|\vet{q}| \simeq \sqrt{2 M T}$.
They are given by the Fourier transforms of the nuclear proton and neutron distributions
and
describe the loss of coherence for
$|\vet{q}| R_{p} \gtrsim 1$
and
$|\vet{q}| R_{n} \gtrsim 1$,
where $R_{p}$ and $R_{n}$ are the corresponding rms radii.
It has been shown in Ref.~\cite{Cadeddu:2017etk}
that different parameterizations of the form factors are practically equivalent in the
analysis of COHERENT data.
Therefore,
we consider only the Helm parameterization~\cite{Helm:1956zz}
\begin{equation}
F(|\vet{q}|^2)
=
3
\,
\dfrac{j_{1}(|\vet{q}| R_{0})}{|\vet{q}| R_{0}}
\,
e^{- |\vet{q}|^2 s^2 / 2}
,
\label{ffHelm}
\end{equation}
where
$
j_{1}(x) = \sin(x) / x^2 - \cos(x) / x
$
is the spherical Bessel function of order one,
$s = 0.9 \, \text{fm}$ \cite{Friedrich:1982esq}
is the surface thickness
and $R_{0}$ is related to the rms radius $R$ by
$ R^2 = 3 R_{0}^2 / 5 + 3 s^2 $.
For the rms radii of the proton distributions of
$^{133}\text{Cs}$ and $^{127}\text{I}$
we adopt the values determined with high accuracy from
muonic atom spectroscopy~\cite{Fricke:1995zz}:
\begin{equation}
R_{p}({}^{133}\text{Cs}) = 4.804 \, \text{fm}
,
\qquad
R_{p}({}^{127}\text{I}) = 4.749 \, \text{fm}
.
\label{Rp}
\end{equation}
On the other hand,
there is no separate measurement of the rms radii of the neutron distributions of
$^{133}\text{Cs}$ and $^{127}\text{I}$.
The average neutron rms radius of CsI has been obtained from the COHERENT data
assuming the absence of non-standard effects~\cite{Cadeddu:2017etk,Papoulias:2019lfi,Huang:2019ene,Papoulias:2019txv,Khan:2019mju,Cadeddu:2019eta}.
Taking into account also atomic parity violation (APV) experimental results~\cite{Cadeddu:2018izq,Cadeddu:2019eta},
the most precise determination of the average neutron rms radius of CsI
from experimental data is~\cite{Cadeddu:2019eta}
\begin{equation}
R_{n} = 5.04 \pm 0.31 \, \text{fm}
.
\label{Rn}
\end{equation}
Taking into account the uncertainties,
this value is compatible with the predictions of nuclear models
(see Table~I in Ref.~\cite{Cadeddu:2017etk}).
Since there are already ten NSI parameters to be determined by the analysis of the COHERENT data
\greenchange{%
(see the discussion at the end of Section~\ref{sec:analysis}),%
}
it is practically advantageous to consider fixed values of
the neutron rms radii of
$^{133}\text{Cs}$ and $^{127}\text{I}$,
instead of considering them free as in Refs.~\cite{Cadeddu:2018dux,Cadeddu:2019eta},
where a smaller number of other parameters have been constrained.
Hence,
we adopt the values
\begin{equation}
R_{n}({}^{133}\text{Cs}) = 5.01 \, \text{fm}
,
\qquad
R_{n}({}^{127}\text{I}) = 4.94 \, \text{fm}
,
\label{RnRMF}
\end{equation}
obtained with the relativistic mean field (RMF) NL-Z2 \cite{Bender:1999yt}
nuclear model calculation in Ref.~\cite{Cadeddu:2017etk},
that are in good agreement with the average value (\ref{Rn}).
We take into account the form factor uncertainties with a 5\% contribution to
$\sigma_{\alpha_{\text{c}}}$
in the least-square functions (\ref{chi-spe}) and (\ref{chi-tim}),
following the COHERENT prescription~\cite{Akimov:2017ade}.

One can note that the NSI contributions of up and down quarks can cancel in $Q_{\alpha}^2$.
A total cancellation happens for
\begin{equation}
\left[ 2 + \frac{N}{Z} \, \frac{F_{N}(|\vet{q}|^2)}{F_{Z}(|\vet{q}|^2)} \right]
\varepsilon_{\alpha\beta}^{uV}
+
\left[ 1 + 2 \, \frac{N}{Z} \, \frac{F_{N}(|\vet{q}|^2)}{F_{Z}(|\vet{q}|^2)} \right]
\varepsilon_{\alpha\beta}^{dV}
=
0
.
\label{canc1}
\end{equation}
\bluechange{%
%\marginnote{\bf\color{red}$\leftarrow$B10}%
If the cancellation were exact,
there would be no constraint on the values of the NSI couplings
when they are all considered as free parameters.%
}
However,
in practice only a partial cancellation is possible,
because:
1) the ratio $F_{N}(|\vet{q}|^2)/F_{Z}(|\vet{q}|^2)$ depends on $|\vet{q}|^2$;
2) in the scattering on different nuclei,
as
$^{133}\text{Cs}$
and
$^{127}\text{I}$
in the case of the COHERENT experiment,
$Z$ and $N$ (and the corresponding form factors) are different.
Nevertheless, since the proton and neutron form factors are not very different for a heavy nucleus
and in the COHERENT case of scattering on CsI
$(N/Z)_{^{133}\text{Cs}} \simeq 1.418$
is not very different of
$(N/Z)_{^{127}\text{I}} \simeq 1.396$,
the cancellation can be strong.
This means that in practice coherent elastic neutrino-nucleus scattering
is not sensitive to small values of
the NSI couplings to $u$ and $d$ quarks if they are both considered as free parameters.
As we will see in Section~\ref{sec:general},
there is only a sensitivity to very large values
of the NSI couplings to $u$ and $d$ quarks,
for which the residuals of the cancellations are significant.

Considering
$(N/Z)_{^{133}\text{Cs}} \simeq (N/Z)_{^{127}\text{I}} \simeq 1.4$
and neglecting the form factors,
the cancellation relation~(\ref{canc1}) becomes
\begin{equation}
\varepsilon_{\alpha\beta}^{dV}
\simeq
- \frac{3.4}{3.8} \, \varepsilon_{\alpha\beta}^{uV}
\simeq
- 0.89 \, \varepsilon_{\alpha\beta}^{uV}
.
\label{canc2}
\end{equation}
We will see in Section~\ref{sec:general}
that the allowed regions of the NSI parameters
obtained from COHERENT data
have a slope close to that in Eq.~(\ref{canc2}) and
lie close to the corresponding line.
This means that there is no indication of a significant NSI signal in the COHERENT data,
taking into account the uncertainties.

From Eq.~(\ref{canc2}) one can see that the cancellation between the $u$ and $d$ couplings
occurs when one is almost equal to the opposite of the other.
This means that the COHERENT data are practically not sensitive to small values of the difference
between
$\varepsilon_{\alpha\beta}^{uV}$ and $\varepsilon_{\alpha\beta}^{dV}$,
but can probe small values of linear combinations of
$\varepsilon_{\alpha\beta}^{uV}$ and $\varepsilon_{\alpha\beta}^{dV}$
that are proportional to a value close to their average.
Hence,
in Section~\ref{sec:general}
we present also the constraints on the up-down averages
\begin{equation}
\overline{\varepsilon}_{\alpha\beta}^{V}
=
\frac{1}{2} \left( \varepsilon_{\alpha\beta}^{uV} + \varepsilon_{\alpha\beta}^{dV} \right)
.
\label{ave}
\end{equation}

\section{COHERENT data analysis}
\label{sec:analysis}

We performed two analyses of the COHERENT data:
one of the spectral data only,
and one of the joint spectral and temporal data.
In this way we can evidence the improvements obtained
by adding the temporal information,
that has been previously used only in Refs.~\cite{Cadeddu:2018dux,Denton:2018xmq,Dutta:2019eml,Cadeddu:2019eta}.

For the analysis of the COHERENT spectral data only,
we considered the least-squares function
\begin{equation}
\chi^2_{\text{S}}
=
\sum_{i=4}^{15}
\left(
\dfrac{
N_{i}^{\text{exp}}
-
\left(1+\alpha_{\text{c}}\right) N_{i}^{\text{th}}
-
\left(1+\beta_{\text{c}}\right) B_{i}
}{ \sigma_{i} }
\right)^2
+
\left( \dfrac{\alpha_{\text{c}}}{\sigma_{\alpha_{\text{c}}}} \right)^2
+
\left( \dfrac{\beta_{\text{c}}}{\sigma_{\beta_{\text{c}}}} \right)^2
+
\left( \dfrac{\eta_{\text{c}}-1}{\sigma_{\eta_{\text{c}}}} \right)^2
.
\label{chi-spe}
\end{equation}
Here,
for each energy bin $i$,
$N_{i}^{\text{exp}}$ is the experimental event number,
with statistical uncertainty $\sigma_{i}$,
\bluechange{%
%\marginnote{\bf\color{red}$\leftarrow$B5}%
taken from Fig.~3A of Ref.~\cite{Akimov:2017ade}%
},
$N_{i}^{\text{th}}$
is the theoretical event number
that depends on the NSI parameters through the cross section~(\ref{cs}),
and
$B_{i}$ is the estimated number of background events
\bluechange{%
%\marginnote{\bf\color{red}$\leftarrow$B3 \& B6}%
extracted from Fig.~S13 of Ref.~\cite{Akimov:2017ade}%
}.
\cyanchange{%
We did not consider bin-to-bin correlated systematic uncertainties
that we assume to be negligible,
since they are not mentioned
in the COHERENT data release~\cite{Akimov:2018vzs}.%
}
We considered only the 12 energy bins from $i=4$ to $i=15$
of the COHERENT spectrum, because they cover the recoil kinetic energy of the new
Chicago-3 quenching factor measurement~\cite{Collar:2019ihs},
where the value of the quenching factor and its uncertainties are more reliable.
In Eq.~(\ref{chi-spe}),
$\alpha_{\text{c}}$, $\beta_{\text{c}}$, and $\eta_{\text{c}}$
are nuisance parameters which quantify,
respectively,
the systematic uncertainties
of the signal rate,
of the background rate,
and of the quenching factor,
with
corresponding standard deviations
$\sigma_{\alpha_{\text{c}}} = 0.12$,
$\sigma_{\beta_{\text{c}}} = 0.25$~\cite{Akimov:2017ade}, and
$\sigma_{\eta_{\text{c}}} = 0.05$~\cite{Collar:2019ihs}.
The value of $\sigma_{\alpha_{\text{c}}}$
has been obtained
by summing in quadrature
a 5\% signal acceptance uncertainty,
a 5\% neutron form factor uncertainty,
and
a 10\% neutron flux uncertainty,
estimated by the COHERENT collaboration~\cite{Akimov:2017ade}.
\bluechange{%
%\marginnote{\bf\color{red}$\leftarrow$B2}%
This allows us to treat separately the uncertainty of the quenching factor,
that was included in the uncertainty of $\alpha_{\text{c}}$ in Ref.~\cite{Akimov:2017ade}.
The quenching factor $ f_{\text{Q}}(T)$
is the ratio between the scintillation light emitted in nuclear and electron recoils.
It determines the relation between the number of detected photoelectrons
$N_{\text{PE}}$
and the nuclear recoil kinetic energy $T$:
\begin{equation}
N_{\text{PE}}
=
\eta_{\text{c}} \, f_{\text{Q}}(T) \, Y_{\text{L}} \, T
,
\label{qfc}
\end{equation}
where
$ Y_{\text{L}} = 13.35 \, N_{\text{PE}} / \text{keV} $
is the light yield of the phototubes
and
the function $ f_{\text{Q}}(T)$
is given in Figure~1 of ref.~\cite{Collar:2019ihs}.
The normalization factor $\eta_{\text{c}}$ has the uncertainty
$\sigma_{\eta_{\text{c}}} = 0.05$~\cite{Collar:2019ihs},
that contributes to the least-squares function.
%\marginnote{\bf\color{red}$\leftarrow$B4}%
The relation~(\ref{qfc}) is necessary for the analysis of the COHERENT data,
that are given as number of events in bins of $N_{\text{PE}}$.

The theoretical event number
$N_{i}^{\text{th}}$
in each energy bin $i$ is given by
\begin{equation}
N_{i}^{\text{th}}
=
N_{\text{CsI}}
\int_{T_{i}}^{T_{i+1}} d T
\int_{E_{\text{min}}} d E
\,
A(T)
\,
\frac{d N_{\nu}}{d E}
\,
\dfrac{d\sigma_{\nu\text{-}\text{CsI}}}{d T}
,
\label{Nth}
\end{equation}
where $N_{\text{CsI}}$
is the number of CsI
in the detector
(given by
$ N_{\text{A}} M_{\text{det}} / M_{\text{CsI}}$,
where
$ N_{\text{A}} $
is the Avogadro number,
$ M_{\text{det}} = 14.6 \,\text{kg} $,
is the detector mass, and
$ M_{\text{CsI}} = 259.8 $
is the molar mass of CsI),
$ E_{\text{min}} = \sqrt{M T / 2} $,
%\marginnote{\bf\color{red}$\leftarrow$B6}%
$A(T)$
is the acceptance function given in the COHERENT data release~\cite{Akimov:2018vzs},
$d N_{\nu} / d E$
is the neutrino flux integrated over the experiment lifetime,
and
$d\sigma_{\nu\text{-}\text{CsI}} / d T$
is the sum of the differential cross sections~(\ref{cs})
for $\mathcal{N}=\text{Cs}$ and $\mathcal{N}=\text{I}$.
%\marginnote{\bf\color{red}$\leftarrow$B4}%
In each bin $i$ of detected photoelectron number,
the integration over the kinetic energy $T$
is performed between the boundaries
$T_{i}$ and $T_{i+1}$
that depend on the quenching factor
through Eq.~(\ref{qfc}).
\cyanchange{%
We neglected energy resolution effects due to the fluctuations of photoelectrons,
that would slow considerably the numerical computation.
We verified that these effects are negligible by calculating the corrections
in the case of standard weak interactions.
Indeed, these effects are not mentioned
in the COHERENT data release~\cite{Akimov:2018vzs}.%
}

Neutrinos arriving at the COHERENT detector
from the Oak Ridge Spallation Neutron Source consist of a prompt component of monochromatic
$\nu_\mu$ from stopped pion decays,
$\pi^+\to \mu^++\nu_\mu$,
and two delayed components of
$\bar{\nu}_\mu$ and $\nu_{e}$
from the subsequent muon decays, $\mu^+\to e^+ + \bar{\nu}_\mu + \nu_{e}$.
The total flux $d N_{\nu} / d E$ is the sum of
\begin{align}
\frac{d N_{\nu_{\mu}}}{d E}
=
\null & \null
\zeta
\,
\delta\!\left(
E - \dfrac{ m_{\pi}^2 - m_{\mu}^2 }{ 2 m_{\pi} }
\right)
,
\label{numu}
\\
\frac{d N_{\nu_{\bar\mu}}}{d E}
=
\null & \null
\zeta
\,
\dfrac{ 64 E^2 }{ m_{\mu}^3 }
\left(
\dfrac{3}{4} - \dfrac{E}{m_{\mu}}
\right)
,
\label{numubar}
\\
\frac{d N_{\nu_{e}}}{d E}
=
\null & \null
\zeta
\,
\dfrac{ 192 E^2 }{ m_{\mu}^3 }
\left(
\dfrac{1}{2} - \dfrac{E}{m_{\mu}}
\right)
,
\label{nue}
\end{align}
for
$E \leq m_{\mu} / 2 \simeq 52.8 \, \text{MeV}$,
with the normalization factor
$ \zeta = r N_{\text{POT}} / 4 \pi L^2 $,
where
$r=0.08$ is the number of neutrinos per flavor
that are produced for each proton on target,
$ N_{\text{POT}} = 1.76\times 10^{23} $
is the number of proton on target
and $ L = 19.3 \, \text{m} $
is the distance between the source and the COHERENT CsI detector~\cite{Akimov:2017ade}.%
}

For the analysis of the joint COHERENT spectral and temporal data,
we considered the least-squares function
\begin{align}
\chi^2_{\text{ST}}
=
\null & \null
2
\sum_{i=4}^{15}
\sum_{j=1}^{12}
\left[
\left( 1 + \alpha_{\text{c}} \right) N_{ij}^{\text{th}}
+
\left( 1 + \beta_{\text{c}} \right) B_{ij}
+
\left( 1 + \gamma_{\text{c}} \right) N_{ij}^{\text{bck}}
-
N_{ij}^{\text{C}}
\vphantom{
\ln\!\left(
\frac{ N_{ij}^{\text{C}} }{
\left( 1 + \alpha_{\text{c}} \right) N_{ij}^{\text{th}}
+
\left( 1 + \beta_{\text{c}} \right) B_{ij}
+
\left( 1 + \gamma_{\text{c}} \right) N_{ij}^{\text{bck}}
}
\right)
}
\right.
\nonumber
\\
\null & \null
\hspace{2cm}
\left.
+
N_{ij}^{\text{C}}
\ln\!\left(
\frac{ N_{ij}^{\text{C}} }{
\left( 1 + \alpha_{\text{c}} \right) N_{ij}^{\text{th}}
+
\left( 1 + \beta_{\text{c}} \right) B_{ij}
+
\left( 1 + \gamma_{\text{c}} \right) N_{ij}^{\text{bck}}
}
\right)
\right]
\nonumber
\\
\null & \null
+
\left( \frac{\alpha_{\text{c}}}{\sigma_{\alpha_{\text{c}}}} \right)^2
+
\left( \frac{\beta_{\text{c}}}{\sigma_{\beta_{\text{c}}}} \right)^2
+
\left( \frac{\gamma_{\text{c}}}{\sigma_{\gamma_{\text{c}}}} \right)^2
+
\left( \dfrac{\eta_{\text{c}}-1}{\sigma_{\eta_{\text{c}}}} \right)^2
,
\label{chi-tim}
\end{align}
that allows us to take into account time-energy bins
with few or zero events.
In Eq.~(\ref{chi-tim}),
$i$ is the index of the energy bins,
$j$ is the index of the time bins,
$N_{ij}^{\text{th}}$ are the theoretical predictions that depend on the NSI parameters,
$N_{ij}^{\text{C}}$ are the coincidence (C) data, which contain signal and background events,
$B_{ij}$ are the estimated neutron-induced backgrounds, and
$N_{ij}^{\text{bck}}$ are the estimated backgrounds obtained from the anti-coincidence (AC) data
given in the COHERENT data release~\cite{Akimov:2018vzs}.
The nuisance parameters $\alpha_{\text{c}}$, $\beta_{\text{c}}$, and $\eta_{\text{c}}$
are the same as in the least-square function in Eq.~(\ref{chi-spe}),
that we used in the analysis of the time-integrated COHERENT data.
The additional nuisance parameter $\gamma_{\text{c}}$
and its uncertainty
$\sigma_{\gamma_{\text{c}}} = 0.05$ quantify the systematic uncertainty of the background estimated from the AC data~\cite{Akimov:2017ade,Akimov:2018vzs}.

\greenchange{%
The flavor content of the neutrino flux arriving at the COHERENT detector
from the Oak Ridge Spallation Neutron Source
determines which NSI parameters can be constrained from the data analysis.
Since the flux is composed by $\nu_{e}$, $\nu_{\mu}$, and $\bar\nu_{\mu}$,
we can determine five NSI couplings for each quark:
$\varepsilon_{ee}^{qV}$,
$\varepsilon_{\mu\mu}^{qV}$,
$\varepsilon_{e\mu}^{qV}=\varepsilon_{\mu e}^{qV*}$,
$\varepsilon_{e\tau}^{qV}$, and
$\varepsilon_{\mu\tau}^{qV}$,
for $q=u,d$.
The diagonal NSI parameters are real, but the off-diagonal ones can be complex.
However,
we can write the effective weak charge in Eq.~(\ref{Qalpha2}) as
\begin{align}
Q_{\alpha}^2
=
\null & \null
\left[
\left( g_{V}^{p} + 2 \varepsilon_{\alpha\alpha}^{uV} + \varepsilon_{\alpha\alpha}^{dV} \right)
Z
F_{Z}(|\vet{q}|^2)
+
\left( g_{V}^{n} + \varepsilon_{\alpha\alpha}^{uV} + 2 \varepsilon_{\alpha\alpha}^{dV} \right)
N
F_{N}(|\vet{q}|^2)
\right]^2
\nonumber
\\
\null & \null
+
\sum_{\beta\neq\alpha}
\left[
\text{Re}\!\left(\varepsilon_{\alpha\beta}^{uV}\right)
\left( 2 Z F_{Z}(|\vet{q}|^2) + N F_{N}(|\vet{q}|^2) \right)
+
\text{Re}\!\left(\varepsilon_{\alpha\beta}^{dV}\right)
\left( Z F_{Z}(|\vet{q}|^2) + 2 N F_{N}(|\vet{q}|^2) \right)
\right]^2
\nonumber
\\
\null & \null
+
\sum_{\beta\neq\alpha}
\left[
\text{Im}\!\left(\varepsilon_{\alpha\beta}^{uV}\right)
\left( 2 Z F_{Z}(|\vet{q}|^2) + N F_{N}(|\vet{q}|^2) \right)
+
\text{Im}\!\left(\varepsilon_{\alpha\beta}^{dV}\right)
\left( Z F_{Z}(|\vet{q}|^2) + 2 N F_{N}(|\vet{q}|^2) \right)
\right]^2
.
\label{Qalpha2reim}
\end{align}
Since the real and imaginary parts of the off-diagonal NSI coupling
contribute in the same way,
they cannot be distinguished and we can consider the off-diagonal NSI coupling as real.
The bounds on a possible imaginary part are the same as those on the real part.
Hence, our general analysis depends on ten real NSI parameters:
$\varepsilon_{ee}^{qV}$,
$\varepsilon_{\mu\mu}^{qV}$,
$\varepsilon_{e\mu}^{qV}$,
$\varepsilon_{e\tau}^{qV}$, and
$\varepsilon_{\mu\tau}^{qV}$,
for $q=u,d$.
Moreover,
since the bounds for each off-diagonal NSI coupling with a quark obtained by marginalizing over
the corresponding off-diagonal NSI coupling with the other quark
are independent of its sign,
we will obtain marginal bounds for
$|\varepsilon_{e\mu}^{qV}|$,
$|\varepsilon_{e\tau}^{qV}|$, and
$|\varepsilon_{\mu\tau}^{qV}|$,
with $q=u,d$.%
}

\section{General COHERENT constraints on NSI}
\label{sec:general}

The general marginalized constraints on the NSI parameters
obtained with the analyses of the spectral and the joint spectral and temporal
COHERENT data are listed in Table~\ref{tab:general}.
\greenchange{%
Note that the marginal bounds on the off-diagonal NSI parameters are given only for their absolute values,
as explained at the end of Section~\ref{sec:analysis}.%
}

From Table~\ref{tab:general},
one can see that, as explained in Section~\ref{sec:nsi},
the marginalized bounds on the individual NSI parameters
$\varepsilon_{\alpha\beta}^{uV}$ and $\varepsilon_{\alpha\beta}^{dV}$
are very weak,
because of the possible cancellations of their effects.
On the other hand,
the up-down averages $\overline{\varepsilon}_{\alpha\beta}^{V}$ in Eq.~(\ref{ave})
are relatively well constrained, with upper values smaller or close to unity
for reasonable values of the confidence level.
However,
these constraints are larger than may be expected,
for example from the analysis in Ref.~\cite{Liao:2017uzy}
that found the 90\% CL bounds
$ -0.16 \leq \overline{\varepsilon}_{ee}^{V} \leq 0.33 $
and
$ -0.11 \leq \overline{\varepsilon}_{ee}^{V} \leq 0.26 $.
The reason why the constraints on the
averages $\overline{\varepsilon}_{\alpha\beta}^{V}$
cannot be so tight is that these averages
do not satisfy well the cancellation constraint in Eq.~(\ref{canc2}),
especially when the individual up and down NSI parameters are large and with opposite signs.
Therefore, we consider also the maximally constrained up-down linear combinations
\begin{equation}
\widetilde{\varepsilon}_{\alpha\beta}^{V}
=
\dfrac{3.4 \, \varepsilon_{\alpha\beta}^{uV} + 3.8  \, \varepsilon_{\alpha\beta}^{dV}}{7.2}
.
\label{tilde}
\end{equation}
Table~\ref{tab:general} shows that the constraints
for these maximally constrained up-down linear combinations
are strong.
In the joint COHERENT spectral and temporal data analysis the
absolute values of all the maximally constrained up-down linear combinations of NSI parameters
are smaller than 0.35 at $3\sigma$.
The larger values of the constraints
on the averages $\overline{\varepsilon}_{\alpha\beta}^{V}$
are due to the fact that
$\varepsilon_{\alpha\beta}^{uV}$
and
$\varepsilon_{\alpha\beta}^{dV}$ can be large and opposite
yielding a very small value of
$\widetilde{\varepsilon}_{\alpha\beta}^{V}$
that corresponds to a much larger value of
$\overline{\varepsilon}_{\alpha\beta}^{V}$.
For example,
if we consider $\varepsilon_{ee}^{uV}=10$,
that is allowed within $1\sigma$ by the limits in Table~\ref{tab:general},
we have $\varepsilon_{ee}^{uV}=0.2$, that is allowed within $1\sigma$,
for
$\varepsilon_{ee}^{dV}=-8.6$, that is also allowed within $1\sigma$.
In this case
$\overline{\varepsilon}_{ee}^{V}=0.7$
and this value must be allowed within $1\sigma$,
in agreement with Table~\ref{tab:general} and
contrary to the limits in Ref.~\cite{Liao:2017uzy}.

It is clear from Table~\ref{tab:general}
that the analysis of the joint COHERENT spectral and temporal data
is more powerful in constraining the NSI parameters than the analysis of the spectral data only.
Therefore,
in the rest of this Section we discuss only the results of the joint spectral and temporal data analysis.

\begin{figure*}[!h]
\centering
\begin{tabular}{cc}
\includegraphics*[width=0.49\linewidth]{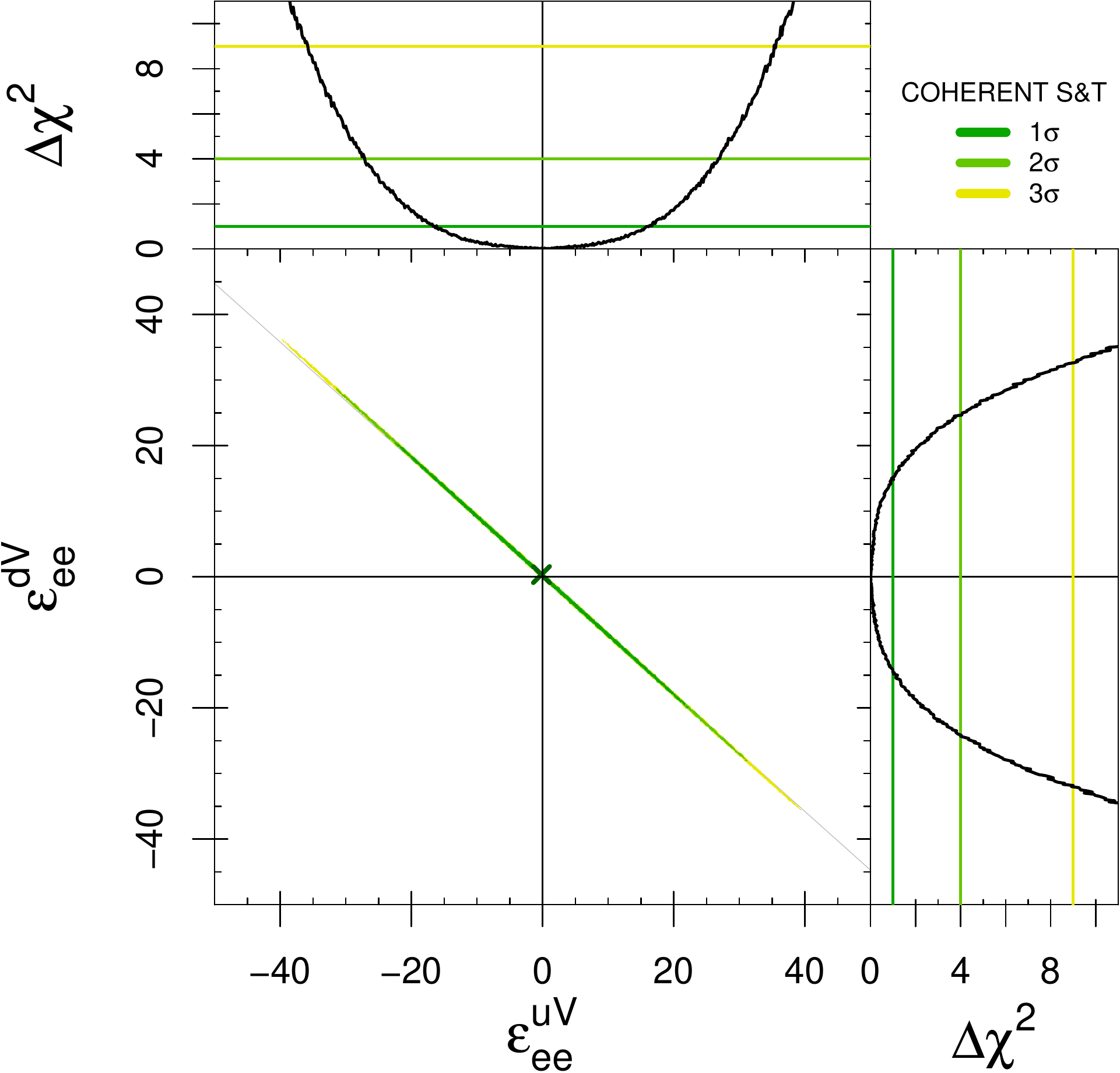}
&
\includegraphics*[width=0.49\linewidth]{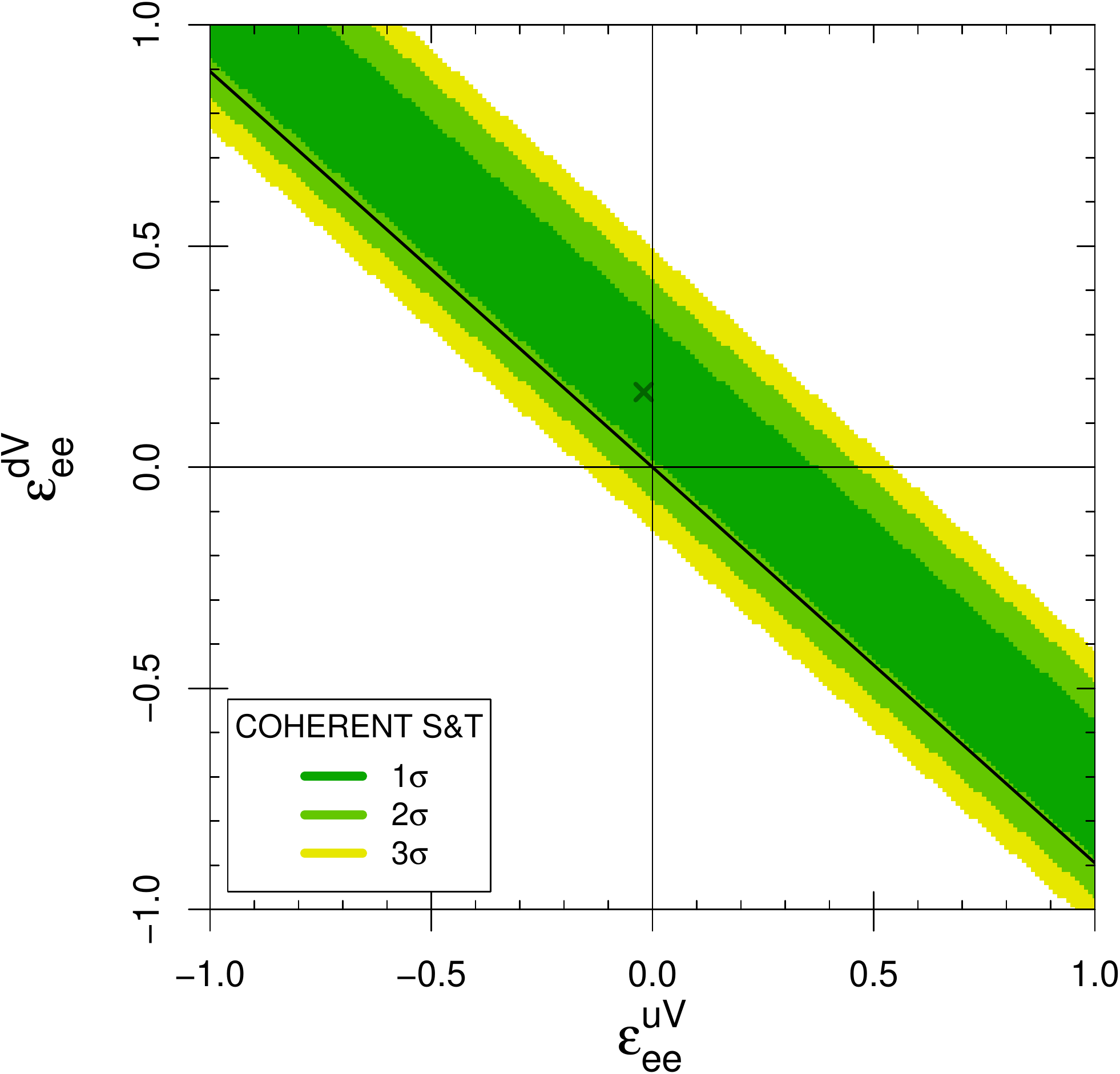}
\end{tabular}
\caption{ \label{fig:eeu-eed}
Left panel:
general marginalized allowed region in the
$(\varepsilon_{ee}^{uV},\varepsilon_{ee}^{dV})$
plane and marginal $\Delta\chi^2$'s
obtained from the analysis of the joint COHERENT spectral and temporal (S\&T) data.
Right panel: enlargement of the area around the origin.
The diagonal gray line in the left panel and the diagonal black line in the right panel represent the cancellation relation~(\ref{canc2}).
The point indicates the best fit.
}
\end{figure*}

\begin{figure*}[!h]
\centering
\begin{tabular}{cc}
\includegraphics*[width=0.49\linewidth]{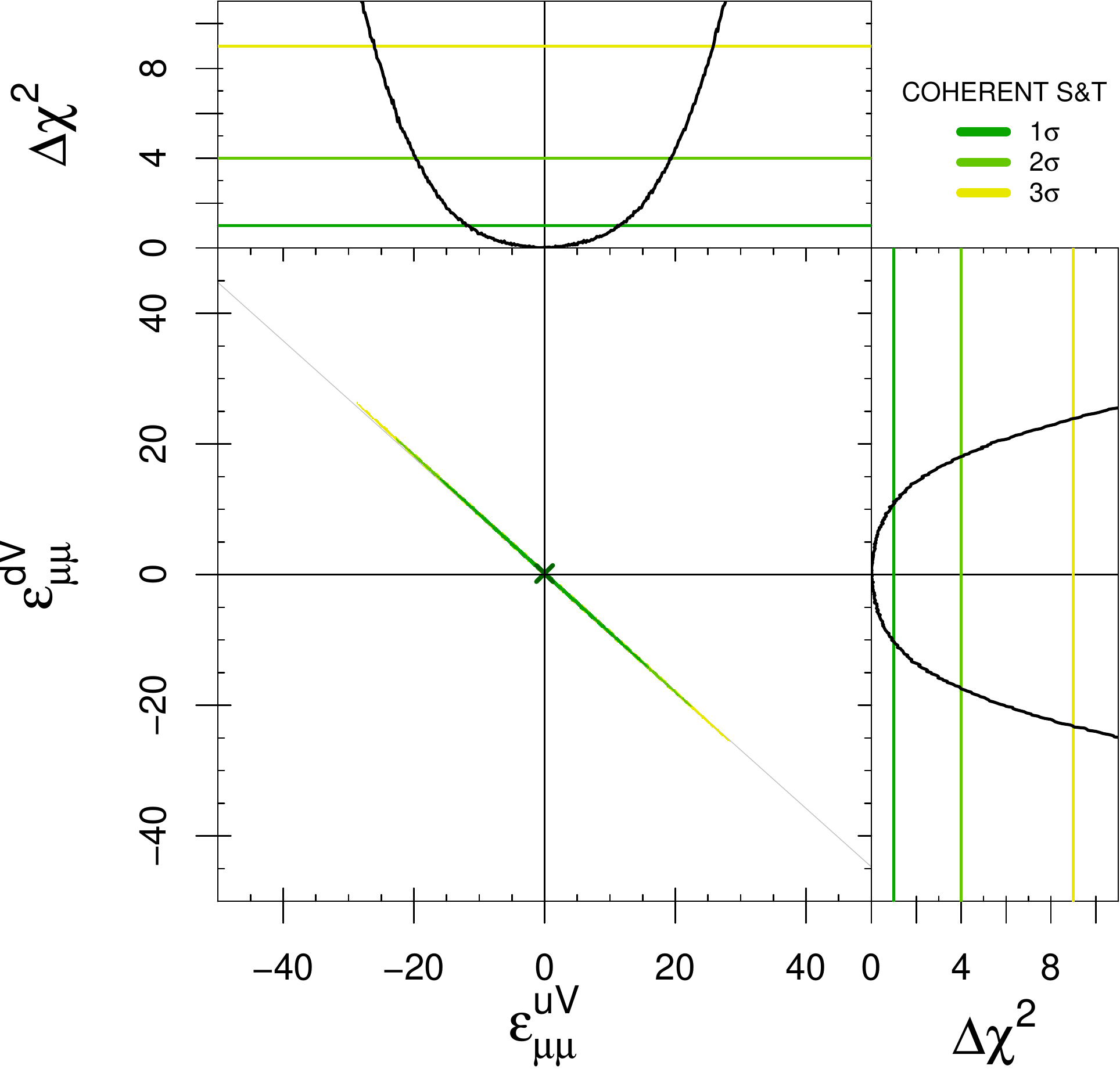}
&
\includegraphics*[width=0.49\linewidth]{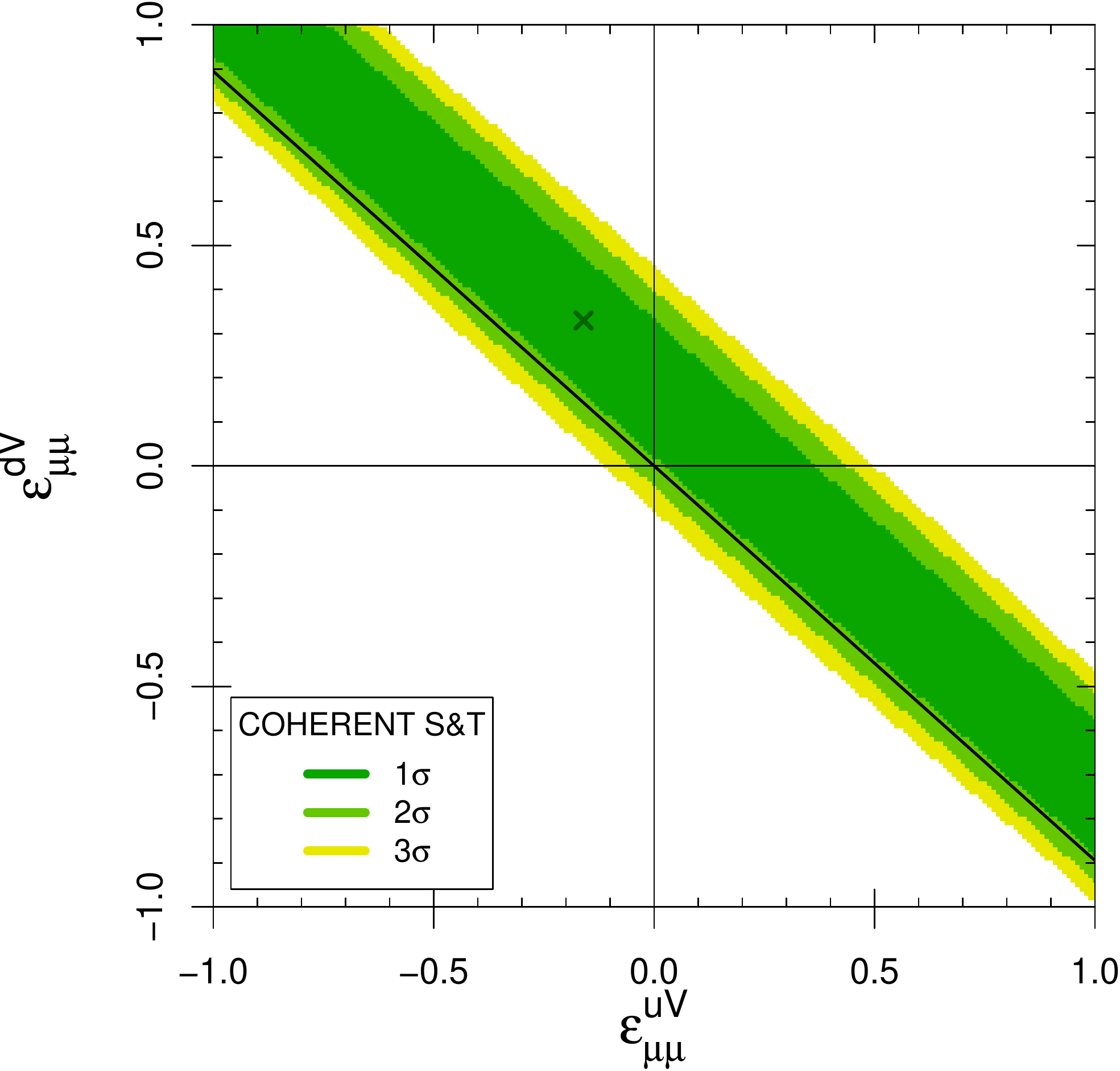}
\end{tabular}
\caption{ \label{fig:mmu-mmd}
Left panel:
general marginalized allowed region in the
$(\varepsilon_{\mu\mu}^{uV},\varepsilon_{\mu\mu}^{dV})$
plane and marginal $\Delta\chi^2$'s
obtained from the analysis of the joint COHERENT spectral and temporal (S\&T) data.
Right panel: enlargement of the area around the origin.
The diagonal gray line in the left panel and the diagonal black line in the right panel represent the cancellation relation~(\ref{canc2}).
The point indicates the best fit.
}
\end{figure*}

\begin{figure*}[!h]
\centering
\begin{tabular}{cc}
\includegraphics*[width=0.49\linewidth]{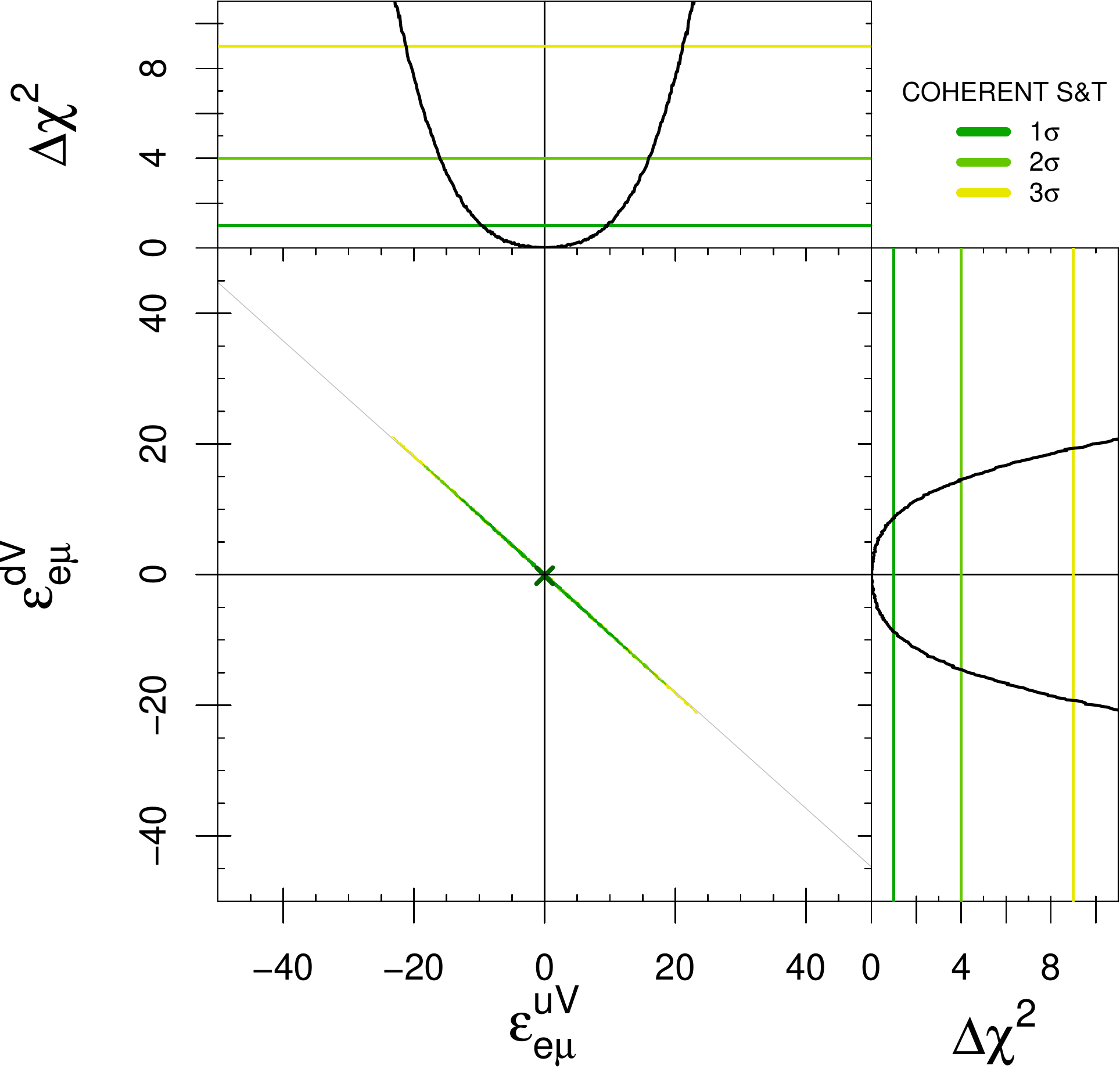}
&
\includegraphics*[width=0.49\linewidth]{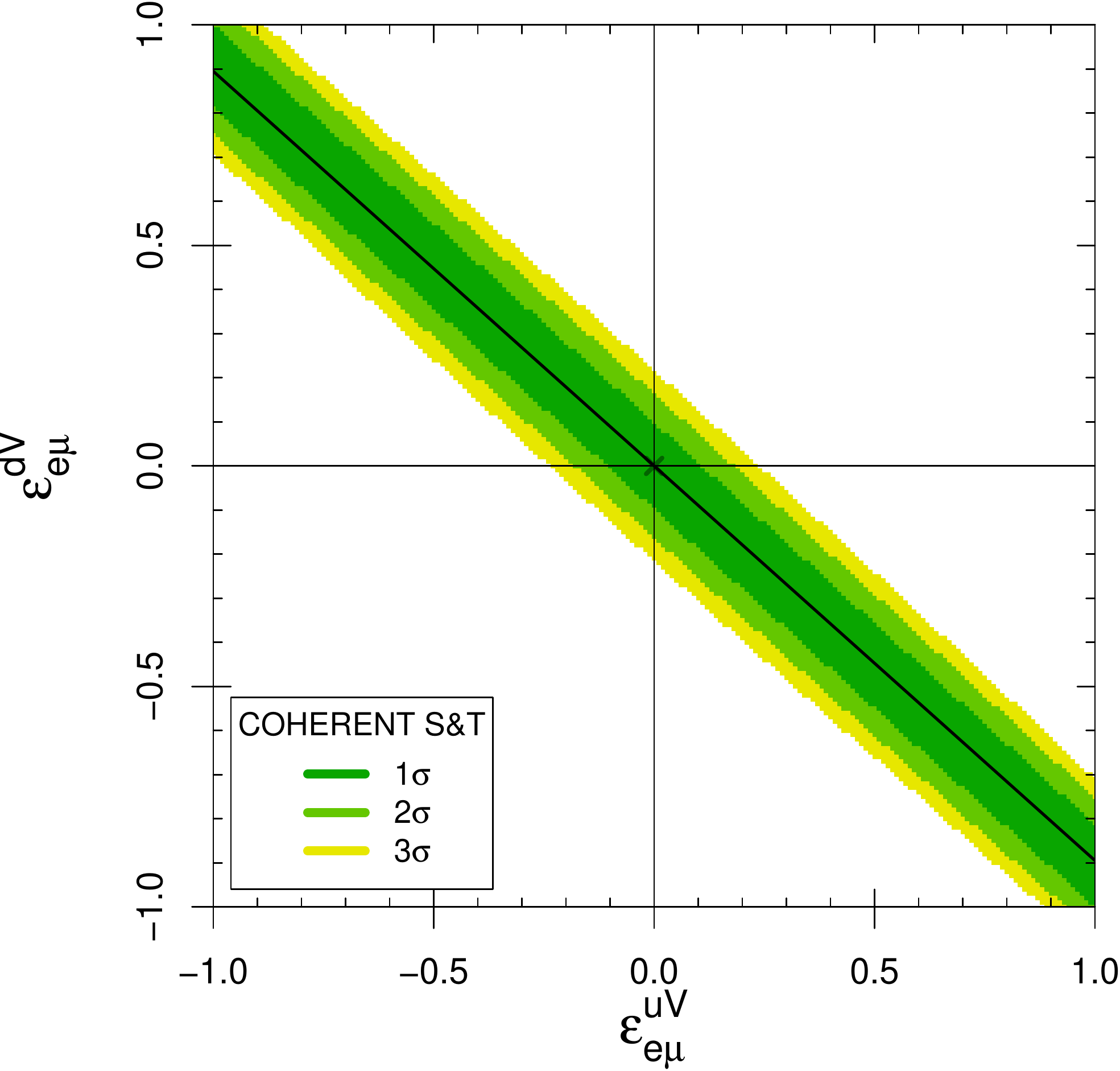}
\end{tabular}
\caption{ \label{fig:emu-emd}
Left panel:
general marginalized allowed region in the real
$(\varepsilon_{e\mu}^{uV},\varepsilon_{e\mu}^{dV})$
plane and marginal $\Delta\chi^2$'s
obtained from the analysis of the joint COHERENT spectral and temporal (S\&T) data.
Right panel: enlargement of the area around the origin.
The diagonal gray line in the left panel and the diagonal black line in the right panel represent the cancellation relation~(\ref{canc2}).
The point indicates the best fit.
}
\end{figure*}

\begin{figure*}[!h]
\centering
\begin{tabular}{cc}
\includegraphics*[width=0.49\linewidth]{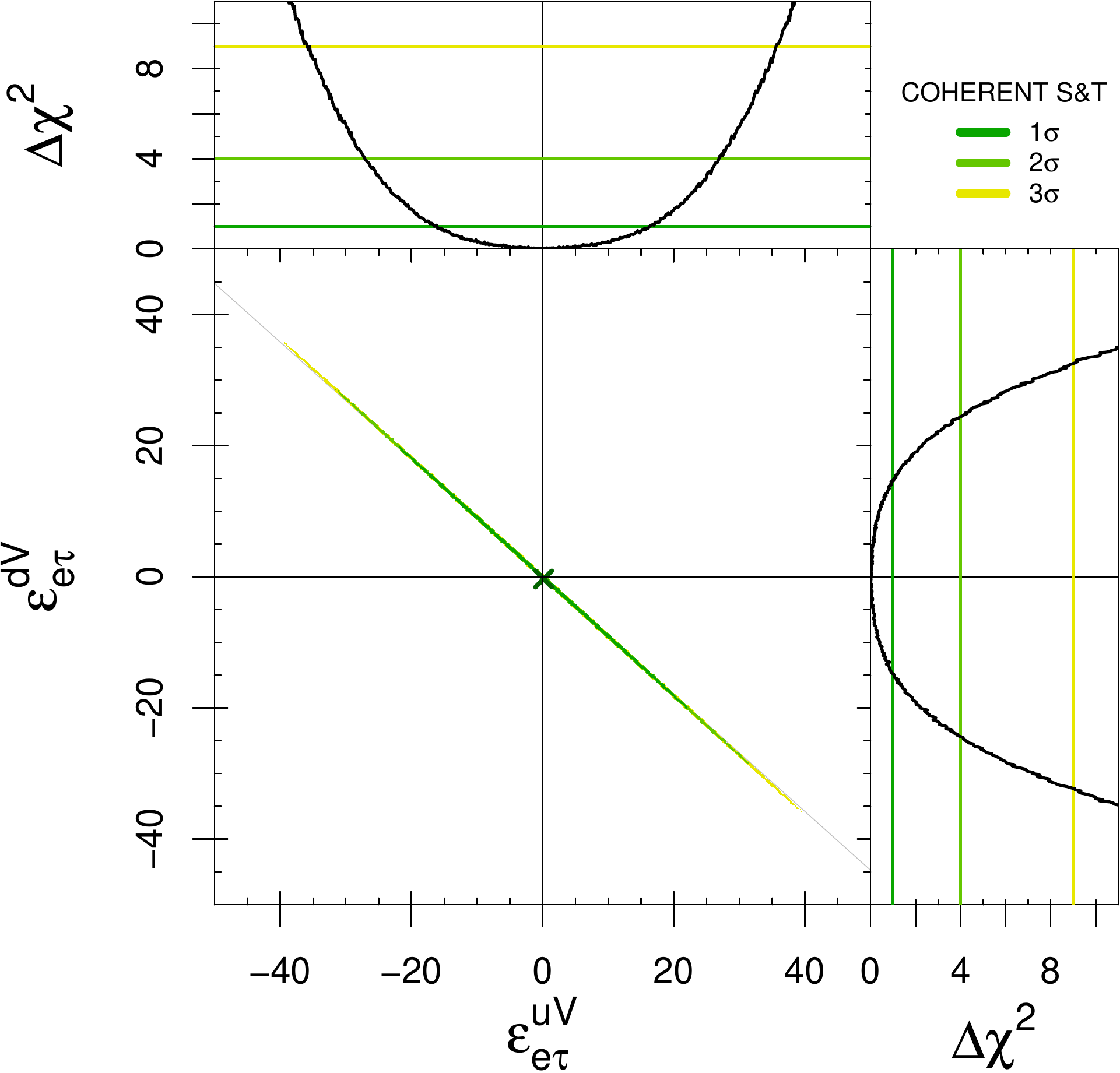}
&
\includegraphics*[width=0.49\linewidth]{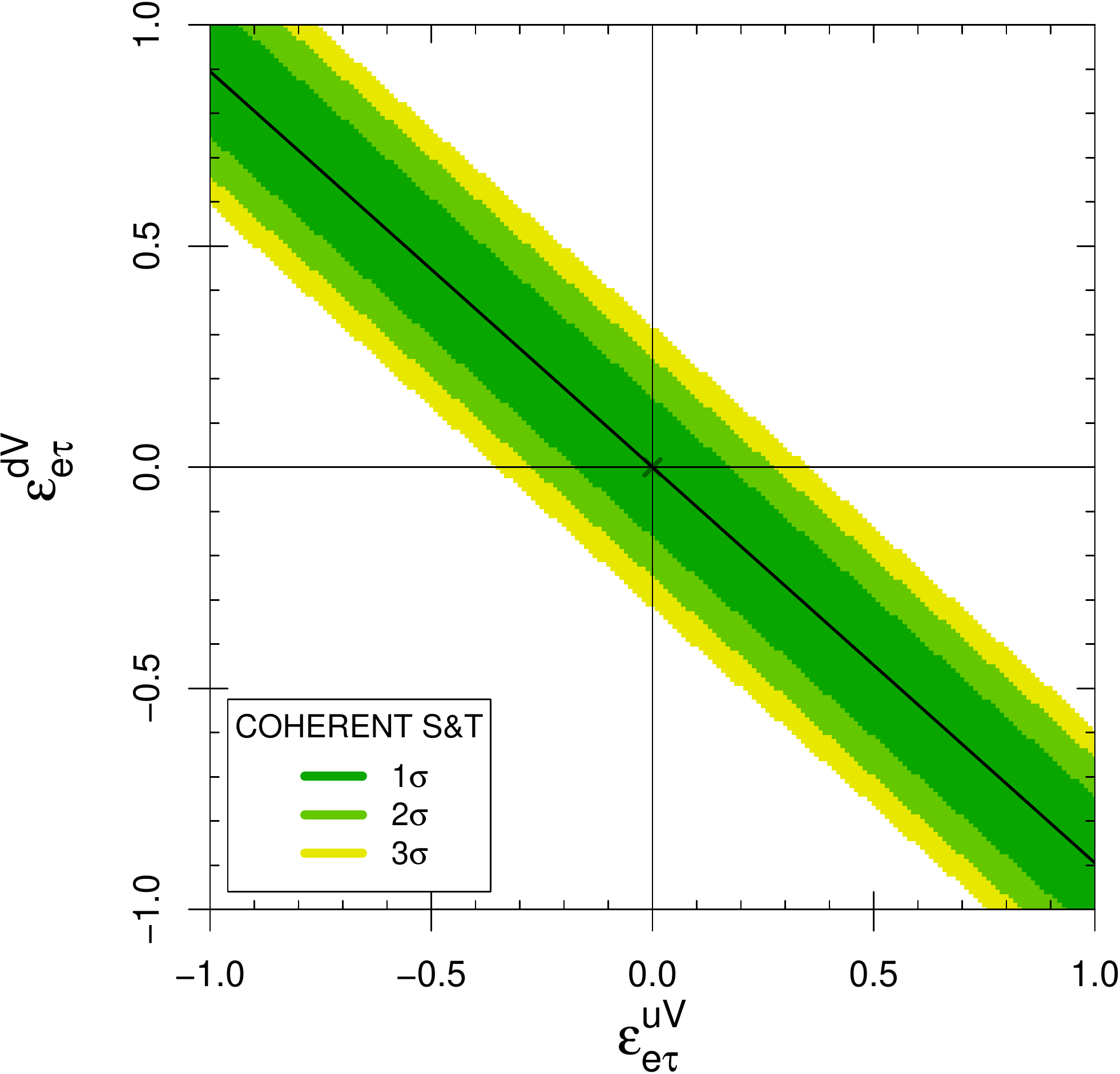}
\end{tabular}
\caption{ \label{fig:etu-etd}
Left panel:
general marginalized allowed region in the real
$(\varepsilon_{e\tau}^{uV},\varepsilon_{e\tau}^{dV})$
plane and marginal $\Delta\chi^2$'s
obtained from the analysis of the joint COHERENT spectral and temporal (S\&T) data.
Right panel: enlargement of the area around the origin.
The diagonal gray line in the left panel and the diagonal black line in the right panel represent the cancellation relation~(\ref{canc2}).
The point indicates the best fit.
}
\end{figure*}

\begin{figure*}[!h]
\centering
\begin{tabular}{cc}
\includegraphics*[width=0.49\linewidth]{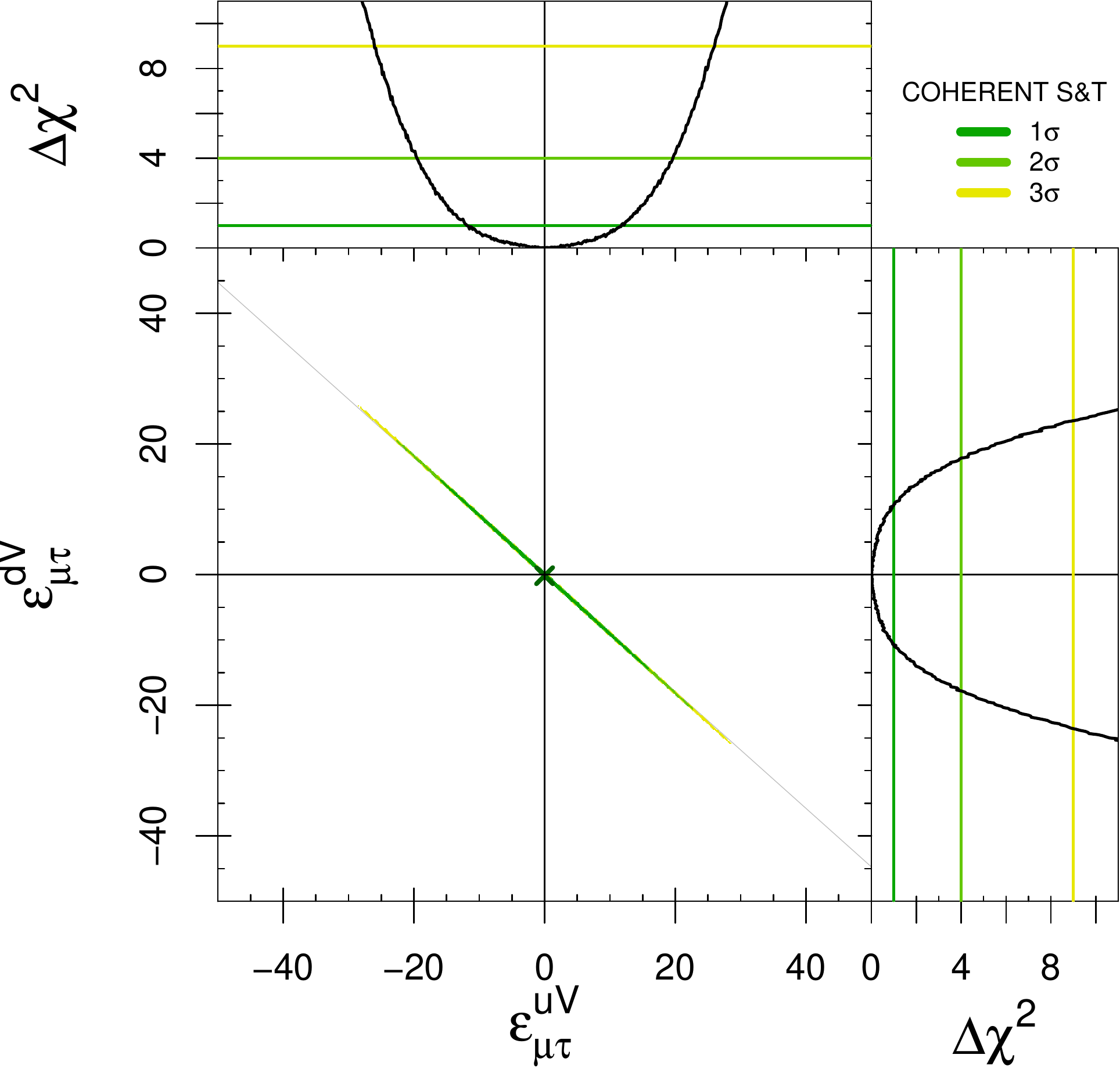}
&
\includegraphics*[width=0.49\linewidth]{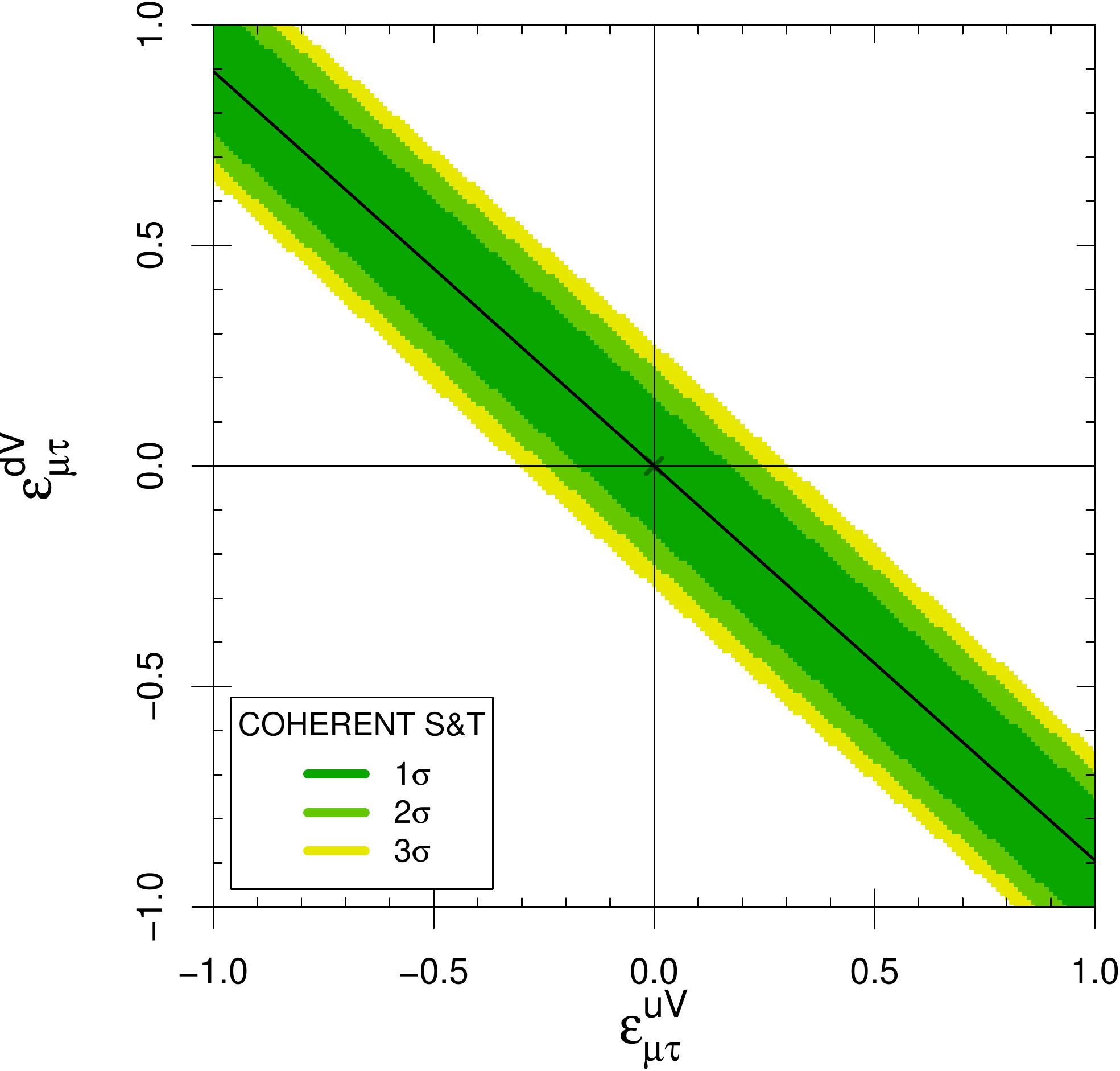}
\end{tabular}
\caption{ \label{fig:mtu-mtd}
Left panel:
general marginalized allowed region in the real
$(\varepsilon_{\mu\tau}^{uV},\varepsilon_{\mu\tau}^{dV})$
plane and marginal $\Delta\chi^2$'s
obtained from the analysis of the joint COHERENT spectral and temporal (S\&T) data.
Right panel: enlargement of the area around the origin.
The diagonal gray line in the left panel and the diagonal black line in the right panel represent the cancellation relation~(\ref{canc2}).
The point indicates the best fit.
}
\end{figure*}

Figures~\ref{fig:eeu-eed}--\ref{fig:mtu-mtd}
show the allowed regions in the planes
$(\varepsilon_{\alpha\beta}^{uV},\varepsilon_{\alpha\beta}^{dV})$
with $\alpha=e,\mu$ and $\beta=e,\mu,\tau$.
For the off-diagonal NSI parameters,
the allowed regions are marginalized in the real planes.
From these figures,
one can see that all the allowed regions are
approximately parallel to the line in Eq.~(\ref{canc2}),
in agreement with the expectation.
The enlargements in the right panels show that the preferred values of the NSI parameters
are close to the approximate cancellation line (\ref{canc2}),
indicating that there is no indication of NSI effects within the uncertainties.
\bluechange{%
%\marginnote{\bf\color{red}$\leftarrow$A6}%
In Figures~\ref{fig:eeu-eed} and~\ref{fig:mmu-mmd}
the allowed region is parallel to the approximate cancellation line,
but not centered on it,
because the data are better fitted with a small contribution
of the diagonal NSI parameters.
However,
since the deviation from the approximate cancellation line
is well within the $2\sigma$ allowed region,
there is no statistically significant indication of NSI effects.%
}

Near the origin the boundaries of the
$1\sigma$,
$2\sigma$, and
$3\sigma$ allowed regions can be parameterized with the lines
\begin{equation}
\varepsilon_{\alpha\beta}^{dV}
=
a \, \varepsilon_{\alpha\beta}^{uV} + b
,
\label{boundaries}
\end{equation}
with the values of the parameters $a$ and $b$ given in Table~\ref{tab:ab}.
One can see that the slopes $a$ are all very close to the one in Eq.~(\ref{canc2}),
as expected.

\begin{figure*}[!h]
\centering
\begin{tabular}{cc}
\includegraphics*[width=0.49\linewidth]{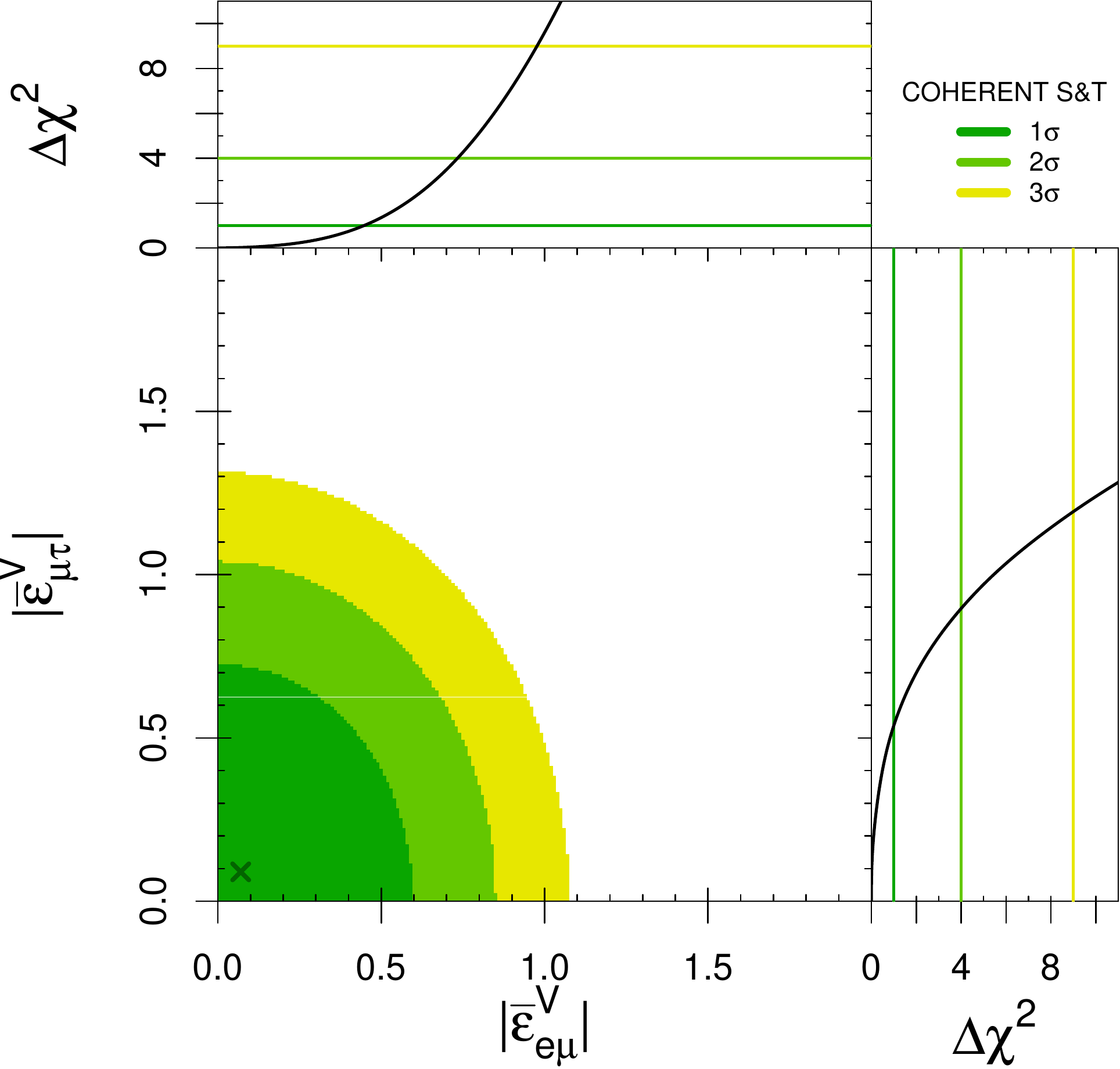}
&
\includegraphics*[width=0.49\linewidth]{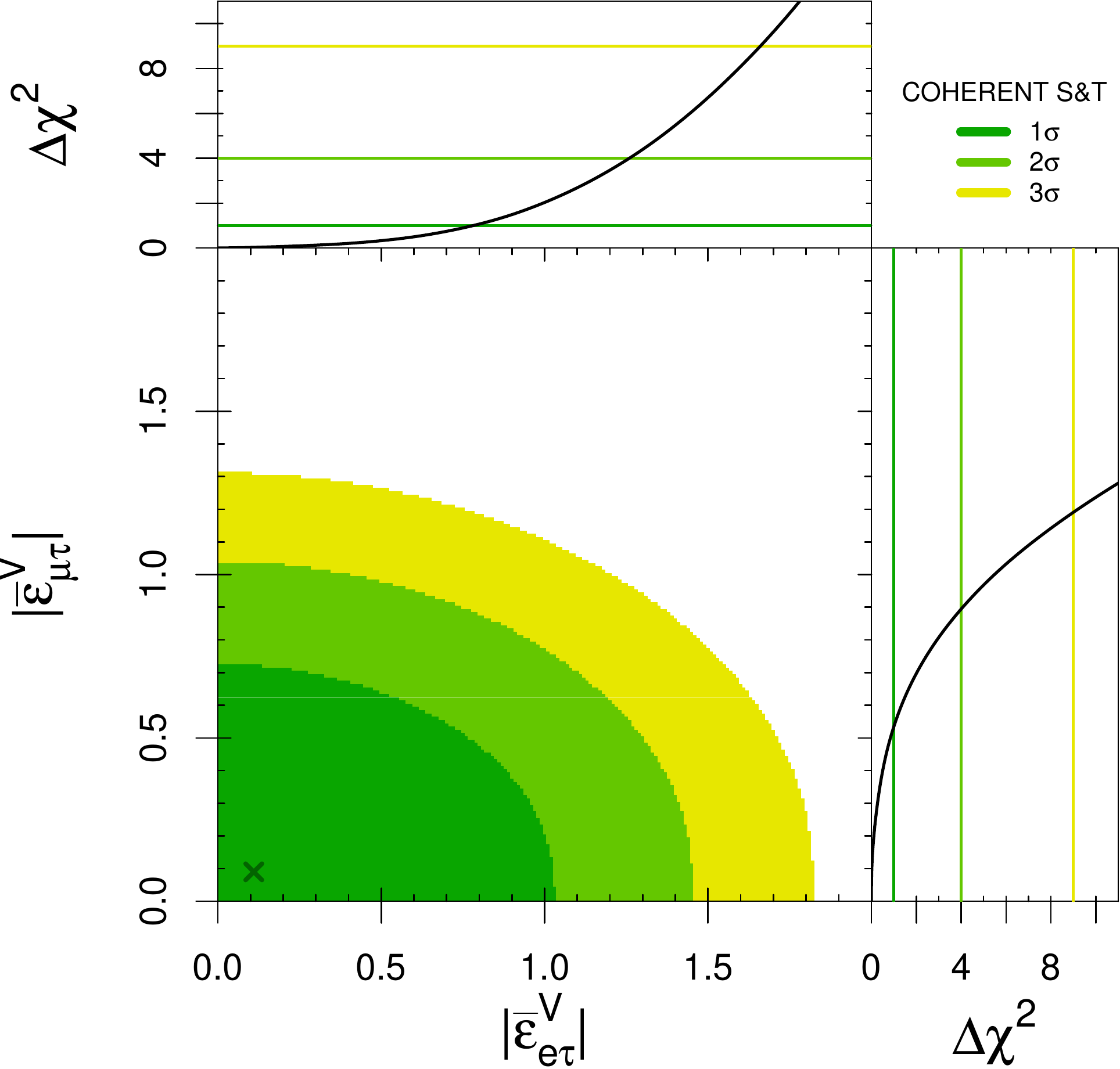}
\\
\includegraphics*[width=0.49\linewidth]{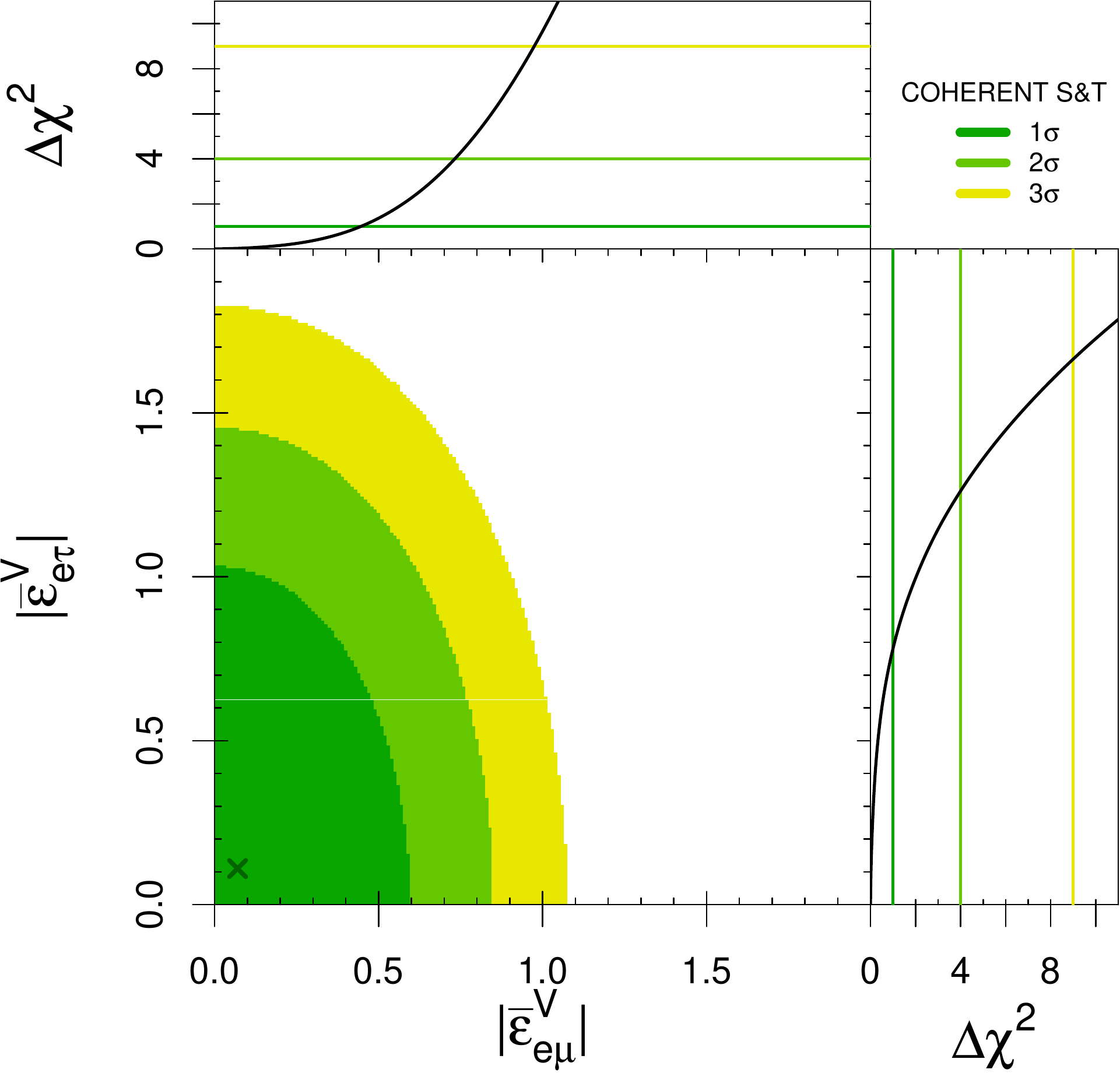}
&
\includegraphics*[width=0.49\linewidth]{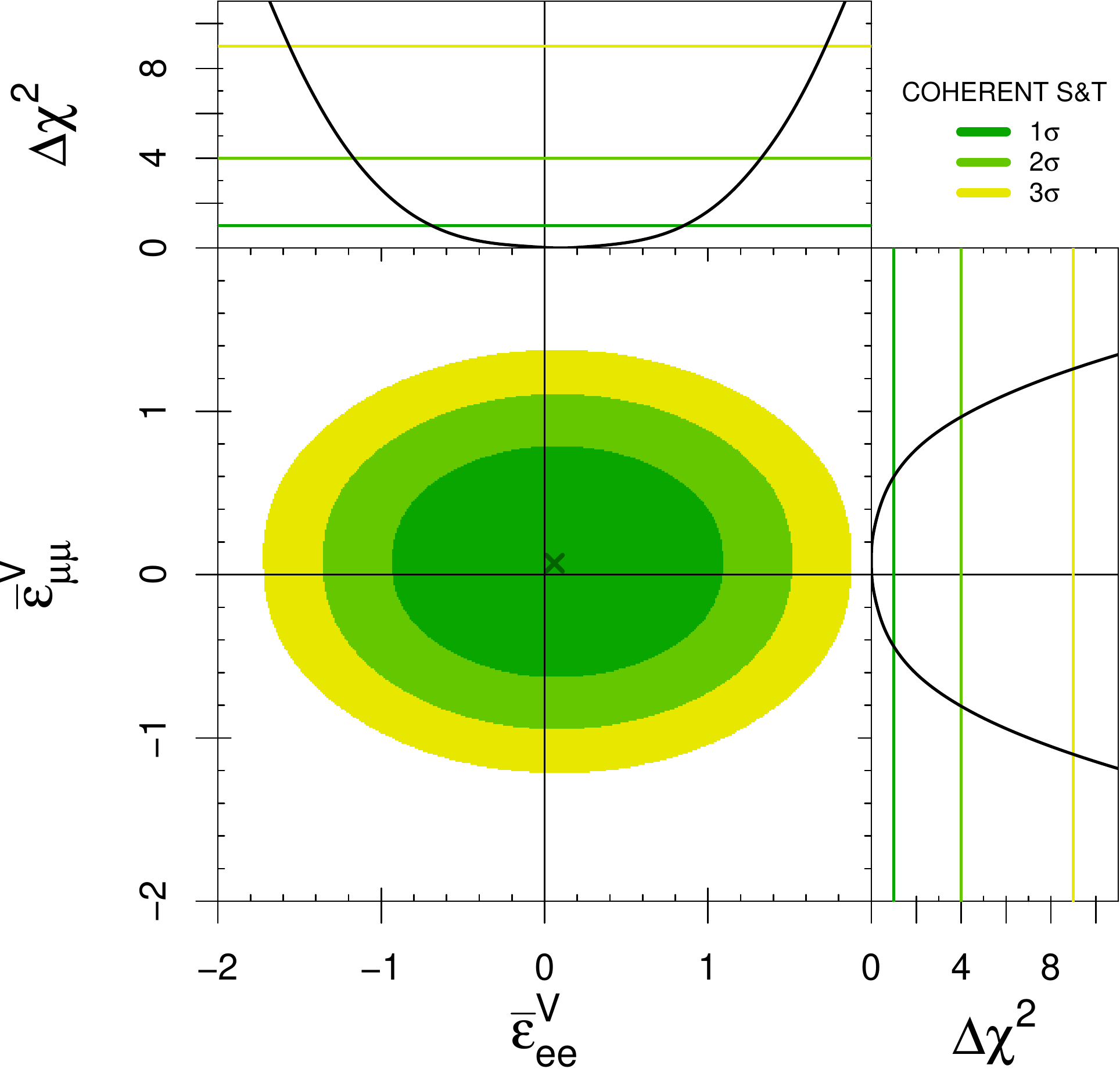}
\end{tabular}
\caption{ \label{fig:ave}
General marginalized allowed regions in different planes of the
up-down average NSI parameters~(\ref{ave})
and marginal $\Delta\chi^2$'s
obtained from the analysis of the joint COHERENT spectral and temporal (S\&T) data.
The points indicate the best-fit values.
}
\end{figure*}

\begin{figure*}[!h]
\centering
\begin{tabular}{cc}
\includegraphics*[width=0.49\linewidth]{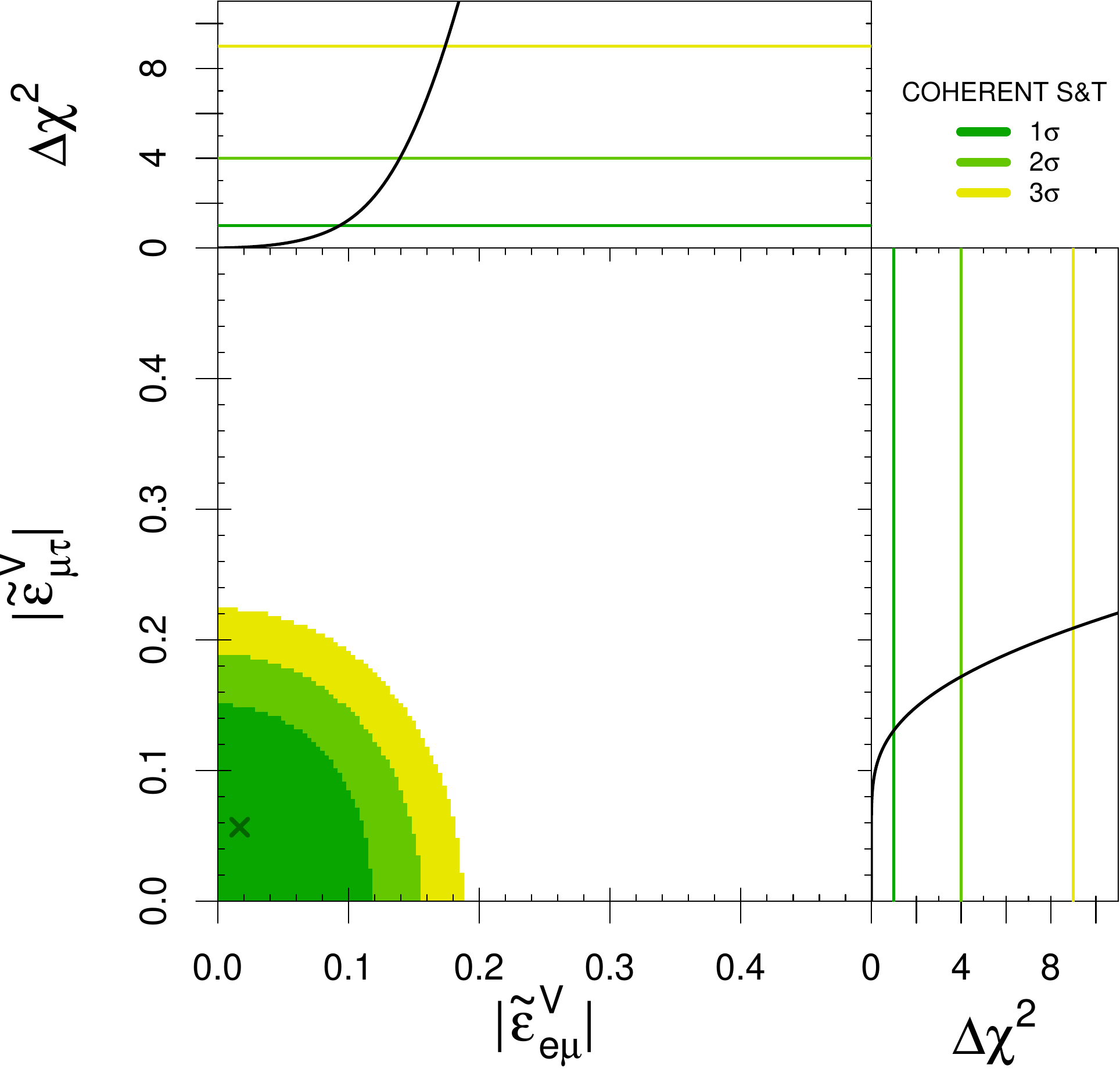}
&
\includegraphics*[width=0.49\linewidth]{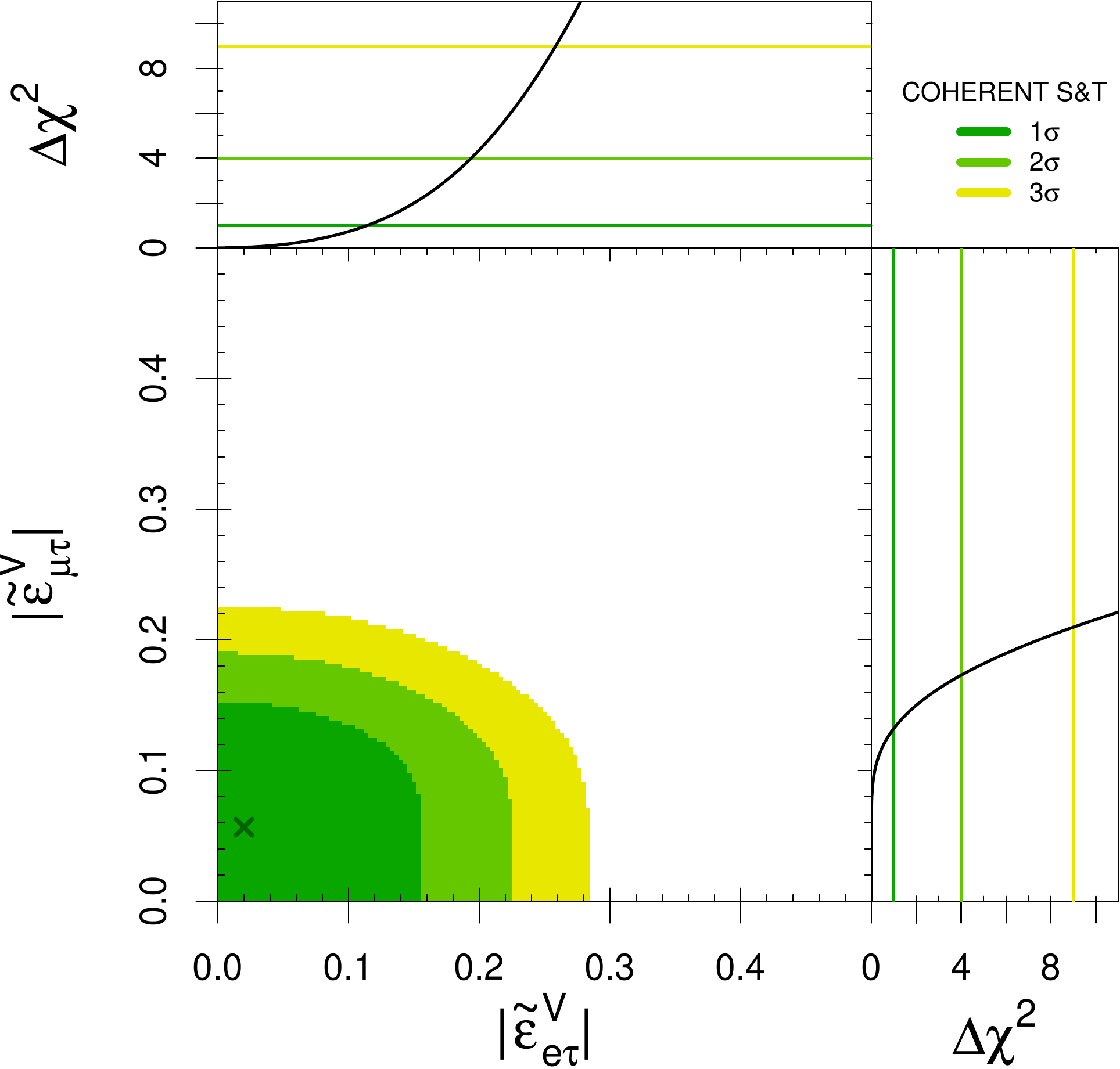}
\\
\includegraphics*[width=0.49\linewidth]{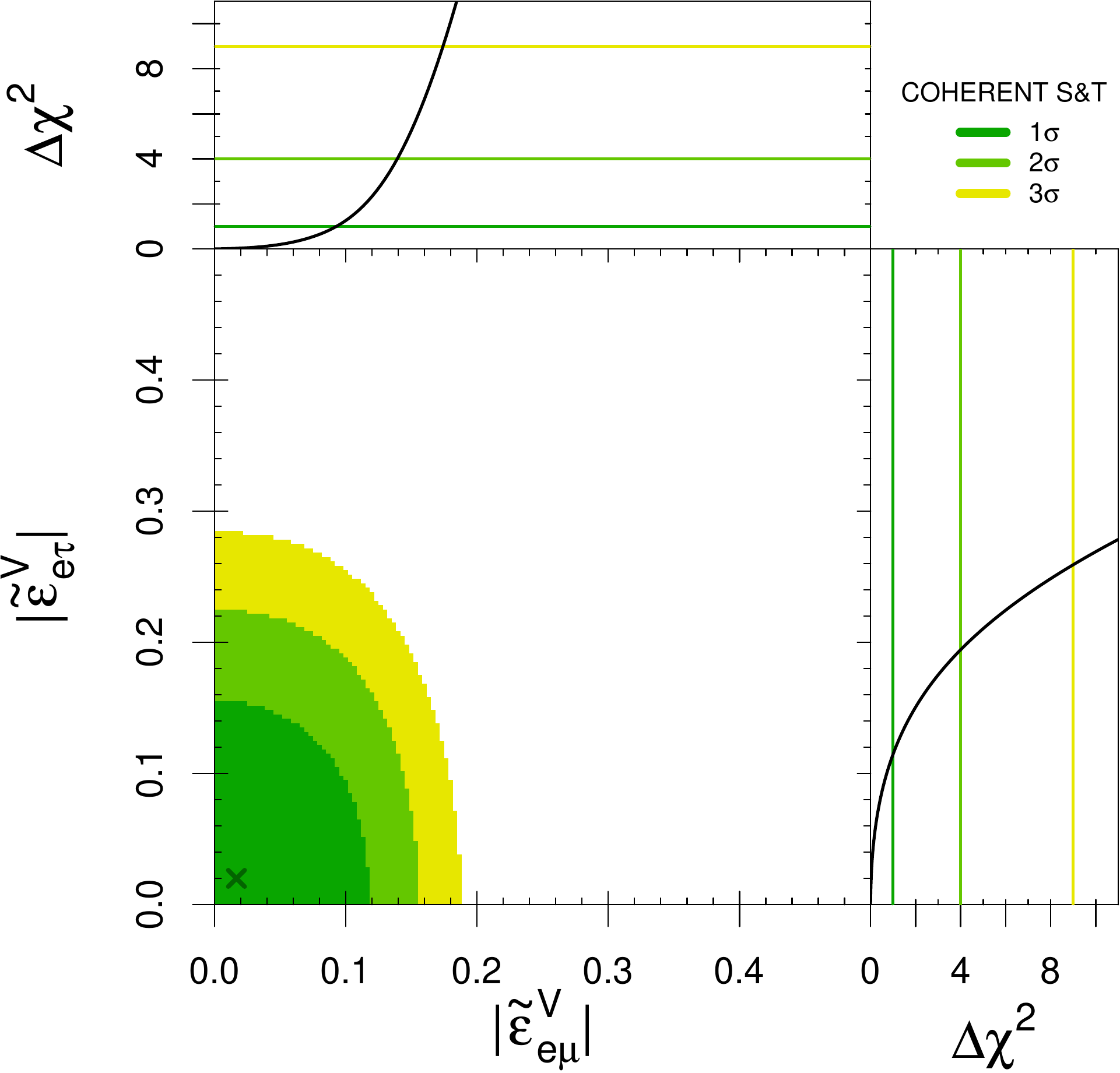}
&
\includegraphics*[width=0.49\linewidth]{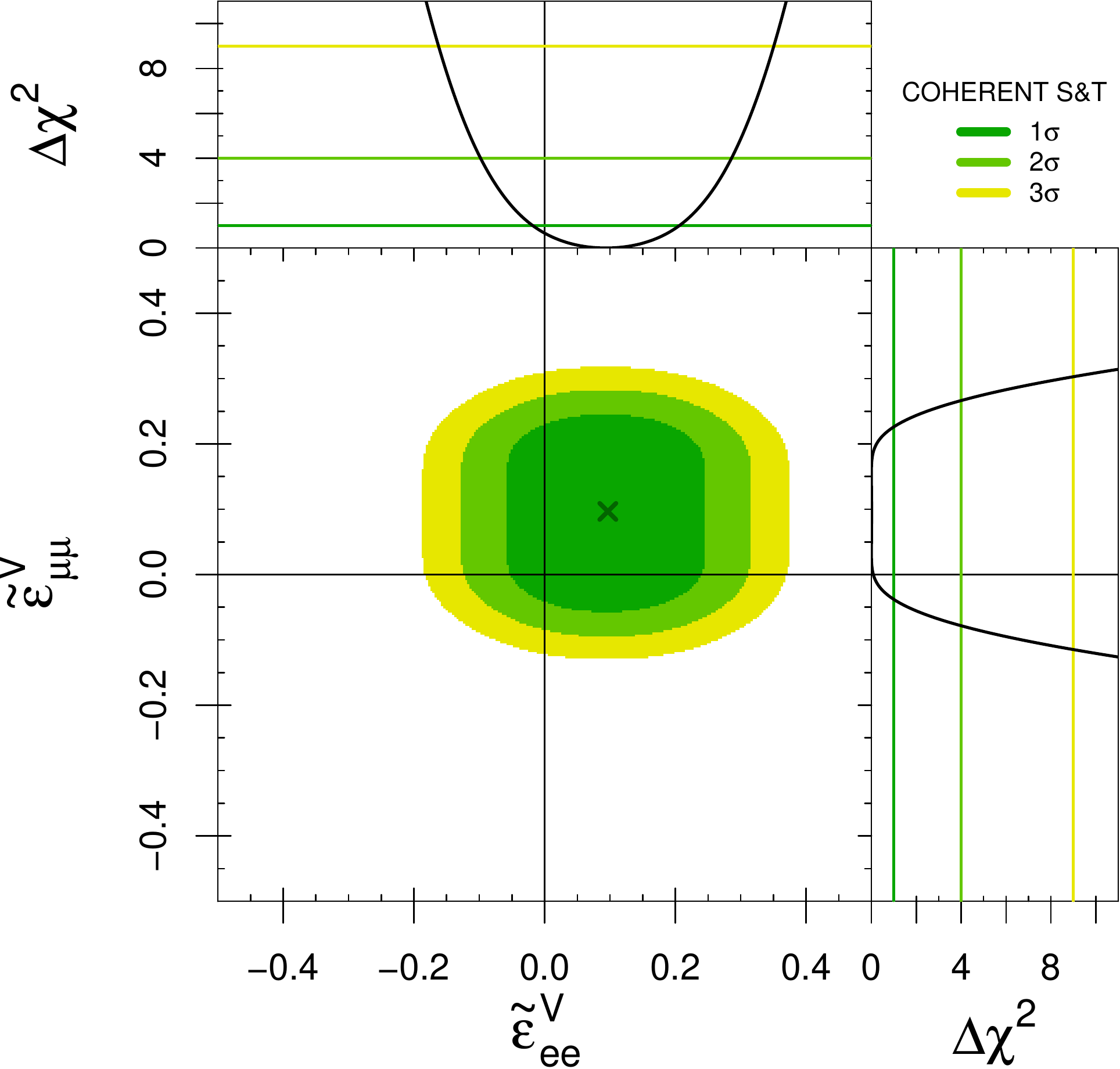}
\end{tabular}
\caption{ \label{fig:tilde}
General marginalized allowed regions in different planes of the
maximally constrained up-down linear combination of NSI parameters~(\ref{tilde})
and marginal $\Delta\chi^2$'s
obtained from the analysis of the joint COHERENT spectral and temporal (S\&T) data.
The points indicate the best-fit values.
}
\end{figure*}

Figure~\ref{fig:ave}
shows the marginal allowed regions in different planes of the
up-down average NSI parameters $\overline{\varepsilon}_{\alpha\beta}^{V}$ in Eq.~(\ref{ave}),
that are relatively well constrained by the COHERENT data.
However,
as shown in Table~\ref{tab:general} and Figure~\ref{fig:tilde},
the strongest constraints are obtained for the maximally constrained up-down linear combinations
$\widetilde{\varepsilon}_{\alpha\beta}^{V}$ in Eq.~(\ref{tilde}).

\section{COHERENT constraints on NSI with either up or down quarks}
\label{sec:ud}

\begin{figure*}[!h]
\centering
\begin{tabular}{cc}
\includegraphics*[width=0.49\linewidth]{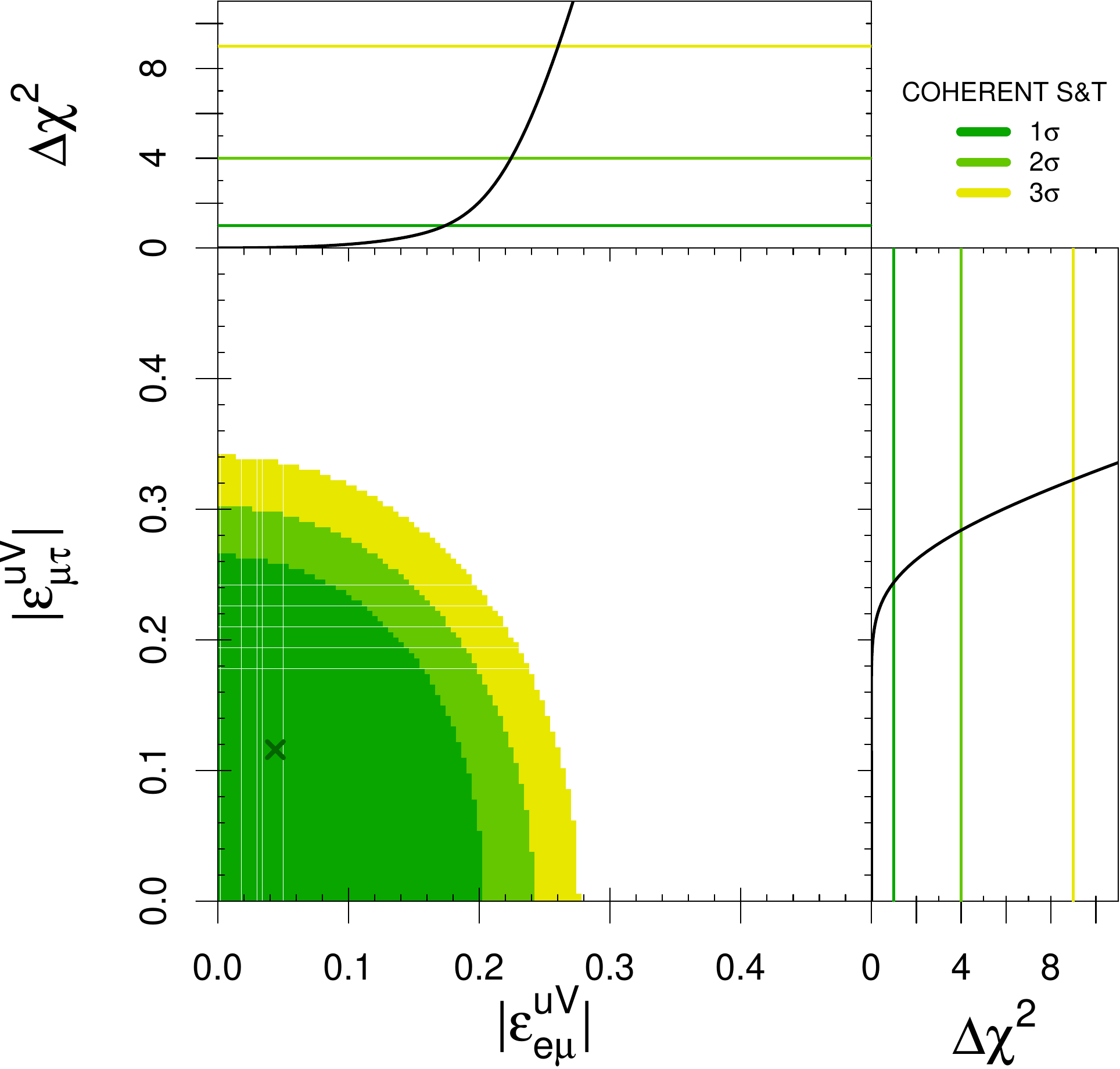}
&
\includegraphics*[width=0.49\linewidth]{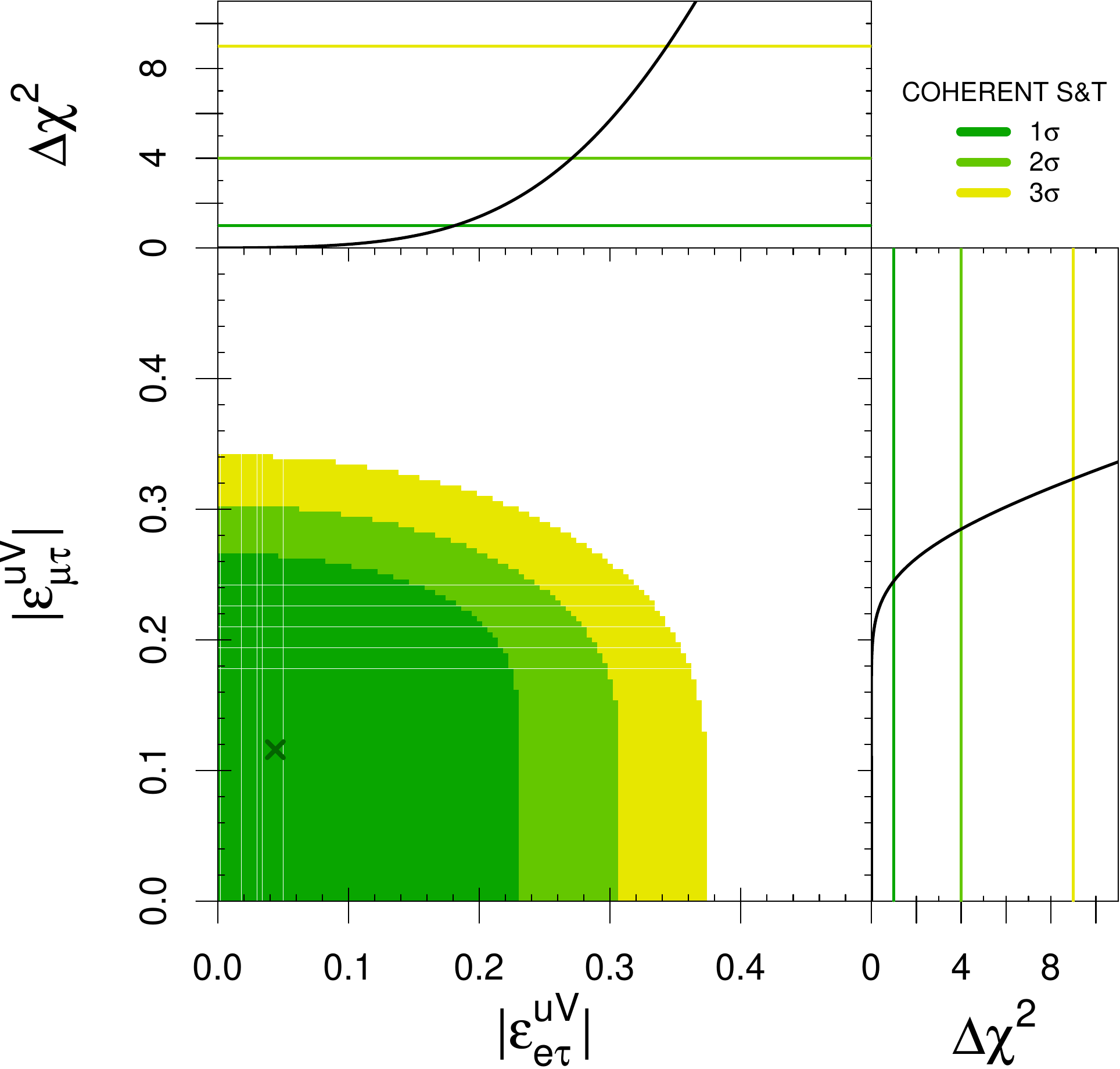}
\\
\includegraphics*[width=0.49\linewidth]{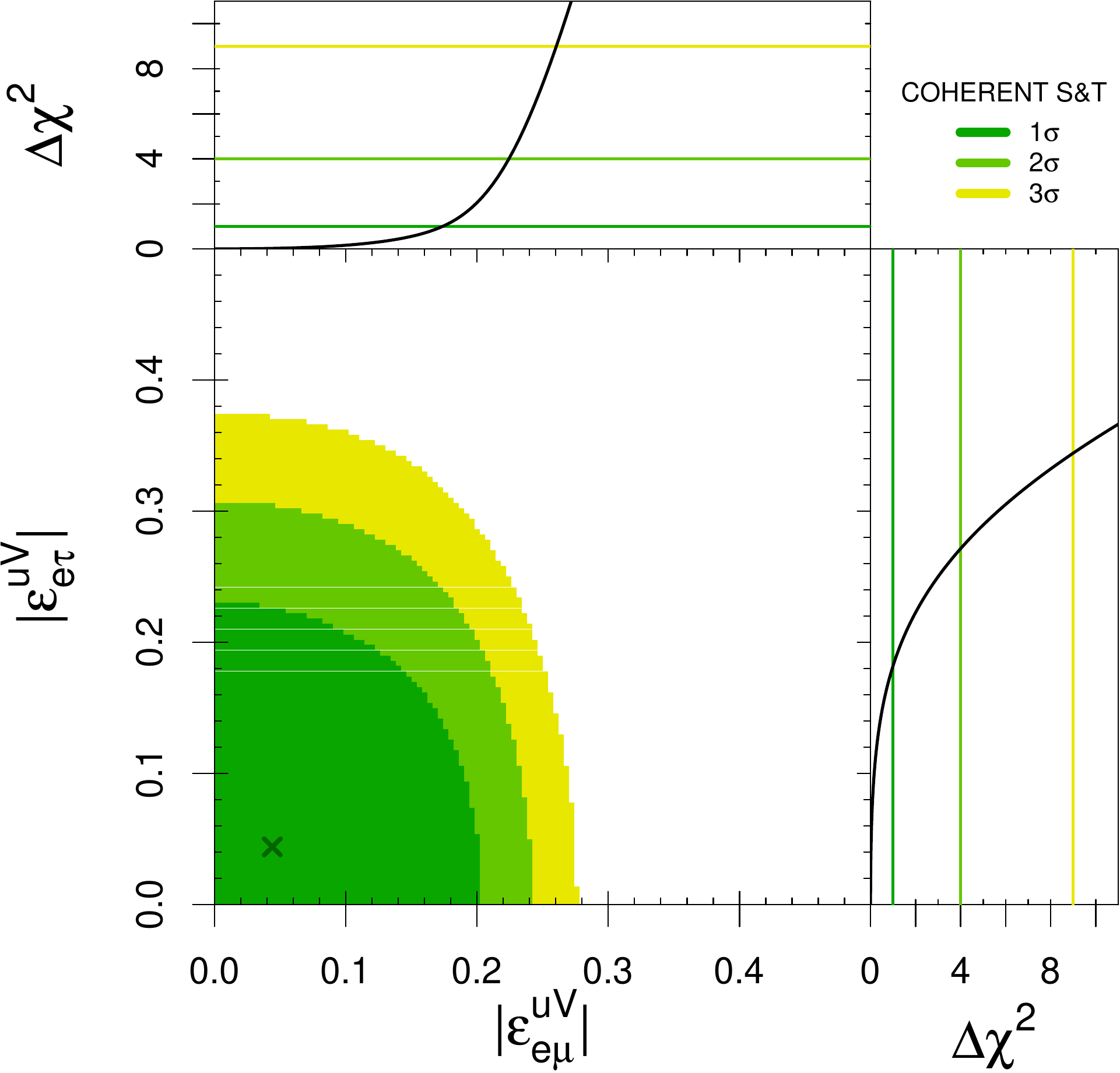}
&
\includegraphics*[width=0.49\linewidth]{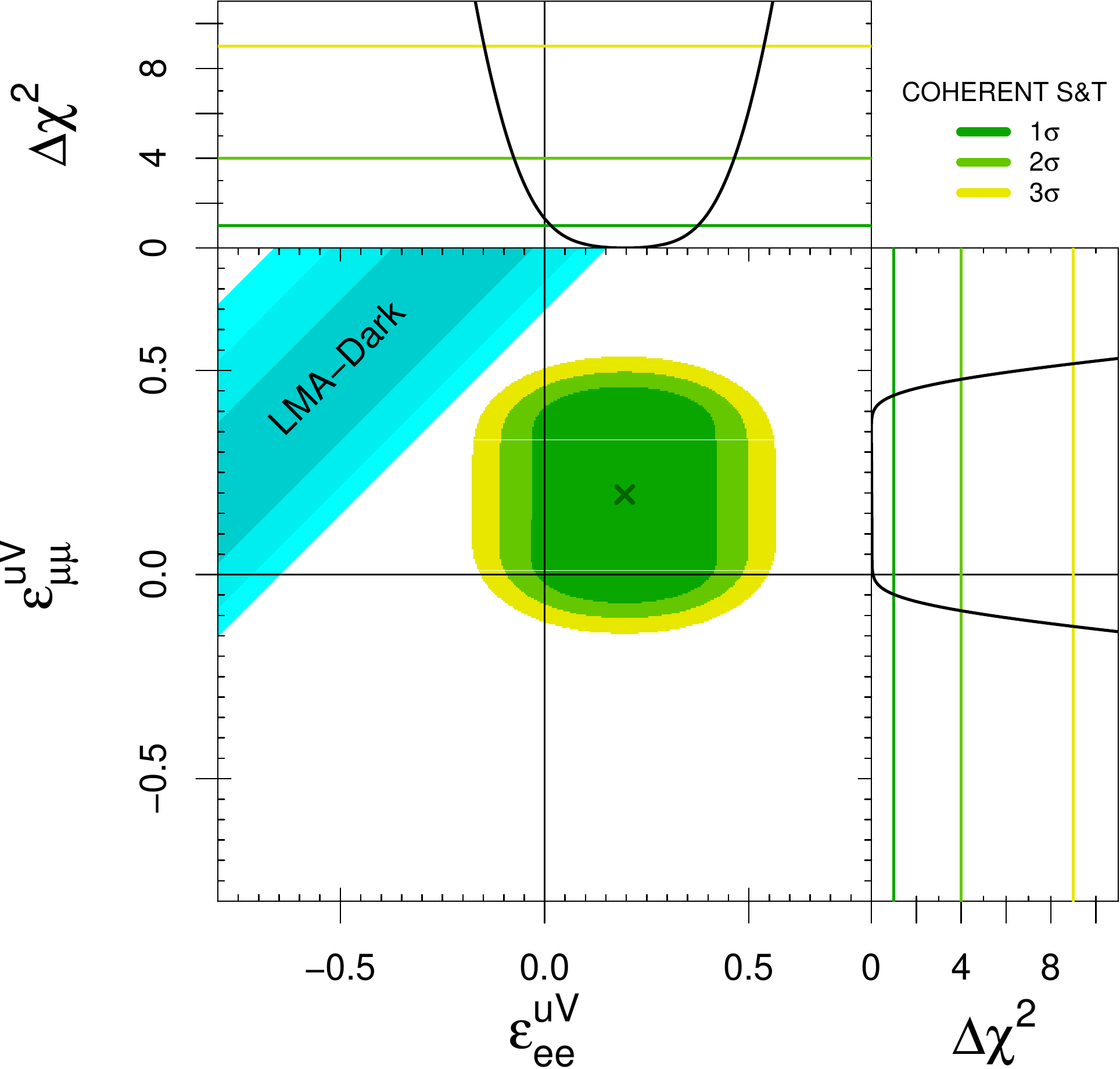}
\end{tabular}
\caption{ \label{fig:5u}
Marginalized allowed regions in different planes of the NSI parameters
and marginal $\Delta\chi^2$'s
obtained from the analysis of the joint COHERENT spectral and temporal (S\&T) data
assuming interactions with up quarks only.
The points indicate the best-fit values.
The diagonal cyan strips in the
$(\varepsilon_{ee}^{uV},\varepsilon_{\mu\mu}^{uV})$
plane are allowed at $1\sigma$, $2\sigma$, and $3\sigma$
by the LMA-Dark fit of solar neutrino data~\cite{Coloma:2017ncl}.
}
\end{figure*}

\begin{figure*}[!h]
\centering
\begin{tabular}{cc}
\includegraphics*[width=0.49\linewidth]{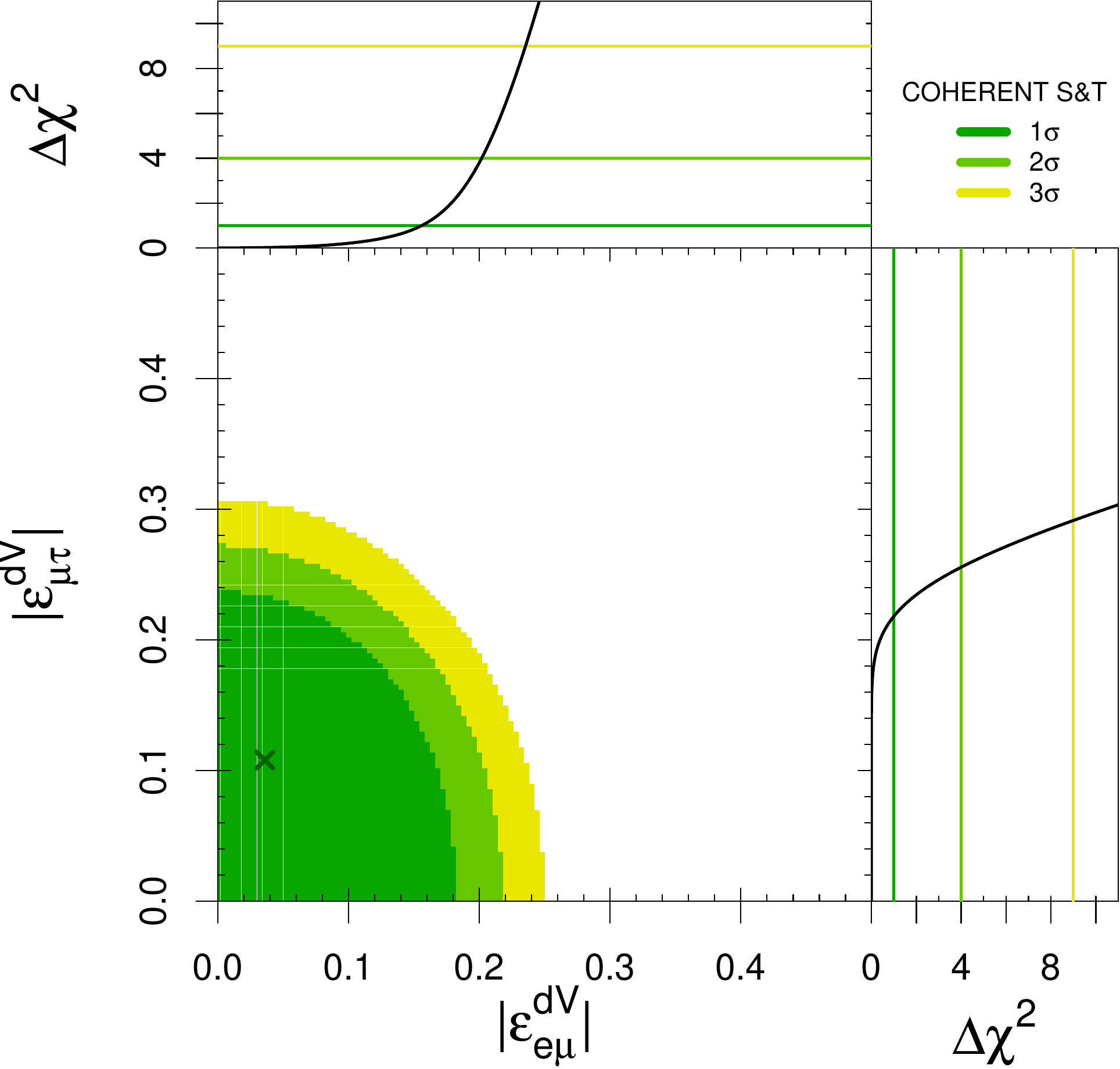}
&
\includegraphics*[width=0.49\linewidth]{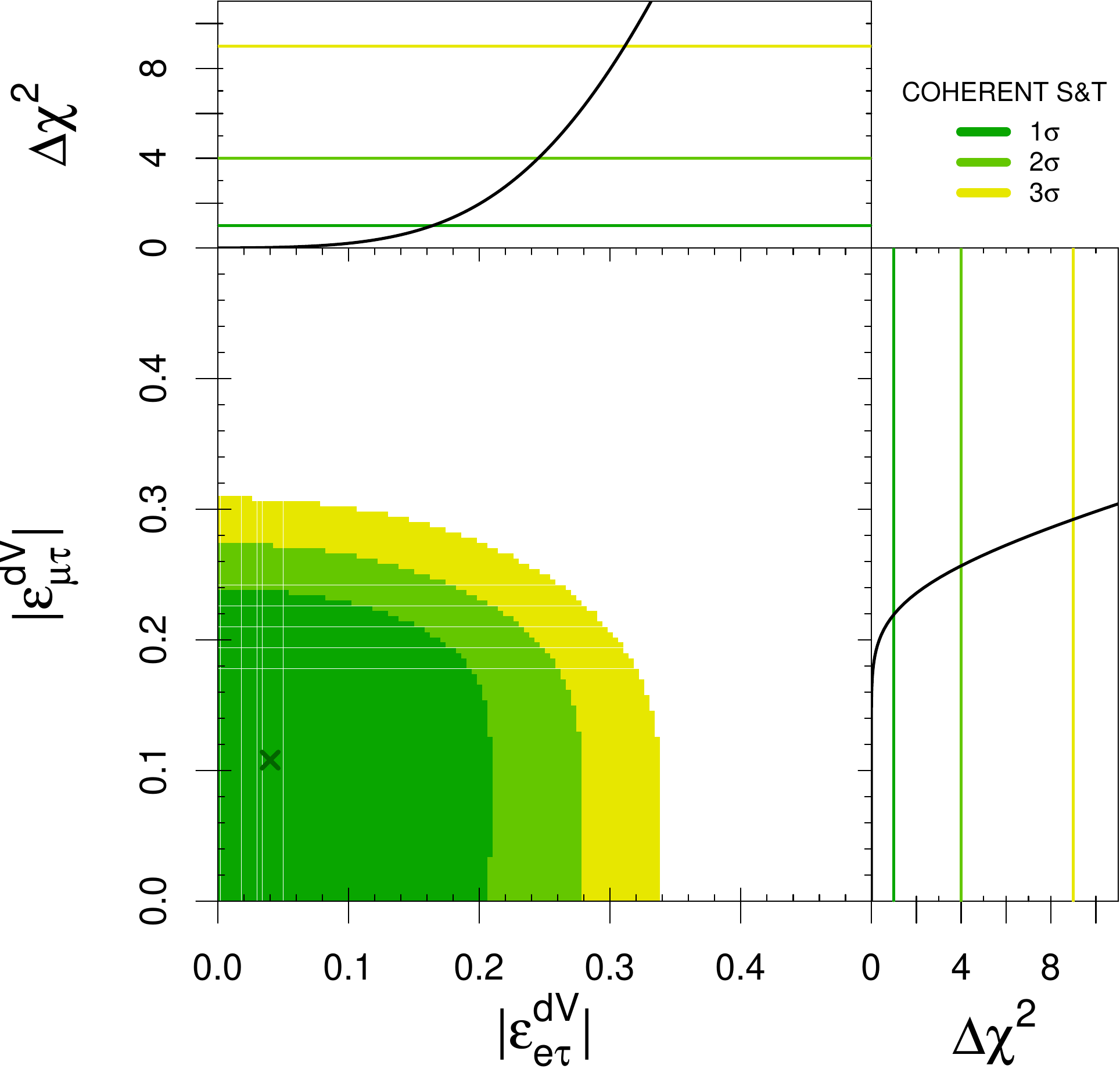}
\\
\includegraphics*[width=0.49\linewidth]{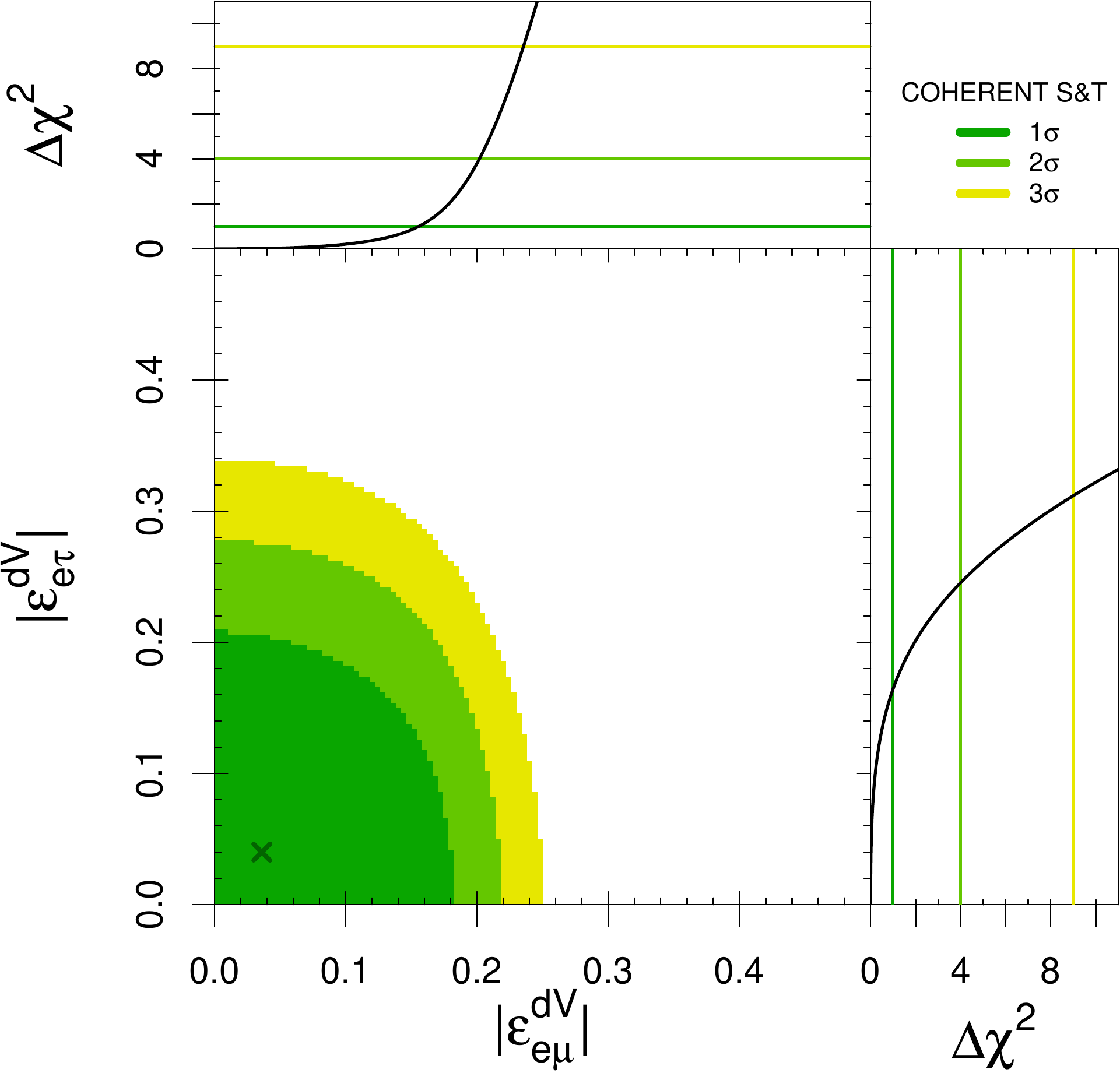}
&
\includegraphics*[width=0.49\linewidth]{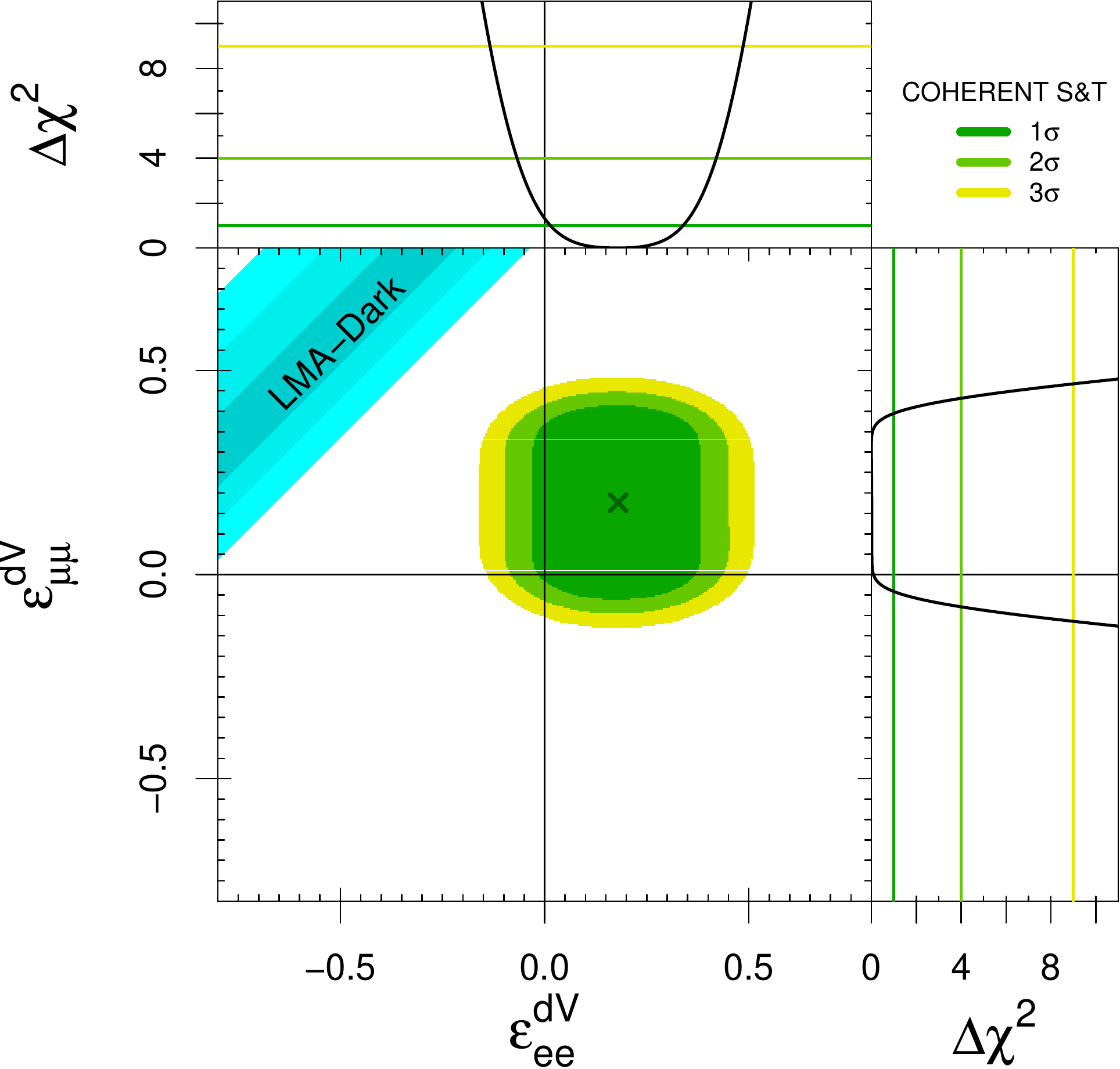}
\end{tabular}
\caption{ \label{fig:5d}
Marginalized allowed regions in different planes of the NSI parameters
and marginal $\Delta\chi^2$'s
obtained from the analysis of the joint COHERENT spectral and temporal (S\&T) data
assuming interactions with down quarks only.
The points indicate the best-fit values.
The diagonal cyan strips in the
$(\varepsilon_{ee}^{dV},\varepsilon_{\mu\mu}^{dV})$
plane are allowed at $1\sigma$, $2\sigma$, and $3\sigma$
by the LMA-Dark fit of solar neutrino data~\cite{Coloma:2017ncl}.
}
\end{figure*}

In this Section we consider the possibility that
neutrino NSI with nuclei are dominated
by either up or down quarks,
with the subdominant quark NSI having negligible effects.
The results of the analysis assuming interactions with up quark only are given in
Table~\ref{tab:up},
and those obtained assuming interactions with down quark only are given in
Table~\ref{tab:down}.
From these tables one can see that with these assumptions
the NSI parameters are well determined to be smaller than one
at more than $3\sigma$
in both the spectral and the joint spectral and temporal analyses.
Since the joint spectral and temporal analysis is more restrictive,
we present only the corresponding correlated allowed regions
in different planes of the NSI parameters
in Figure~\ref{fig:5u} for interactions with up quarks only
and in Figure~\ref{fig:5d} for interactions with down quarks only.

Figures~\ref{fig:5u} and~\ref{fig:5d} show the comparison of the
COHERENT allowed regions in the
$(\varepsilon_{ee}^{uV},\varepsilon_{\mu\mu}^{uV})$
and
$(\varepsilon_{ee}^{dV},\varepsilon_{\mu\mu}^{dV})$
planes,
respectively,
with the solar LMA-Dark~\cite{Miranda:2004nb}
allowed regions reported in Ref.~\cite{Coloma:2017ncl}.
One can see that the allowed regions are incompatible at more than $3\sigma$,
disfavoring the LMA-Dark fit of solar neutrino data more than
in Ref.~\cite{Coloma:2017ncl},
where the analysis of the COHERENT data was performed
considering only the total number of events
\bluechange{%
%\marginnote{\bf\color{red}$\leftarrow$B7}%
and with the large 25\% uncertainty
of the original COHERENT quenching factor~\cite{Akimov:2017ade}.%
}

Combining the $\chi^2$'s of the analyses of COHERENT spectral data
with the marginal $\Delta\chi^2$'s of the LMA-Dark and LMA
fits of solar neutrino data in Fig.~1 of Ref.~\cite{Coloma:2017ncl},
we found a $\chi^2_{\text{min}}$ difference between LMA-Dark and LMA of
$24.5$
and
$46.5$,
respectively,
for NSI with up and down quark.
Therefore,
LMA-Dark is excluded at
$4.9\sigma$
and
$6.8\sigma$,
respectively,
for NSI with up and down quark,
for one degree of freedom.
These exclusions are already much stronger than the
$3.1\sigma$
and
$3.6\sigma$
obtained in Ref.~\cite{Coloma:2017ncl}.
We further improved the comparison between
LMA-Dark and LMA by considering the COHERENT spectral and temporal data,
that lead to the exclusion of LMA-Dark at
$5.6\sigma$
($\Delta\chi^2_{\text{min}} = 31.3$)
and
$7.2\sigma$
($\Delta\chi^2_{\text{min}} = 52.6$),
respectively,
for NSI with up and down quark.

One can note that our allowed region from COHERENT data in the
$(\varepsilon_{ee}^{uV},\varepsilon_{\mu\mu}^{uV})$
plane has a different shape than that in Figure~2 of Ref.~\cite{Coloma:2017ncl},
that has a hole around about $(0.2,0.2)$.
The only explanation that we found of this difference is that
in Ref.~\cite{Coloma:2017ncl}
the allowed region was not calculated marginalizing over
the remaining off-diagonal NSI parameters of the interaction with up quarks.
Indeed,
a hole in the allowed region of
$(\varepsilon_{ee}^{uV},\varepsilon_{\mu\mu}^{uV})$
appears around about $(0.2,0.2)$
if only the first line in Eq.~(\ref{Qalpha2}) is considered
and corresponds to its suppression.
Neglecting
$g_{V}^{p} \simeq 0.023$
and the form factors,
for $\varepsilon_{\alpha\alpha}^{dV}=0$
the first line in Eq.~(\ref{Qalpha2}) vanishes for
\begin{equation}
\varepsilon_{\alpha\alpha}^{uV}
\simeq
\dfrac{N}{2 \left( 2 Z + N \right)}
\simeq
0.21
\qquad
(\alpha=e,\mu)
,
\label{uhole}
\end{equation}
considering the average CsI values $Z=54$ and $N=76$.
If one does not consider the effects of the
off-diagonal NSI parameters of the interaction with up quarks
in the second line of Eq.~(\ref{Qalpha2})
the cross section is suppressed around
$(\varepsilon_{ee}^{uV},\varepsilon_{\mu\mu}^{uV}) \simeq (0.2,0.2)$
generating a hole in the allowed region.
However, appropriate values of the off-diagonal NSI parameters of the interaction with up quarks
can compensate the suppression of the first line in Eq.~(\ref{Qalpha2}),
filling the hole.
That is why our allowed region in Figure~\ref{fig:5u} has no hole.

\begin{figure*}[!h]
\centering
\begin{tabular}{cc}
\includegraphics*[width=0.49\linewidth]{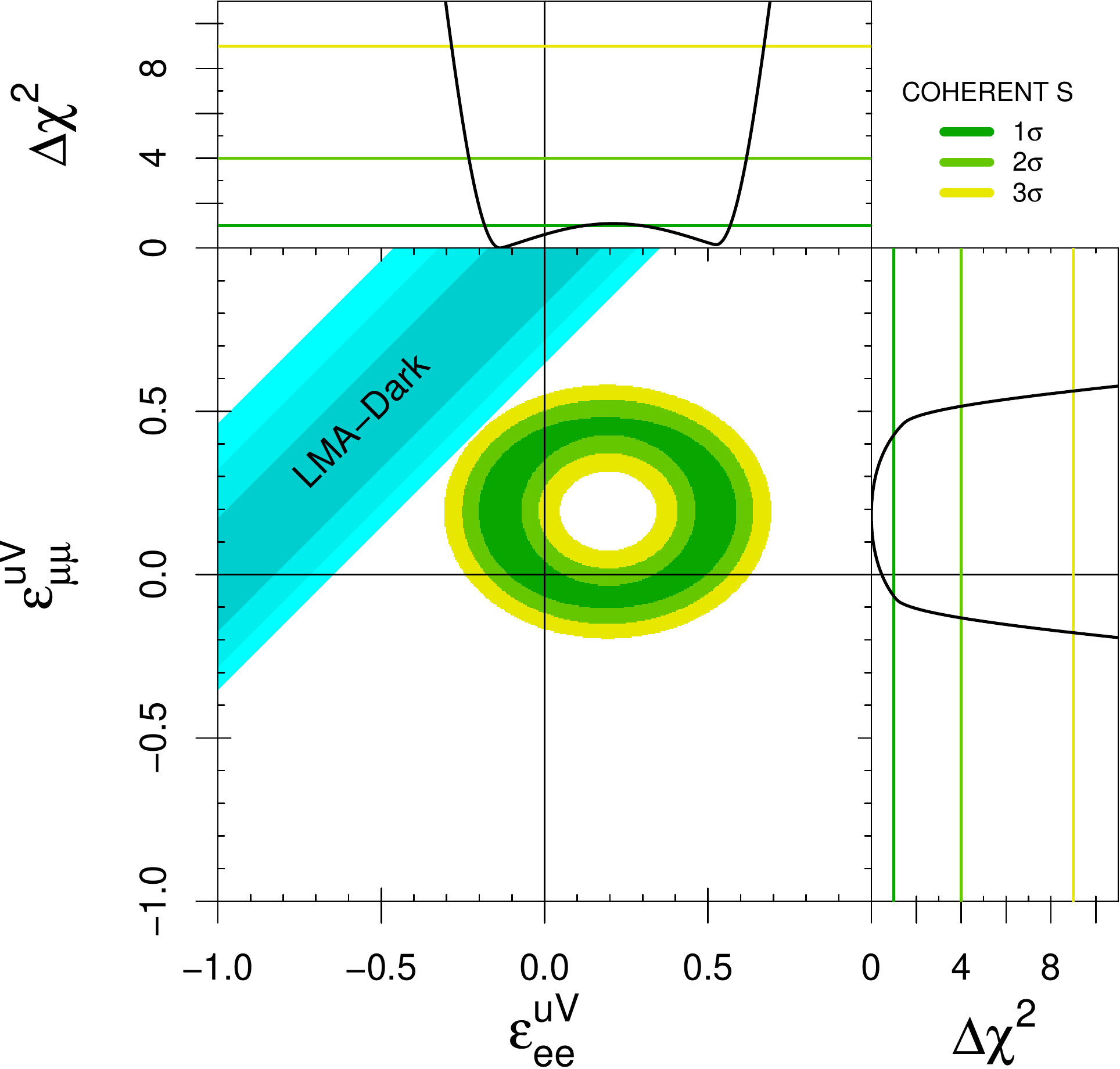}
&
\includegraphics*[width=0.49\linewidth]{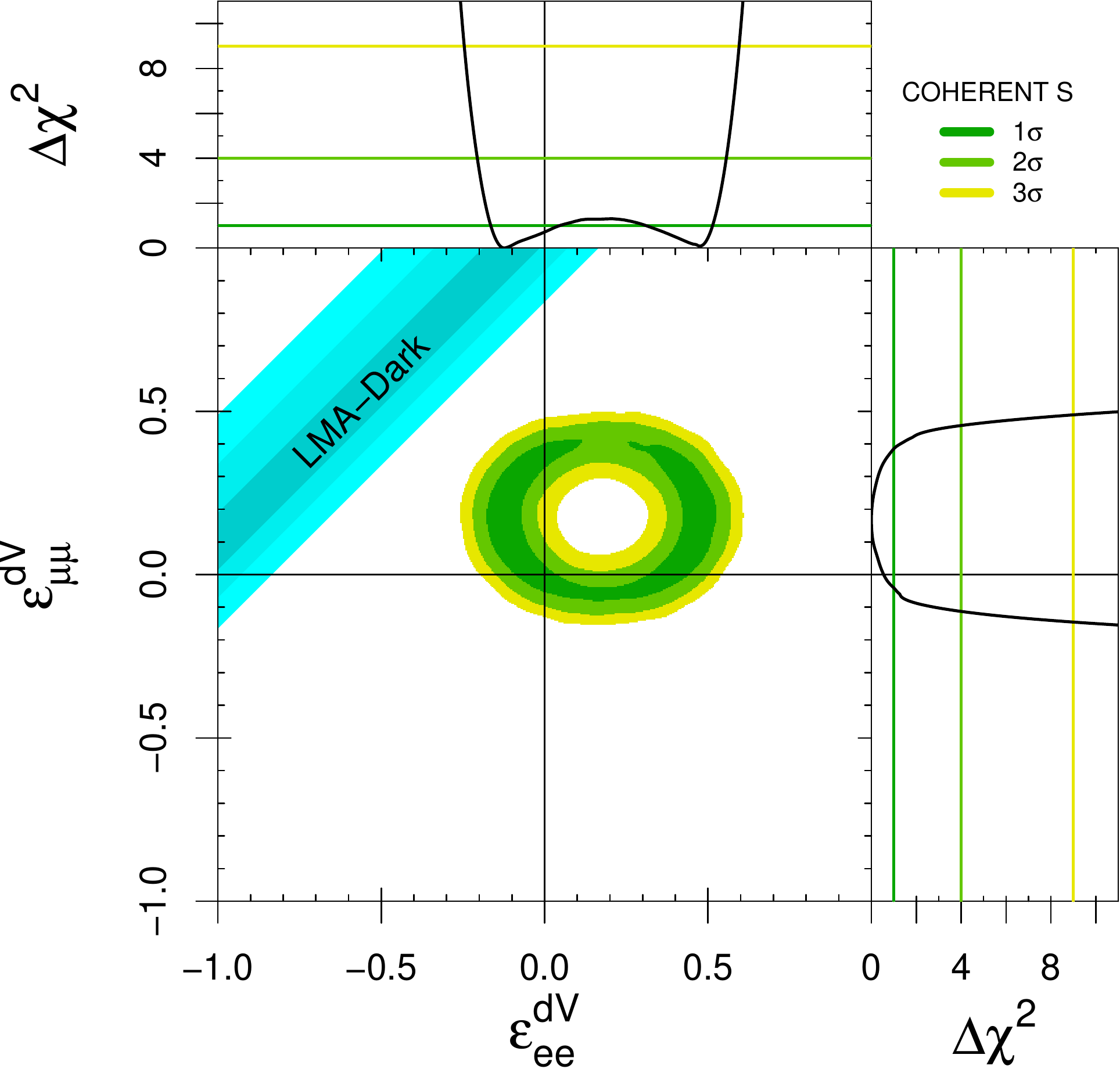}
\end{tabular}
\caption{ \label{fig:spe2}
Allowed regions and marginal $\Delta\chi^2$'s
in the plane
$(\varepsilon_{ee}^{uV},\varepsilon_{\mu\mu}^{uV})$
assuming that only these two NSI parameters are non-vanishing (left)
and
in the plane
$(\varepsilon_{ee}^{dV},\varepsilon_{\mu\mu}^{dV})$
assuming that only these two NSI parameters are non-vanishing (right).
Results obtained from the analysis of the COHERENT spectral (S) data alone.
The diagonal cyan strips
are allowed at $1\sigma$, $2\sigma$, and $3\sigma$
by the LMA-Dark fit of solar neutrino data~\cite{Coloma:2017ncl}.
}
\end{figure*}

\begin{figure*}[!h]
\centering
\begin{tabular}{cc}
\includegraphics*[width=0.49\linewidth]{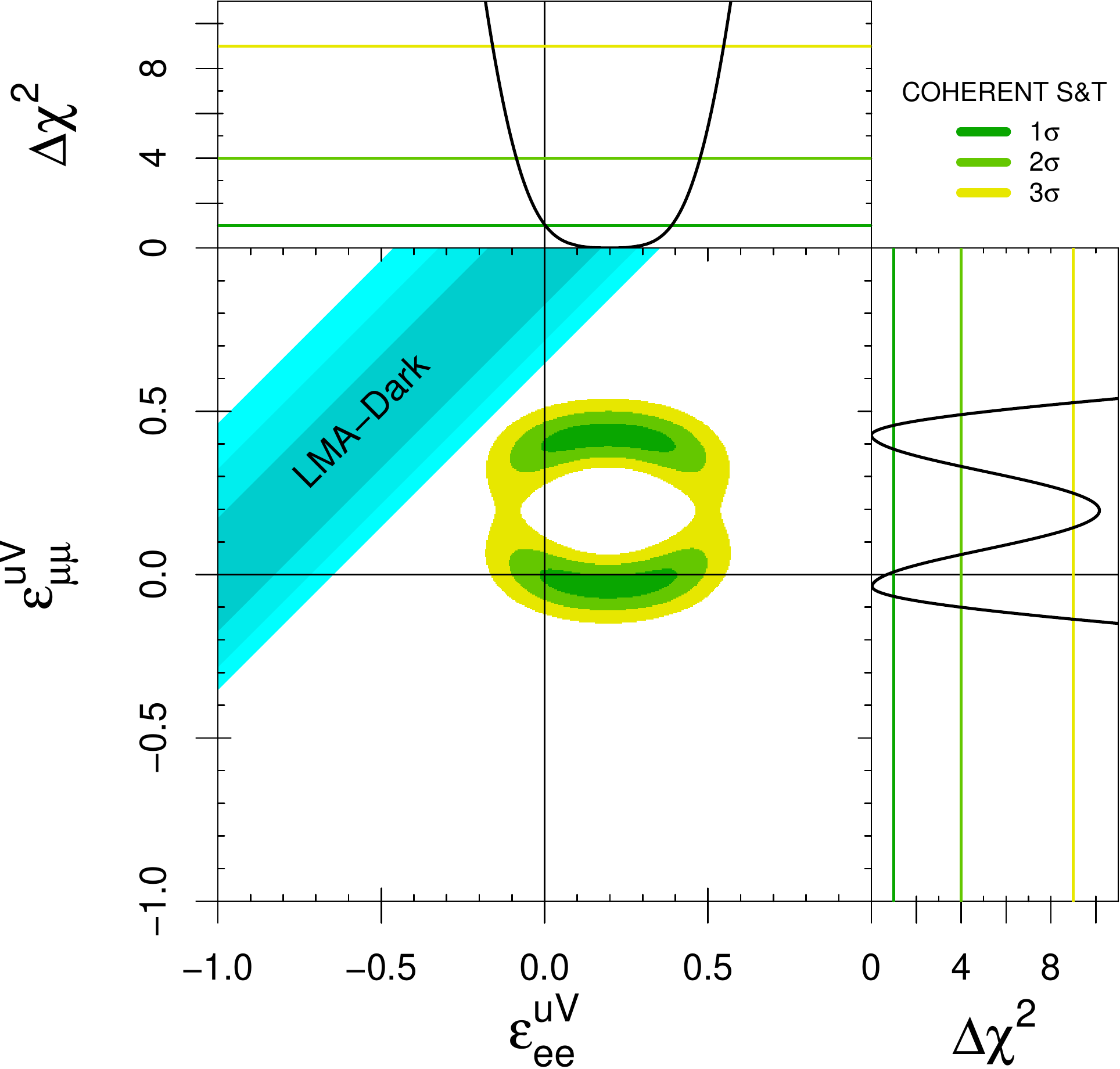}
&
\includegraphics*[width=0.49\linewidth]{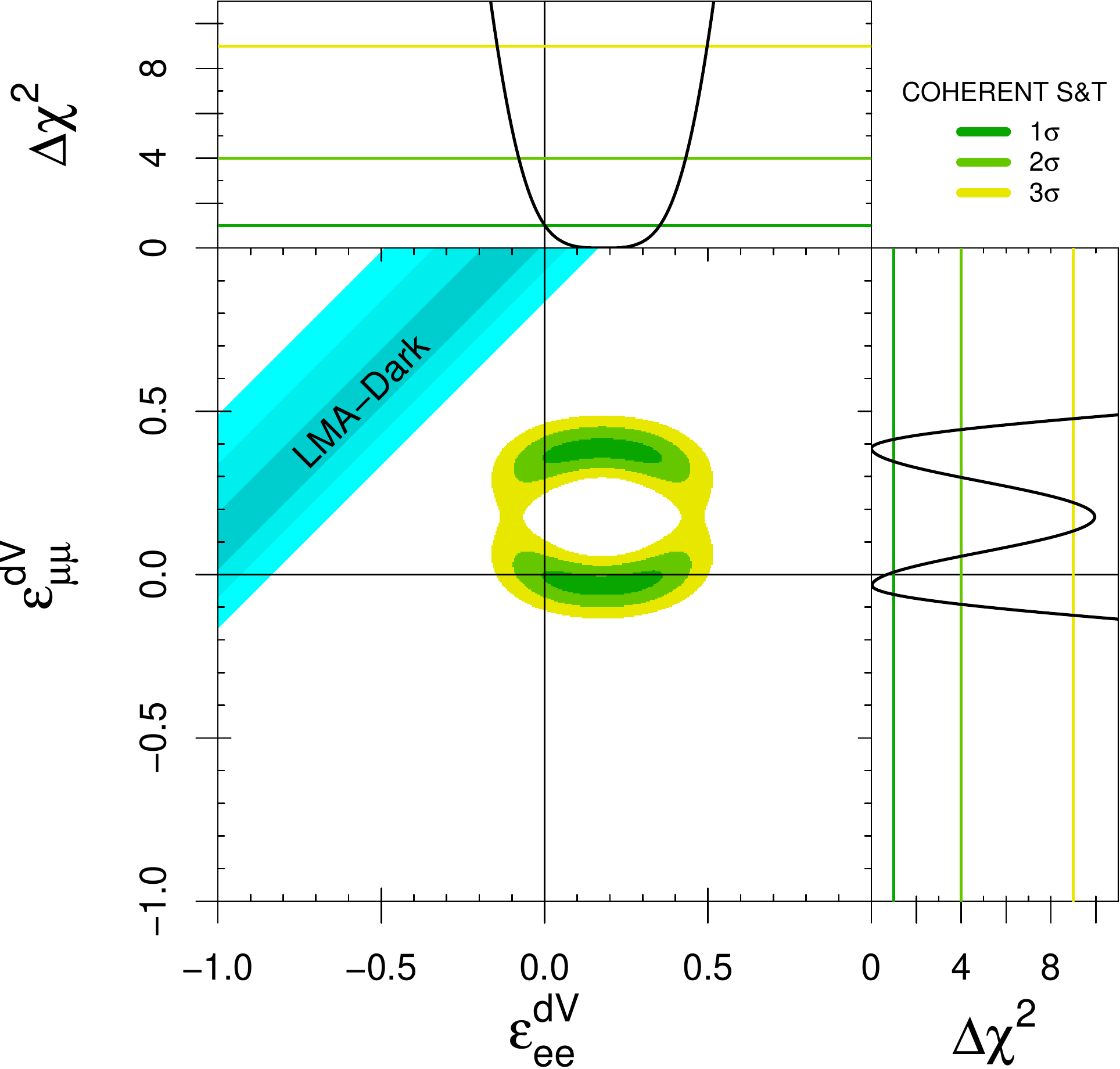}
\end{tabular}
\caption{ \label{fig:tim2}
Allowed regions and marginal $\Delta\chi^2$'s
in the plane
$(\varepsilon_{ee}^{uV},\varepsilon_{\mu\mu}^{uV})$
assuming that only these two NSI parameters are non-vanishing (left)
and
in the plane
$(\varepsilon_{ee}^{dV},\varepsilon_{\mu\mu}^{dV})$
assuming that only these two NSI parameters are non-vanishing (right).
Results obtained from the analysis of the joint COHERENT spectral and temporal (S\&T) data.
The diagonal cyan strips
are allowed at $1\sigma$, $2\sigma$, and $3\sigma$
by the LMA-Dark fit of solar neutrino data~\cite{Coloma:2017ncl}.
}
\end{figure*}

As a further check,
we present in the left panel of Figure~\ref{fig:spe2}
the allowed region in the
$(\varepsilon_{ee}^{uV},\varepsilon_{\mu\mu}^{uV})$
plane that we obtained assuming that only these two NSI parameters are non-vanishing
and fitting the COHERENT spectral data alone.
One can see that there is a hole
around
$(0.2,0.2)$
and the shape is similar to that in Ref.~\cite{Coloma:2017ncl}.
The left panel in Figure~\ref{fig:tim2}
shows that the allowed region reduces significantly in the analysis
of the joint COHERENT spectral and temporal data,
increasing the tension with the LMA-Dark fit of solar neutrino data.

The allowed region in the
$(\varepsilon_{ee}^{uV},\varepsilon_{\mu\mu}^{uV})$
plane assuming only these two non-vanishing NSI parameters
was obtained also in Ref.~\cite{Khan:2019mju}
by fitting the COHERENT spectral data.
Our allowed region in the left panel of Figure~\ref{fig:spe2}
\bluechange{%
%\marginnote{\bf\color{red}$\leftarrow$B8}%
is similar to that in the right panel of Fig.~7 of Ref.~\cite{Khan:2019mju},
confirming the validity of the analysis.%
}

For completeness, we present in the right panels of Figure~\ref{fig:spe2} and~\ref{fig:tim2}
also the allowed regions in the
$(\varepsilon_{ee}^{dV},\varepsilon_{\mu\mu}^{dV})$
plane that we obtained assuming that only these two NSI parameters are non-vanishing.
In this case the hole corresponding to the suppression of
the first line in Eq.~(\ref{Qalpha2})
occurs for
\begin{equation}
\varepsilon_{\alpha\alpha}^{dV}
\simeq
\dfrac{N}{2 \left( Z + 2 N \right)}
\simeq
0.18
\qquad
(\alpha=e,\mu)
,
\label{dhole}
\end{equation}
considering the average CsI values $Z=54$ and $N=76$.
The figures show also the strong tension with the LMA-Dark fit of solar neutrino data.

\section{COHERENT constraints on dominant individual NSI parameters}
\label{sec:dominant}

\begin{figure*}[!h]
\centering
\begin{tabular}{c}
\includegraphics*[width=0.99\linewidth]{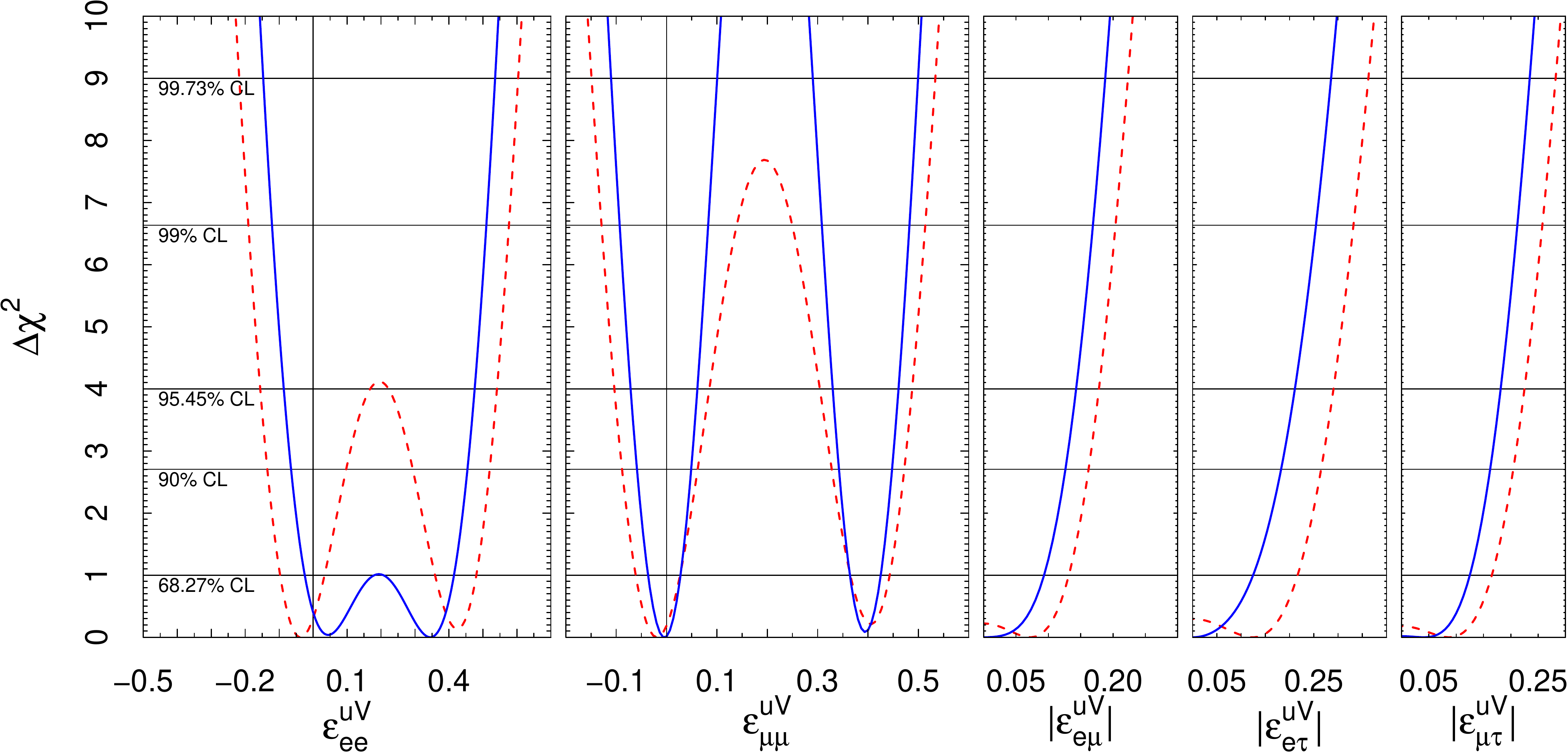}
\\
\includegraphics*[width=0.99\linewidth]{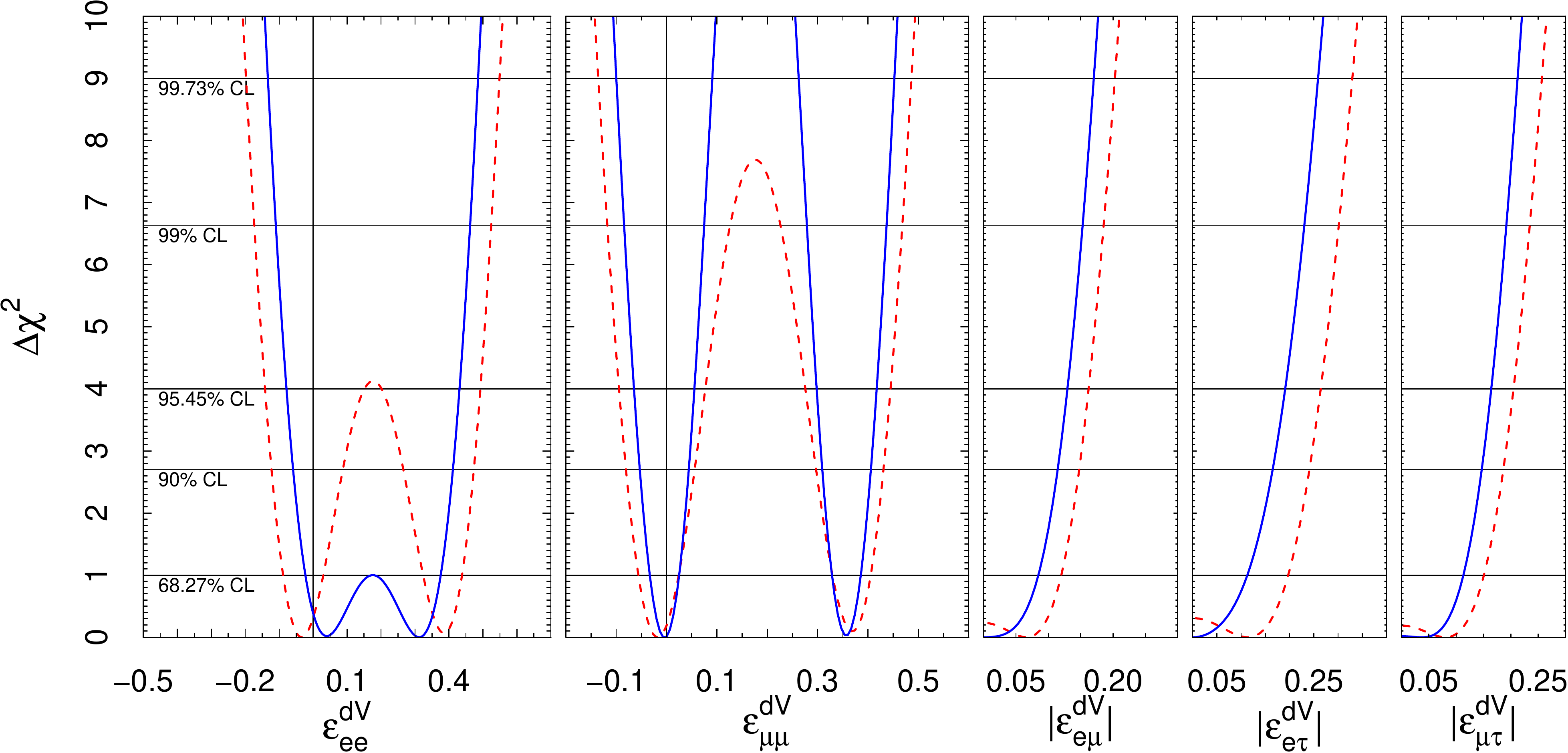}
\end{tabular}
\caption{ \label{fig:1}
$\Delta\chi^2 = \chi^2 - \chi^2_{\text{min}}$
for each of the NSI parameters
assuming it to be the only non-vanishing one.
The dashed red and solid blue curves correspond,
respectively,
to the analysis of spectral and joint spectral and temporal COHERENT data.
}
\end{figure*}

In this Section we present the results of the analyses of COHERENT data
assuming that only one of the NSI parameters is dominant and the others have negligible effects.
The allowed intervals for each parameter are listed in Table~\ref{tab:1}.
Figure~\ref{fig:1} shows the $\Delta\chi^2 = \chi^2 - \chi^2_{\text{min}}$
for each parameter.
One can see that for the diagonal NSI parameters and some confidence levels there are
disconnected allowed intervals,
because the $\Delta\chi^2$ is not parabolic,
but has a local central maximum.
This occurs for the same reason of the hole in the
two-dimensional plots in Figures~\ref{fig:spe2} and~\ref{fig:tim2},
because there is a cancellation in the first line of Eq.~(\ref{Qalpha2})
that suppresses the cross section and gives a bad fit of the data.
Indeed,
the local central maximum occurs at a value of about $0.2$,
in agreement with Eqs.~(\ref{uhole}) and~(\ref{dhole}).

Table~\ref{tab:1} and Figure~\ref{fig:1}
show that the individual NSI parameters are better determined
with the analysis
of the joint COHERENT spectral and temporal data
than with the analysis of the spectral data alone.
The resulting bounds are more stringent than those
obtained recently in Refs.~\cite{Papoulias:2019txv,Khan:2019mju}.

In particular,
the diagonal $\nu_{\mu}$ NSI parameters
$\varepsilon_{\mu\mu}^{uV}$
and
$\varepsilon_{\mu\mu}^{dV}$
are well constrained in two disconnected intervals
at $3\sigma$
with the joint spectral and temporal analysis.
In general,
the constraints on the $\nu_{\mu}$ NSI parameters
are more stringent than those on the $\nu_{e}$ NSI parameters
because there are two $\nu_{\mu}$ fluxes,
one from $\pi^{+}$ decay and one, of $\bar\nu_{\mu}$, from $\mu^{+}$ decay,
whereas there is only one flux of $\nu_{e}$
from $\mu^{+}$ decay.
One can also note that
$\varepsilon_{e\mu}^{uV}$
and
$\varepsilon_{e\mu}^{dV}$
are more constrained than the other off-diagonal
NSI parameters,
because they contribute to all the interactions of
$\nu_{\mu}$,
$\bar\nu_{\mu}$, and
$\nu_{e}$.
The less constrained off-diagonal NSI parameters are
$\varepsilon_{e\tau}^{uV}$
and
$\varepsilon_{e\tau}^{dV}$,
that contribute only to the interactions of the $\nu_{e}$ flux.

\section{Conclusions}
\label{sec:conclusions}

In this work we performed a systematic study of the
constraints on neutrino neutral-current non-standard interactions
that can be obtained from the
analysis of the COHERENT spectral and temporal data.
We have shown that the joint analysis of the COHERENT spectral and temporal data
gives more information on the NSI parameters
than the analysis of the COHERENT spectral data alone.
This is a general feature for quantities
that depend on the neutrino flavor,
as already emphasized in Refs.~\cite{Cadeddu:2018dux,Cadeddu:2019eta}.

First, we considered the general case in which all the ten
neutral-current NSI parameters are considered as free.
We have shown that in this case the analysis of the COHERENT data give very weak constraints
the NSI parameters,
because the contributions of the NSI parameters with up and down quarks can almost entirely cancel
each other.
The up-down average parameters in Eq.~(\ref{ave})
are relatively well constrained,
but the strongest constraints are obtained for the
maximally constrained up-down linear combination of NSI parameters in Eq.~(\ref{tilde}).

We also considered the case in which there are only NSI with either up or down quarks,
that was considered in Ref.~\cite{Coloma:2017ncl}
in order to test the LMA-Dark fit of solar neutrino data.
In this case, we obtained very stringent constraints
on the NSI parameters,
that exclude LMA-Dark at
$5.6\sigma$
and
$7.2\sigma$
respectively,
for NSI with up and down quark.
These exclusions are much stronger than the
$3.1\sigma$
and
$3.6\sigma$
obtained in Ref.~\cite{Coloma:2017ncl}.

We finally considered also the case of only one dominant
NSI parameter,
assuming that the effects of the others is negligible.
This is the simplest analysis that has been performed recently also by other
authors~\cite{Papoulias:2019txv,Khan:2019mju}.
We obtained more stringent constraints on each individual NSI parameter
through the joint analysis of the COHERENT spectral and temporal data.

%\bibliographystyle{apsrev4-1}
%\bibliography{nsi}

\begin{thebibliography}{33}%
\makeatletter
\providecommand \@ifxundefined [1]{%
\@ifx{#1\undefined}
}%
\providecommand \@ifnum [1]{%
\ifnum #1\expandafter \@firstoftwo
\else \expandafter \@secondoftwo
\fi
}%
\providecommand \@ifx [1]{%
\ifx #1\expandafter \@firstoftwo
\else \expandafter \@secondoftwo
\fi
}%
\providecommand \natexlab [1]{#1}%
\providecommand \enquote [1]{``#1''}%
\providecommand \bibnamefont [1]{#1}%
\providecommand \bibfnamefont [1]{#1}%
\providecommand \citenamefont [1]{#1}%
\providecommand \href@noop [0]{\@secondoftwo}%
\providecommand \href [0]{\begingroup \@sanitize@url \@href}%
\providecommand \@href[1]{\@@startlink{#1}\@@href}%
\providecommand \@@href[1]{\endgroup#1\@@endlink}%
\providecommand \@sanitize@url [0]{\catcode `\\12\catcode `\$12\catcode
`\&12\catcode `\#12\catcode `\^12\catcode `\_12\catcode `\%12\relax}%
\providecommand \@@startlink[1]{}%
\providecommand \@@endlink[0]{}%
\providecommand \url [0]{\begingroup\@sanitize@url \@url }%
\providecommand \@url [1]{\endgroup\@href {#1}{\urlprefix }}%
\providecommand \urlprefix [0]{URL }%
\providecommand \Eprint [0]{\href }%
\providecommand \doibase [0]{http://dx.doi.org/}%
\providecommand \selectlanguage [0]{\@gobble}%
\providecommand \bibinfo [0]{\@secondoftwo}%
\providecommand \bibfield [0]{\@secondoftwo}%
\providecommand \translation [1]{[#1]}%
\providecommand \BibitemOpen [0]{}%
\providecommand \bibitemStop [0]{}%
\providecommand \bibitemNoStop [0]{.\EOS\space}%
\providecommand \EOS [0]{\spacefactor3000\relax}%
\providecommand \BibitemShut [1]{\csname bibitem#1\endcsname}%
\let\auto@bib@innerbib\@empty
%</preamble>
\bibitem [{\citenamefont {Akimov}\ \emph {et~al.}(2017)\citenamefont {Akimov}
\emph {et~al.}}]{Akimov:2017ade}%
\BibitemOpen
\bibfield {author} {\bibinfo {author} {\bibfnamefont {D.}~\bibnamefont
{Akimov}} \emph {et~al.} (\bibinfo {collaboration} {COHERENT}),\ }\href
{\doibase 10.1126/science.aao0990} {\bibfield {journal} {\bibinfo {journal}
{Science}\ }\textbf {\bibinfo {volume} {357}},\ \bibinfo {pages} {1123}
(\bibinfo {year} {2017})},\ \Eprint {http://arxiv.org/abs/arXiv:1708.01294}
{arXiv:1708.01294 [nucl-ex]} \BibitemShut {NoStop}%
\bibitem [{\citenamefont {Freedman}(1974)}]{Freedman:1973yd}%
\BibitemOpen
\bibfield {author} {\bibinfo {author} {\bibfnamefont {D.~Z.}\ \bibnamefont
{Freedman}},\ }\href {\doibase 10.1103/PhysRevD.9.1389} {\bibfield {journal}
{\bibinfo {journal} {Phys. Rev.}\ }\textbf {\bibinfo {volume} {D9}},\
\bibinfo {pages} {1389} (\bibinfo {year} {1974})}\BibitemShut {NoStop}%
\bibitem [{\citenamefont {Freedman}\ \emph {et~al.}(1977)\citenamefont
{Freedman}, \citenamefont {Schramm},\ and\ \citenamefont
{Tubbs}}]{Freedman:1977xn}%
\BibitemOpen
\bibfield {author} {\bibinfo {author} {\bibfnamefont {D.~Z.}\ \bibnamefont
{Freedman}}, \bibinfo {author} {\bibfnamefont {D.~N.}\ \bibnamefont
{Schramm}}, \ and\ \bibinfo {author} {\bibfnamefont {D.~L.}\ \bibnamefont
{Tubbs}},\ }\href {\doibase 10.1146/annurev.ns.27.120177.001123} {\bibfield
{journal} {\bibinfo {journal} {Ann. Rev. Nucl. Part. Sci.}\ }\textbf
{\bibinfo {volume} {27}},\ \bibinfo {pages} {167} (\bibinfo {year}
{1977})}\BibitemShut {NoStop}%
\bibitem [{\citenamefont {Drukier}\ and\ \citenamefont
{Stodolsky}(1984)}]{Drukier:1983gj}%
\BibitemOpen
\bibfield {author} {\bibinfo {author} {\bibfnamefont {A.}~\bibnamefont
{Drukier}}\ and\ \bibinfo {author} {\bibfnamefont {L.}~\bibnamefont
{Stodolsky}},\ }\href {\doibase 10.1103/PhysRevD.30.2295} {\bibfield
{journal} {\bibinfo {journal} {Phys. Rev.}\ }\textbf {\bibinfo {volume}
{D30}},\ \bibinfo {pages} {2295} (\bibinfo {year} {1984})}\BibitemShut
{NoStop}%
\bibitem [{\citenamefont {Cadeddu}\ \emph
{et~al.}(2018{\natexlab{a}})\citenamefont {Cadeddu}, \citenamefont {Giunti},
\citenamefont {Li},\ and\ \citenamefont {Zhang}}]{Cadeddu:2017etk}%
\BibitemOpen
\bibfield {author} {\bibinfo {author} {\bibfnamefont {M.}~\bibnamefont
{Cadeddu}}, \bibinfo {author} {\bibfnamefont {C.}~\bibnamefont {Giunti}},
\bibinfo {author} {\bibfnamefont {Y.~F.}\ \bibnamefont {Li}}, \ and\ \bibinfo
{author} {\bibfnamefont {Y.~Y.}\ \bibnamefont {Zhang}},\ }\href@noop {}
{\bibfield {journal} {\bibinfo {journal} {Phys.Rev.Lett.}\ }\textbf
{\bibinfo {volume} {120}},\ \bibinfo {pages} {072501} (\bibinfo {year}
{2018}{\natexlab{a}})},\ \Eprint {http://arxiv.org/abs/arXiv:1710.02730}
{arXiv:1710.02730 [hep-ph]} \BibitemShut {NoStop}%
\bibitem [{\citenamefont {Papoulias}\ \emph {et~al.}(2020)\citenamefont
{Papoulias}, \citenamefont {Kosmas}, \citenamefont {Sahu}, \citenamefont
{Kota},\ and\ \citenamefont {Hota}}]{Papoulias:2019lfi}%
\BibitemOpen
\bibfield {author} {\bibinfo {author} {\bibfnamefont {D.~K.}\ \bibnamefont
{Papoulias}}, \bibinfo {author} {\bibfnamefont {T.~S.}\ \bibnamefont
{Kosmas}}, \bibinfo {author} {\bibfnamefont {R.}~\bibnamefont {Sahu}},
\bibinfo {author} {\bibfnamefont {V.~K.~B.}\ \bibnamefont {Kota}}, \ and\
\bibinfo {author} {\bibfnamefont {M.}~\bibnamefont {Hota}},\ }\href@noop {}
{\bibfield {journal} {\bibinfo {journal} {Phys.Lett.}\ }\textbf {\bibinfo
{volume} {B800}},\ \bibinfo {pages} {135133} (\bibinfo {year} {2020})},\
\Eprint {http://arxiv.org/abs/arXiv:1903.03722} {arXiv:1903.03722 [hep-ph]}
\BibitemShut {NoStop}%
\bibitem [{\citenamefont {Huang}\ and\ \citenamefont
{Chen}(2019)}]{Huang:2019ene}%
\BibitemOpen
\bibfield {author} {\bibinfo {author} {\bibfnamefont {X.-R.}\ \bibnamefont
{Huang}}\ and\ \bibinfo {author} {\bibfnamefont {L.-W.}\ \bibnamefont
{Chen}},\ }\href@noop {} {\bibfield {journal} {\bibinfo {journal}
{Phys.Rev.}\ }\textbf {\bibinfo {volume} {D100}},\ \bibinfo {pages} {071301}
(\bibinfo {year} {2019})},\ \Eprint {http://arxiv.org/abs/arXiv:1902.07625}
{arXiv:1902.07625 [hep-ph]} \BibitemShut {NoStop}%
\bibitem [{\citenamefont {Papoulias}()}]{Papoulias:2019txv}%
\BibitemOpen
\bibfield {author} {\bibinfo {author} {\bibfnamefont {D.~K.}\ \bibnamefont
{Papoulias}},\ }\href@noop {} {\ }\Eprint
{http://arxiv.org/abs/arXiv:1907.11644} {arXiv:1907.11644 [hep-ph]}
\BibitemShut {NoStop}%
\bibitem [{\citenamefont {Khan}\ and\ \citenamefont
{Rodejohann}()}]{Khan:2019mju}%
\BibitemOpen
\bibfield {author} {\bibinfo {author} {\bibfnamefont {A.~N.}\ \bibnamefont
{Khan}}\ and\ \bibinfo {author} {\bibfnamefont {W.}~\bibnamefont
{Rodejohann}},\ }\href@noop {} {\ }\Eprint
{http://arxiv.org/abs/arXiv:1907.12444} {arXiv:1907.12444 [hep-ph]}
\BibitemShut {NoStop}%
\bibitem [{\citenamefont {Cadeddu}\ \emph {et~al.}()\citenamefont {Cadeddu},
\citenamefont {Dordei}, \citenamefont {Giunti}, \citenamefont {Li},\ and\
\citenamefont {Zhang}}]{Cadeddu:2019eta}%
\BibitemOpen
\bibfield {author} {\bibinfo {author} {\bibfnamefont {M.}~\bibnamefont
{Cadeddu}}, \bibinfo {author} {\bibfnamefont {F.}~\bibnamefont {Dordei}},
\bibinfo {author} {\bibfnamefont {C.}~\bibnamefont {Giunti}}, \bibinfo
{author} {\bibfnamefont {Y.}~\bibnamefont {Li}}, \ and\ \bibinfo {author}
{\bibfnamefont {Y.}~\bibnamefont {Zhang}},\ }\href@noop {} {\ }\Eprint
{http://arxiv.org/abs/arXiv:1908.06045} {arXiv:1908.06045 [hep-ph]}
\BibitemShut {NoStop}%
\bibitem [{\citenamefont {Coloma}\ \emph {et~al.}(2017)\citenamefont {Coloma},
\citenamefont {Gonzalez-Garcia}, \citenamefont {Maltoni},\ and\ \citenamefont
{Schwetz}}]{Coloma:2017ncl}%
\BibitemOpen
\bibfield {author} {\bibinfo {author} {\bibfnamefont {P.}~\bibnamefont
{Coloma}}, \bibinfo {author} {\bibfnamefont {M.~C.}\ \bibnamefont
{Gonzalez-Garcia}}, \bibinfo {author} {\bibfnamefont {M.}~\bibnamefont
{Maltoni}}, \ and\ \bibinfo {author} {\bibfnamefont {T.}~\bibnamefont
{Schwetz}},\ }\href@noop {} {\bibfield {journal} {\bibinfo {journal}
{Phys.Rev.}\ }\textbf {\bibinfo {volume} {D96}},\ \bibinfo {pages} {115007}
(\bibinfo {year} {2017})},\ \Eprint {http://arxiv.org/abs/arXiv:1708.02899}
{arXiv:1708.02899 [hep-ph]} \BibitemShut {NoStop}%
\bibitem [{\citenamefont {Liao}\ and\ \citenamefont
{Marfatia}(2017)}]{Liao:2017uzy}%
\BibitemOpen
\bibfield {author} {\bibinfo {author} {\bibfnamefont {J.}~\bibnamefont
{Liao}}\ and\ \bibinfo {author} {\bibfnamefont {D.}~\bibnamefont
{Marfatia}},\ }\href@noop {} {\bibfield {journal} {\bibinfo {journal}
{Phys.Lett.}\ }\textbf {\bibinfo {volume} {B775}},\ \bibinfo {pages} {54}
(\bibinfo {year} {2017})},\ \Eprint {http://arxiv.org/abs/arXiv:1708.04255}
{arXiv:1708.04255 [hep-ph]} \BibitemShut {NoStop}%
\bibitem [{\citenamefont {Papoulias}\ and\ \citenamefont
{Kosmas}(2018)}]{Kosmas:2017tsq}%
\BibitemOpen
\bibfield {author} {\bibinfo {author} {\bibfnamefont {D.~K.}\ \bibnamefont
{Papoulias}}\ and\ \bibinfo {author} {\bibfnamefont {T.~S.}\ \bibnamefont
{Kosmas}},\ }\href@noop {} {\bibfield {journal} {\bibinfo {journal}
{Phys.Rev.}\ }\textbf {\bibinfo {volume} {D97}},\ \bibinfo {pages} {033003}
(\bibinfo {year} {2018})},\ \Eprint {http://arxiv.org/abs/arXiv:1711.09773}
{arXiv:1711.09773 [hep-ph]} \BibitemShut {NoStop}%
\bibitem [{\citenamefont {Denton}\ \emph {et~al.}(2018)\citenamefont {Denton},
\citenamefont {Farzan},\ and\ \citenamefont {Shoemaker}}]{Denton:2018xmq}%
\BibitemOpen
\bibfield {author} {\bibinfo {author} {\bibfnamefont {P.~B.}\ \bibnamefont
{Denton}}, \bibinfo {author} {\bibfnamefont {Y.}~\bibnamefont {Farzan}}, \
and\ \bibinfo {author} {\bibfnamefont {I.~M.}\ \bibnamefont {Shoemaker}},\
}\href@noop {} {\bibfield {journal} {\bibinfo {journal} {JHEP}\ }\textbf
{\bibinfo {volume} {1807}},\ \bibinfo {pages} {037} (\bibinfo {year}
{2018})},\ \Eprint {http://arxiv.org/abs/arXiv:1804.03660} {arXiv:1804.03660
[hep-ph]} \BibitemShut {NoStop}%
\bibitem [{\citenamefont {Aristizabal~Sierra}\ \emph
{et~al.}(2018)\citenamefont {Aristizabal~Sierra}, \citenamefont {De~Romeri},\
and\ \citenamefont {Rojas}}]{AristizabalSierra:2018eqm}%
\BibitemOpen
\bibfield {author} {\bibinfo {author} {\bibfnamefont {D.}~\bibnamefont
{Aristizabal~Sierra}}, \bibinfo {author} {\bibfnamefont {V.}~\bibnamefont
{De~Romeri}}, \ and\ \bibinfo {author} {\bibfnamefont {N.}~\bibnamefont
{Rojas}},\ }\href@noop {} {\bibfield {journal} {\bibinfo {journal}
{Phys.Rev.}\ }\textbf {\bibinfo {volume} {D98}},\ \bibinfo {pages} {075018}
(\bibinfo {year} {2018})},\ \Eprint {http://arxiv.org/abs/arXiv:1806.07424}
{arXiv:1806.07424 [hep-ph]} \BibitemShut {NoStop}%
\bibitem [{\citenamefont {Cadeddu}\ \emph
{et~al.}(2018{\natexlab{b}})\citenamefont {Cadeddu}, \citenamefont {Giunti},
\citenamefont {Kouzakov}, \citenamefont {Li}, \citenamefont {Studenikin},\
and\ \citenamefont {Zhang}}]{Cadeddu:2018dux}%
\BibitemOpen
\bibfield {author} {\bibinfo {author} {\bibfnamefont {M.}~\bibnamefont
{Cadeddu}}, \bibinfo {author} {\bibfnamefont {C.}~\bibnamefont {Giunti}},
\bibinfo {author} {\bibfnamefont {K.}~\bibnamefont {Kouzakov}}, \bibinfo
{author} {\bibfnamefont {Y.~F.}\ \bibnamefont {Li}}, \bibinfo {author}
{\bibfnamefont {A.}~\bibnamefont {Studenikin}}, \ and\ \bibinfo {author}
{\bibfnamefont {Y.~Y.}\ \bibnamefont {Zhang}},\ }\href@noop {} {\bibfield
{journal} {\bibinfo {journal} {Phys.Rev.}\ }\textbf {\bibinfo {volume}
{D98}},\ \bibinfo {pages} {113010} (\bibinfo {year} {2018}{\natexlab{b}})},\
\Eprint {http://arxiv.org/abs/arXiv:1810.05606} {arXiv:1810.05606 [hep-ph]}
\BibitemShut {NoStop}%
\bibitem [{\citenamefont {Esteban}\ \emph {et~al.}(2018)\citenamefont
{Esteban}, \citenamefont {Gonzalez-Garcia}, \citenamefont {Maltoni},
\citenamefont {Martinez-Soler},\ and\ \citenamefont
{Salvado}}]{Esteban:2018ppq}%
\BibitemOpen
\bibfield {author} {\bibinfo {author} {\bibfnamefont {I.}~\bibnamefont
{Esteban}}, \bibinfo {author} {\bibfnamefont {M.}~\bibnamefont
{Gonzalez-Garcia}}, \bibinfo {author} {\bibfnamefont {M.}~\bibnamefont
{Maltoni}}, \bibinfo {author} {\bibfnamefont {I.}~\bibnamefont
{Martinez-Soler}}, \ and\ \bibinfo {author} {\bibfnamefont {J.}~\bibnamefont
{Salvado}},\ }\href@noop {} {\bibfield {journal} {\bibinfo {journal}
{JHEP}\ }\textbf {\bibinfo {volume} {1808}},\ \bibinfo {pages} {180}
(\bibinfo {year} {2018})},\ \Eprint {http://arxiv.org/abs/arXiv:1805.04530}
{arXiv:1805.04530 [hep-ph]} \BibitemShut {NoStop}%
\bibitem [{\citenamefont {Cadeddu}\ and\ \citenamefont
{Dordei}(2019)}]{Cadeddu:2018izq}%
\BibitemOpen
\bibfield {author} {\bibinfo {author} {\bibfnamefont {M.}~\bibnamefont
{Cadeddu}}\ and\ \bibinfo {author} {\bibfnamefont {F.}~\bibnamefont
{Dordei}},\ }\href@noop {} {\bibfield {journal} {\bibinfo {journal}
{Phys.Rev.}\ }\textbf {\bibinfo {volume} {D99}},\ \bibinfo {pages} {033010}
(\bibinfo {year} {2019})},\ \Eprint {http://arxiv.org/abs/arXiv:1808.10202}
{arXiv:1808.10202 [hep-ph]} \BibitemShut {NoStop}%
\bibitem [{\citenamefont {Dutta}\ \emph {et~al.}(2019)\citenamefont {Dutta},
\citenamefont {Liao}, \citenamefont {Sinha},\ and\ \citenamefont
{Strigari}}]{Dutta:2019eml}%
\BibitemOpen
\bibfield {author} {\bibinfo {author} {\bibfnamefont {B.}~\bibnamefont
{Dutta}}, \bibinfo {author} {\bibfnamefont {S.}~\bibnamefont {Liao}},
\bibinfo {author} {\bibfnamefont {S.}~\bibnamefont {Sinha}}, \ and\ \bibinfo
{author} {\bibfnamefont {L.~E.}\ \bibnamefont {Strigari}},\ }\href@noop {}
{\bibfield {journal} {\bibinfo {journal} {Phys.Rev.Lett.}\ }\textbf
{\bibinfo {volume} {123}},\ \bibinfo {pages} {061801} (\bibinfo {year}
{2019})},\ \Eprint {http://arxiv.org/abs/arXiv:1903.10666} {arXiv:1903.10666
[hep-ph]} \BibitemShut {NoStop}%
\bibitem [{\citenamefont {Dutta}\ \emph {et~al.}()\citenamefont {Dutta},
\citenamefont {Kim}, \citenamefont {Liao}, \citenamefont {Park},
\citenamefont {Shin},\ and\ \citenamefont {Strigari}}]{Dutta:2019nbn}%
\BibitemOpen
\bibfield {author} {\bibinfo {author} {\bibfnamefont {B.}~\bibnamefont
{Dutta}}, \bibinfo {author} {\bibfnamefont {D.}~\bibnamefont {Kim}}, \bibinfo
{author} {\bibfnamefont {S.}~\bibnamefont {Liao}}, \bibinfo {author}
{\bibfnamefont {J.-C.}\ \bibnamefont {Park}}, \bibinfo {author}
{\bibfnamefont {S.}~\bibnamefont {Shin}}, \ and\ \bibinfo {author}
{\bibfnamefont {L.~E.}\ \bibnamefont {Strigari}},\ }\href@noop {} {\ }\Eprint
{http://arxiv.org/abs/arXiv:1906.10745} {arXiv:1906.10745 [hep-ph]}
\BibitemShut {NoStop}%
\bibitem [{\citenamefont {Collar}\ \emph {et~al.}(2019)\citenamefont {Collar},
\citenamefont {Kavner},\ and\ \citenamefont {Lewis}}]{Collar:2019ihs}%
\BibitemOpen
\bibfield {author} {\bibinfo {author} {\bibfnamefont {J.~I.}\ \bibnamefont
{Collar}}, \bibinfo {author} {\bibfnamefont {A.~R.~L.}\ \bibnamefont
{Kavner}}, \ and\ \bibinfo {author} {\bibfnamefont {C.~M.}\ \bibnamefont
{Lewis}},\ }\href@noop {} {\bibfield {journal} {\bibinfo {journal}
{Phys.Rev.}\ }\textbf {\bibinfo {volume} {D100}},\ \bibinfo {pages} {033003}
(\bibinfo {year} {2019})},\ \Eprint {http://arxiv.org/abs/arXiv:1907.04828}
{arXiv:1907.04828 [nucl-ex]} \BibitemShut {NoStop}%
\bibitem [{\citenamefont {Akimov}\ \emph {et~al.}()\citenamefont {Akimov} \emph
{et~al.}}]{Akimov:2018vzs}%
\BibitemOpen
\bibfield {author} {\bibinfo {author} {\bibfnamefont {D.}~\bibnamefont
{Akimov}} \emph {et~al.} (\bibinfo {collaboration} {COHERENT}),\ }\href@noop
{} {\ }\Eprint {http://arxiv.org/abs/arXiv:1804.09459} {arXiv:1804.09459
[nucl-ex]} \BibitemShut {NoStop}%
\bibitem [{\citenamefont {Miranda}\ \emph {et~al.}(2006)\citenamefont
{Miranda}, \citenamefont {Tortola},\ and\ \citenamefont
{Valle}}]{Miranda:2004nb}%
\BibitemOpen
\bibfield {author} {\bibinfo {author} {\bibfnamefont {O.~G.}\ \bibnamefont
{Miranda}}, \bibinfo {author} {\bibfnamefont {M.~A.}\ \bibnamefont
{Tortola}}, \ and\ \bibinfo {author} {\bibfnamefont {J.~W.~F.}\ \bibnamefont
{Valle}},\ }\href@noop {} {\bibfield {journal} {\bibinfo {journal} {JHEP}\
}\textbf {\bibinfo {volume} {10}},\ \bibinfo {pages} {008} (\bibinfo {year}
{2006})},\ \Eprint {http://arxiv.org/abs/hep-ph/0406280} {hep-ph/0406280}
\BibitemShut {NoStop}%
\bibitem [{\citenamefont {Ohlsson}(2013)}]{Ohlsson:2012kf}%
\BibitemOpen
\bibfield {author} {\bibinfo {author} {\bibfnamefont {T.}~\bibnamefont
{Ohlsson}},\ }\href@noop {} {\bibfield {journal} {\bibinfo {journal}
{Rept.Prog.Phys.}\ }\textbf {\bibinfo {volume} {76}},\ \bibinfo {pages}
{044201} (\bibinfo {year} {2013})},\ \Eprint
{http://arxiv.org/abs/arXiv:1209.2710} {arXiv:1209.2710 [hep-ph]}
\BibitemShut {NoStop}%
\bibitem [{\citenamefont {Miranda}\ and\ \citenamefont
{Nunokawa}(2015)}]{Miranda:2015dra}%
\BibitemOpen
\bibfield {author} {\bibinfo {author} {\bibfnamefont {O.}~\bibnamefont
{Miranda}}\ and\ \bibinfo {author} {\bibfnamefont {H.}~\bibnamefont
{Nunokawa}},\ }\href@noop {} {\bibfield {journal} {\bibinfo {journal} {New
J. Phys.}\ }\textbf {\bibinfo {volume} {17}},\ \bibinfo {pages} {095002}
(\bibinfo {year} {2015})},\ \Eprint {http://arxiv.org/abs/arXiv:1505.06254}
{arXiv:1505.06254 [hep-ph]} \BibitemShut {NoStop}%
\bibitem [{\citenamefont {Farzan}\ and\ \citenamefont
{Tortola}(2018)}]{Farzan:2017xzy}%
\BibitemOpen
\bibfield {author} {\bibinfo {author} {\bibfnamefont {Y.}~\bibnamefont
{Farzan}}\ and\ \bibinfo {author} {\bibfnamefont {M.}~\bibnamefont
{Tortola}},\ }\href@noop {} {\bibfield {journal} {\bibinfo {journal}
{Front.in Phys.}\ }\textbf {\bibinfo {volume} {6}},\ \bibinfo {pages} {10}
(\bibinfo {year} {2018})},\ \Eprint {http://arxiv.org/abs/arXiv:1710.09360}
{arXiv:1710.09360 [hep-ph]} \BibitemShut {NoStop}%
\bibitem [{\citenamefont {Dev}\ \emph {et~al.}()\citenamefont {Dev} \emph
{et~al.}}]{Dev:2019anc}%
\BibitemOpen
\bibfield {author} {\bibinfo {author} {\bibfnamefont {P.~S.~B.}\
\bibnamefont {Dev}} \emph {et~al.},\ }\href@noop {} {\ }\Eprint
{http://arxiv.org/abs/arXiv:1907.00991} {arXiv:1907.00991 [hep-ph]}
\BibitemShut {NoStop}%
\bibitem [{\citenamefont {Barranco}\ \emph {et~al.}(2005)\citenamefont
{Barranco}, \citenamefont {Miranda},\ and\ \citenamefont
{Rashba}}]{Barranco:2005yy}%
\BibitemOpen
\bibfield {author} {\bibinfo {author} {\bibfnamefont {J.}~\bibnamefont
{Barranco}}, \bibinfo {author} {\bibfnamefont {O.~G.}\ \bibnamefont
{Miranda}}, \ and\ \bibinfo {author} {\bibfnamefont {T.~I.}\ \bibnamefont
{Rashba}},\ }\href@noop {} {\bibfield {journal} {\bibinfo {journal} {JHEP}\
}\textbf {\bibinfo {volume} {0512}},\ \bibinfo {pages} {021} (\bibinfo {year}
{2005})},\ \Eprint {http://arxiv.org/abs/hep-ph/0508299} {hep-ph/0508299}
\BibitemShut {NoStop}%
\bibitem [{\citenamefont {Tanabashi}\ \emph {et~al.}(2018)\citenamefont
{Tanabashi} \emph {et~al.}}]{Tanabashi:2018oca}%
\BibitemOpen
\bibfield {author} {\bibinfo {author} {\bibfnamefont {M.}~\bibnamefont
{Tanabashi}} \emph {et~al.} (\bibinfo {collaboration} {Particle Data
Group}),\ }\href {\doibase 10.1103/PhysRevD.98.030001} {\bibfield {journal}
{\bibinfo {journal} {Phys. Rev.}\ }\textbf {\bibinfo {volume} {D98}},\
\bibinfo {pages} {030001} (\bibinfo {year} {2018})}\BibitemShut {NoStop}%
\bibitem [{\citenamefont {Helm}(1956)}]{Helm:1956zz}%
\BibitemOpen
\bibfield {author} {\bibinfo {author} {\bibfnamefont {R.~H.}\ \bibnamefont
{Helm}},\ }\href {\doibase 10.1103/PhysRev.104.1466} {\bibfield {journal}
{\bibinfo {journal} {Phys. Rev.}\ }\textbf {\bibinfo {volume} {104}},\
\bibinfo {pages} {1466} (\bibinfo {year} {1956})}\BibitemShut {NoStop}%
\bibitem [{\citenamefont {Friedrich}\ and\ \citenamefont
{Voegler}(1982)}]{Friedrich:1982esq}%
\BibitemOpen
\bibfield {author} {\bibinfo {author} {\bibfnamefont {J.}~\bibnamefont
{Friedrich}}\ and\ \bibinfo {author} {\bibfnamefont {N.}~\bibnamefont
{Voegler}},\ }\href {\doibase 10.1016/0375-9474(82)90147-6} {\bibfield
{journal} {\bibinfo {journal} {Nucl. Phys.}\ }\textbf {\bibinfo {volume}
{A373}},\ \bibinfo {pages} {192} (\bibinfo {year} {1982})}\BibitemShut
{NoStop}%
\bibitem [{\citenamefont {Fricke}\ \emph {et~al.}(1995)\citenamefont {Fricke},
\citenamefont {Bernhardt}, \citenamefont {Heilig}, \citenamefont {Schaller},
\citenamefont {Schellenberg}, \citenamefont {Shera},\ and\ \citenamefont
{de~Jager}}]{Fricke:1995zz}%
\BibitemOpen
\bibfield {author} {\bibinfo {author} {\bibfnamefont {G.}~\bibnamefont
{Fricke}}, \bibinfo {author} {\bibfnamefont {C.}~\bibnamefont {Bernhardt}},
\bibinfo {author} {\bibfnamefont {K.}~\bibnamefont {Heilig}}, \bibinfo
{author} {\bibfnamefont {L.~A.}\ \bibnamefont {Schaller}}, \bibinfo {author}
{\bibfnamefont {L.}~\bibnamefont {Schellenberg}}, \bibinfo {author}
{\bibfnamefont {E.~B.}\ \bibnamefont {Shera}}, \ and\ \bibinfo {author}
{\bibfnamefont {C.~W.}\ \bibnamefont {de~Jager}},\ }\href {\doibase
10.1006/adnd.1995.1007} {\bibfield {journal} {\bibinfo {journal} {Atom.
Data Nucl. Data Tabl.}\ }\textbf {\bibinfo {volume} {60}},\ \bibinfo {pages}
{177} (\bibinfo {year} {1995})}\BibitemShut {NoStop}%
\bibitem [{\citenamefont {Bender}\ \emph {et~al.}(1999)\citenamefont {Bender},
\citenamefont {Rutz}, \citenamefont {Reinhard}, \citenamefont {Maruhn},\ and\
\citenamefont {Greiner}}]{Bender:1999yt}%
\BibitemOpen
\bibfield {author} {\bibinfo {author} {\bibfnamefont {M.}~\bibnamefont
{Bender}}, \bibinfo {author} {\bibfnamefont {K.}~\bibnamefont {Rutz}},
\bibinfo {author} {\bibfnamefont {P.~G.}\ \bibnamefont {Reinhard}}, \bibinfo
{author} {\bibfnamefont {J.~A.}\ \bibnamefont {Maruhn}}, \ and\ \bibinfo
{author} {\bibfnamefont {W.}~\bibnamefont {Greiner}},\ }\href {\doibase
10.1103/PhysRevC.60.034304} {\bibfield {journal} {\bibinfo {journal} {Phys.
Rev.}\ }\textbf {\bibinfo {volume} {C60}},\ \bibinfo {pages} {034304}
(\bibinfo {year} {1999})},\ \Eprint {http://arxiv.org/abs/nucl-th/9906030}
{nucl-th/9906030 [nucl-th]} \BibitemShut {NoStop}%
\end{thebibliography}

% Tables

\begin{table*}[!h]
\begin{center}
\renewcommand{\arraystretch}{1.2}
\begin{tabular}{c|cccc|cccc}
&
\multicolumn{4}{c|}{Spectrum}
&
\multicolumn{4}{c}{Spectrum and Time}
\\
&
Best Fit
&
$1\sigma$
&
$2\sigma$
&
$3\sigma$
&
Best Fit
&
$1\sigma$
&
$2\sigma$
&
$3\sigma$
\\
\hline
$\varepsilon_{ee}^{uV}$
&
$-0.2$
&
$ -29 \div 29 $
&
$ -40 \div 39 $
&
$ -48 \div 48 $
&
$0$
&
$ -16 \div 16 $
&
$ -27 \div 27 $
&
$ -36 \div 36 $
\\
$\varepsilon_{ee}^{dV}$
&
$0.4$
&
$ -26 \div 27 $
&
$ -35 \div 37 $
&
$ -43 \div 44 $
&
$0.2$
&
$ -14 \div 15 $
&
$ -24 \div 25 $
&
$ -32 \div 33 $
\\
\hline
$\varepsilon_{\mu\mu}^{uV}$
&
$0.6$
&
$ -26 \div 26 $
&
$ -33 \div 33 $
&
$ -40 \div 40 $
&
$-0.2$
&
$ -11 \div 11 $
&
$ -20 \div 19 $
&
$ -26 \div 26 $
\\
$\varepsilon_{\mu\mu}^{dV}$
&
$-0.1$
&
$ -23 \div 24 $
&
$ -30 \div 30 $
&
$ -36 \div 37 $
&
$0.3$
&
$ -10 \div 11 $
&
$ -17 \div 18 $
&
$ -23 \div 24 $
\\
\hline
$|\varepsilon_{e\mu}^{uV}|$
&
0%$0$
&
$<19$
&
$<25$
&
$<31$
&
0%$0$
&
$<9$
&
$<16$
&
$<21$
\\
$|\varepsilon_{e\mu}^{dV}|$
&
0%$0$
&
$<17$
&
$<23$
&
$<28$
&
0%$0$
&
$<9$
&
$<15$
&
$<19$
\\
\hline
$|\varepsilon_{e\tau}^{uV}|$
&
0%$0$
&
$<30$
&
$<39$
&
$<49$
&
0%$0$
&
$<16$
&
$<27$
&
$<36$
\\
$|\varepsilon_{e\tau}^{dV}|$
&
0%$0$
&
$<27$
&
$<36$
&
$<44$
&
0%$0$
&
$<15$
&
$<24$
&
$<32$
\\
\hline
$|\varepsilon_{\mu\tau}^{uV}|$
&
0%$0$
&
$<26$
&
$<33$
&
$<40$
&
0%$0$
&
$<12$
&
$<19$
&
$<26$
\\
$|\varepsilon_{\mu\tau}^{dV}|$
&
0%$0$
&
$<24$
&
$<30$
&
$<36$
&
0%$0$
&
$<10$
&
$<18$
&
$<24$
\\
\hline
$\overline{\varepsilon}_{ee}^{V}$
&
$0.1$
&
$ -1.2 \div 1.4 $
&
$ -1.7 \div 1.9 $
&
$ -2.1 \div 2.3 $
&
$0.1$
&
$ -0.7 \div 0.9 $
&
$ -1.2 \div 1.4 $
&
$ -1.6 \div 1.8 $
\\
%\hline
$\overline{\varepsilon}_{\mu\mu}^{V}$
&
$0.1$
&
$ -1.1 \div 1.2 $
&
$ -1.4 \div 1.6 $
&
$ -1.7 \div 1.9 $
&
$0.1$
&
$ -0.4 \div 0.6 $
&
$ -0.8 \div 1.0 $
&
$ -1.1 \div 1.3 $
\\
%\hline
$|\overline{\varepsilon}_{e\mu}^{V}|$
&
$0.2$
&
$ < 0.9 $
&
$ < 1.1 $
&
$ < 1.4 $
&
$0.1$
&
$ < 0.5 $
&
$ < 0.7 $
&
$ < 1.0 $
\\
%\hline
$|\overline{\varepsilon}_{e\tau}^{V}|$
&
$0.2$
&
$ < 1.3 $
&
$ < 1.8 $
&
$ < 2.2 $
&
$0.1$
&
$ < 0.8 $
&
$ < 1.3 $
&
$ < 1.7 $
\\
%\hline
$|\overline{\varepsilon}_{\mu\tau}^{V}|$
&
$0.4$
&
$ < 1.2 $
&
$ < 1.5 $
&
$ < 1.8 $
&
$0.1$
&
$ < 0.6 $
&
$ < 0.9 $
&
$ < 1.2 $
\\
\hline
$\widetilde{\varepsilon}_{ee}^{V}$
&
$0.10$
&
$ -0.18 \div 0.37 $
&
$ -0.24 \div 0.43 $
&
$ -0.29 \div 0.48 $
&
$0.10$
&
$ -0.02 \div 0.21 $
&
$ -0.10 \div 0.29 $
&
$ -0.17 \div 0.35 $
\\
%\hline
$\widetilde{\varepsilon}_{\mu\mu}^{V}$
&
$0.11$
&
$ -0.16 \div 0.35 $
&
$ -0.20 \div 0.39 $
&
$ -0.24 \div 0.44 $
&
$0.10$
&
$ -0.04 \div 0.23 $
&
$ -0.08 \div 0.27 $
&
$ -0.12 \div 0.30 $
\\
%\hline
$|\widetilde{\varepsilon}_{e\mu}^{V}|$
&
$0.04$
&
$ < 0.19 $
&
$ < 0.22 $
&
$ < 0.26 $
&
$0.02$
&
$ < 0.10 $
&
$ < 0.14 $
&
$ < 0.18 $
\\
%\hline
$|\widetilde{\varepsilon}_{e\tau}^{V}|$
&
$0.07$
&
$ < 0.27 $
&
$ < 0.33 $
&
$ < 0.39 $
&
$0.02$
&
$ < 0.12 $
&
$ < 0.20 $
&
$ < 0.26 $
\\
%\hline
$|\widetilde{\varepsilon}_{\mu\tau}^{V}|$
&
$0.10$
&
$ < 0.25 $
&
$ < 0.30 $
&
$ < 0.34 $
&
$0.06$
&
$ < 0.14 $
&
$ < 0.18 $
&
$ < 0.21 $
%\\
%\hline
\end{tabular}
\end{center}
\caption{ \label{tab:general}
General marginalized constraints on the NSI parameters
obtained with the analyses of the spectral and the joint spectral and temporal
COHERENT data.
}
\end{table*}

\begin{table*}[!h]
\begin{center}
\renewcommand{\arraystretch}{1.2}
\begin{tabular}{cc|cc|cc|cc}
&
&
\multicolumn{2}{c|}{$1\sigma$}
&
\multicolumn{2}{c|}{$2\sigma$}
&
\multicolumn{2}{c}{$3\sigma$}
\\
&
&
$a$
&
$b$
&
$a$
&
$b$
&
$a$
&
$b$
\\
\hline
$(\varepsilon_{ee}^{uV},\varepsilon_{ee}^{dV})$
&
LB
&
$-0.9$
&
$0$
&
$-0.9$
&
$-0.1$
&
$-0.9$
&
$-0.2$
\\
&
UB
&
$-0.9$
&
$0.4$
&
$-0.9$
&
$0.5$
&
$-0.9$
&
$0.5$
\\
\hline
$(\varepsilon_{\mu\mu}^{uV},\varepsilon_{\mu\mu}^{dV})$
&
LB
&
$-0.9$
&
$-0.1$
&
$-0.9$
&
$-0.1$
&
$-0.9$
&
$-0.1$
\\
&
UB
&
$-0.9$
&
$0.4$
&
$-0.9$
&
$0.4$
&
$-0.9$
&
$0.5$
\\
\hline
$(\varepsilon_{e\mu}^{uV},\varepsilon_{e\mu}^{dV})$
&
LB
&
$-0.9$
&
$-0.2$
&
$-0.9$
&
$-0.2$
&
$-0.9$
&
$-0.2$
\\
&
UB
&
$-0.9$
&
$0.2$
&
$-0.9$
&
$0.2$
&
$-0.9$
&
$0.2$
\\
\hline
$(\varepsilon_{e\tau}^{uV},\varepsilon_{e\tau}^{dV})$
&
LB
&
$-0.9$
&
$-0.2$
&
$-0.9$
&
$-0.3$
&
$-0.9$
&
$-0.3$
\\
&
UB
&
$-0.9$
&
$0.2$
&
$-0.9$
&
$0.3$
&
$-0.9$
&
$0.3$
\\
\hline
$(\varepsilon_{\mu\tau}^{uV},\varepsilon_{\mu\tau}^{dV})$
&
LB
&
$-0.9$
&
$-0.2$
&
$-0.9$
&
$-0.3$
&
$-0.9$
&
$-0.3$
\\
&
UB
&
$-0.9$
&
$0.2$
&
$-0.9$
&
$0.3$
&
$-0.9$
&
$0.3$
%\\
%\hline
\end{tabular}
\end{center}
\caption{ \label{tab:ab}
Values of the parameters $a$ and $b$ in Eq.~(\ref{boundaries}) of the lower (LB) and upper (UB)
boundaries of the allowed regions in the right panels of Figures~\ref{fig:eeu-eed}--\ref{fig:mtu-mtd}.
}
\end{table*}

\begin{table*}[!h]
\begin{center}
\renewcommand{\arraystretch}{1.2}
\begin{tabular}{c|cccc|cccc}
&
\multicolumn{4}{c|}{Spectrum}
&
\multicolumn{4}{c}{Spectrum and Time}
\\
&
Best Fit
&
$1\sigma$
&
$2\sigma$
&
$3\sigma$
&
Best Fit
&
$1\sigma$
&
$2\sigma$
&
$3\sigma$
\\
\hline
$\varepsilon_{ee}^{uV}$
&
$0.02$
&
$ -0.18 \div 0.56 $
&
$ -0.23 \div 0.62 $
&
$ -0.28 \div 0.66 $
&
$0.20$
&
$ 0.01 \div 0.38 $
&
$ -0.08 \div 0.47 $
&
$ -0.15 \div 0.54 $
\\
%\hline
$\varepsilon_{\mu\mu}^{uV}$
&
$0.18$
&
$ -0.08 \div 0.44 $
&
$ -0.13 \div 0.51 $
&
$ -0.17 \div 0.56 $
&
$0.20$
&
$ -0.06 \div 0.45 $
&
$ -0.10 \div 0.48 $
&
$ -0.13 \div 0.52 $
\\
%\hline
$|\varepsilon_{e\mu}^{uV}|$
&
$0.04$
&
$ < 0.22 $
&
$ < 0.26 $
&
$ < 0.29 $
&
$0.04$
&
$ < 0.19 $
&
$ < 0.23 $
&
$ < 0.26 $
\\
%\hline
$|\varepsilon_{e\tau}^{uV}|$
&
$0.16$
&
$ < 0.37 $
&
$ < 0.42 $
&
$ < 0.47 $
&
$0.04$
&
$ < 0.19 $
&
$ < 0.28 $
&
$ < 0.35 $
\\
%\hline
$|\varepsilon_{\mu\tau}^{uV}|$
&
$0.04$
&
$ < 0.26 $
&
$ < 0.32 $
&
$ < 0.37 $
&
$0.12$
&
$ < 0.26 $
&
$ < 0.29 $
&
$ < 0.32 $
%\\
%\hline
\end{tabular}
\end{center}
\caption{ \label{tab:up}
Marginalized constraints on the NSI parameters
obtained with the analyses of the spectral and the joint spectral and temporal
COHERENT data assuming NSI with the up quark only.
}
\end{table*}

\begin{table*}[!h]
\begin{center}
\renewcommand{\arraystretch}{1.2}
\begin{tabular}{c|cccc|cccc}
&
\multicolumn{4}{c|}{Spectrum}
&
\multicolumn{4}{c}{Spectrum and Time}
\\
&
Best Fit
&
$1\sigma$
&
$2\sigma$
&
$3\sigma$
&
Best Fit
&
$1\sigma$
&
$2\sigma$
&
$3\sigma$
\\
\hline
$\varepsilon_{ee}^{dV}$
&
$0.17$
&
$ -0.16 \div 0.52 $
&
$ -0.20 \div 0.56 $
&
$ -0.25 \div 0.60 $
&
$0.18$
&
$ 0.01 \div 0.34 $
&
$ -0.07 \div 0.42 $
&
$ -0.14 \div 0.49 $
\\
%\hline
$\varepsilon_{\mu\mu}^{dV}$
&
$0.17$
&
$ -0.07 \div 0.41 $
&
$ -0.12 \div 0.47 $
&
$ -0.16 \div 0.51 $
&
$0.18$
&
$ -0.06 \div 0.41 $
&
$ -0.08 \div 0.44 $
&
$ -0.12 \div 0.47 $
\\
%\hline
$|\varepsilon_{e\mu}^{dV}|$
&
$0.04$
&
$ < 0.20 $
&
$ < 0.23 $
&
$ < 0.26 $
&
$0.04$
&
$ < 0.17 $
&
$ < 0.21 $
&
$ < 0.24 $
\\
%\hline
$|\varepsilon_{e\tau}^{dV}|$
&
$0.16$
&
$ < 0.34 $
&
$ < 0.38 $
&
$ < 0.43 $
&
$0.04$
&
$ < 0.17 $
&
$ < 0.25 $
&
$ < 0.31 $
\\
%\hline
$|\varepsilon_{\mu\tau}^{dV}|$
&
$0.04$
&
$ < 0.24 $
&
$ < 0.30 $
&
$ < 0.33 $
&
$0.11$
&
$ < 0.23 $
&
$ < 0.26 $
&
$ < 0.29 $
%\\
%\hline
\end{tabular}
\end{center}
\caption{ \label{tab:down}
Marginalized constraints on the NSI parameters
obtained with the analyses of the spectral and the joint spectral and temporal
COHERENT data assuming NSI with the down quark only.
}
\end{table*}

\begin{table*}[!h]
\begin{center}
\renewcommand{\arraystretch}{1.2}
\begin{tabular}{c|ccc|ccc}
&
\multicolumn{3}{c|}{Spectrum}
&
\multicolumn{3}{c}{Spectrum and Time}
\\
&
$1\sigma$
&
$2\sigma$
&
$3\sigma$
&
$1\sigma$
&
$2\sigma$
&
$3\sigma$
\\
\hline
$\varepsilon_{ee}^{uV}$
&
$
\left(
\begin{array}{c}
-0.09\div0.03
\\
0.36\div0.48
\end{array}
\right)
$
&
$
\left(
\begin{array}{c}
-0.15\div0.17
\\
0.23\div0.54
\end{array}
\right)
$
&
$-0.21\div0.60$
&
$
\left(
\begin{array}{c}
-0.02\div0.18
\\
0.21\div0.41
\end{array}
\right)
$
&
$-0.08\div0.47$
&
$-0.15\div0.53$
\\
%\hline
$\varepsilon_{\mu\mu}^{uV}$
&
$
\left(
\begin{array}{c}
-0.06\div0.03
\\
0.37\div0.44
\end{array}
\right)
$
&
$
\left(
\begin{array}{c}
-0.10\div0.08
\\
0.31\div0.49
\end{array}
\right)
$
&
$-0.15\div0.53$
&
$
\left(
\begin{array}{c}
-0.03\div0.03
\\
0.37\div0.42
\end{array}
\right)
$
&
$
\left(
\begin{array}{c}
-0.07\div0.06
\\
0.33\div0.46
\end{array}
\right)
$
&
$
\left(
\begin{array}{c}
-0.11\div0.10
\\
0.29\div0.49
\end{array}
\right)
$
\\
%\hline
$|\varepsilon_{e\mu}^{uV}|$
&
$<0.13$
&
$<0.17$
&
$<0.22$
&
$<0.09$
&
$<0.14$
&
$<0.19$
\\
%\hline
$|\varepsilon_{e\tau}^{uV}|$
&
$<0.21$
&
$<0.29$
&
$<0.36$
&
$<0.12$
&
$<0.21$
&
$<0.28$
\\
%\hline
$|\varepsilon_{\mu\tau}^{uV}|$
&
$<0.16$
&
$<0.22$
&
$<0.28$
&
$<0.12$
&
$<0.18$
&
$<0.23$
\\
%\hline
$\varepsilon_{ee}^{dV}$
&
$
\left(
\begin{array}{c}
-0.09\div0.03
\\
0.33\div0.43
\end{array}
\right)
$
&
$
\left(
\begin{array}{c}
-0.14\div0.15
\\
0.21\div0.49
\end{array}
\right)
$
&
$-0.19\div0.55$
&
$
\left(
\begin{array}{c}
-0.02\div0.17
\\
0.18\div0.37
\end{array}
\right)
$
&
$-0.07\div0.43$
&
$-0.13\div0.48$
\\
%\hline
$\varepsilon_{\mu\mu}^{dV}$
&
$
\left(
\begin{array}{c}
-0.05\div0.02
\\
0.33\div0.40
\end{array}
\right)
$
&
$
\left(
\begin{array}{c}
-0.09\div0.07
\\
0.28\div0.44
\end{array}
\right)
$
&
$-0.13\div0.48$
&
$
\left(
\begin{array}{c}
-0.03\div0.02
\\
0.33\div0.38
\end{array}
\right)
$
&
$
\left(
\begin{array}{c}
-0.06\div0.05
\\
0.30\div0.41
\end{array}
\right)
$
&
$
\left(
\begin{array}{c}
-0.09\div0.09
\\
0.27\div0.45
\end{array}
\right)
$
\\
%\hline
$|\varepsilon_{e\mu}^{dV}|$
&
$<0.12$
&
$<0.16$
&
$<0.20$
&
$<0.08$
&
$<0.13$
&
$<0.17$
\\
%\hline
$|\varepsilon_{e\tau}^{dV}|$
&
$<0.19$
&
$<0.26$
&
$<0.33$
&
$<0.11$
&
$<0.19$
&
$<0.25$
\\
%\hline
$|\varepsilon_{\mu\tau}^{dV}|$
&
$<0.15$
&
$<0.20$
&
$<0.25$
&
$<0.11$
&
$<0.16$
&
$<0.21$
%\\
%\hline
\end{tabular}
\end{center}
\caption{ \label{tab:1}
Allowed intervals for each of the NSI parameter
assuming it to be the only non-vanishing one.
Disconnected intervals are grouped in parentheses.
}
\end{table*}

% Figures

\end{document}